\documentclass[pdflatex,sn-mathphys-ay]{sn-jnl}

\usepackage{graphicx}%
\usepackage{multirow}%
\usepackage{amsmath,amssymb,amsfonts}%
\usepackage{amsthm}%
\usepackage{mathrsfs}%
\usepackage[title]{appendix}%
\usepackage{xcolor}%
\usepackage{textcomp}%
\usepackage{manyfoot}%
\usepackage{booktabs}%
\usepackage{algorithm}%
\usepackage{algorithmicx}%
\usepackage{algpseudocode}%
\usepackage{listings}%
\usepackage{braket}
\usepackage{caption} 
\usepackage{subcaption} 
\usepackage{mathtools}
\usepackage{hhline}
\usepackage{tikz}
\usepackage{graphicx}  
\usepackage{bm}
\usepackage{ulem}
\usepackage{tabularx} 

\usetikzlibrary{arrows.meta}

\theoremstyle{thmstylethree}%

\newcommand{\bigtimes}{\mathop{\times}\displaylimits}
\renewcommand{\ket}[1]{{| #1 \rangle}}
\renewcommand{\bra}[1]{ \langle #1 |}
\newcommand{\ketbra}[2]{| #1 \rangle\!\langle #2 |}

\newtheorem{corollary}{Corollary}
\newtheorem{proposition}{Proposition}
\newtheorem{example}{Example}
\newtheorem{lemma}{Lemma}
\newtheorem{definition}{Definition}

\newcounter{pr}
\newtheorem{proofprop}[pr]{Proof of Proposition}

\newtheorem{prooflemma}[pr]{Proof of Lemma}

\newtheorem*{remark}{Remark}

\renewcommand{\cite}{\citep}

\begin{document}

\title[Article Title]{Quantum Machine Learning in Multi-Qubit Phase-Space Part I: Foundations}



\author*[1,2]{\fnm{Timothy} \sur{Heightman}}\email{timothy.heightman@icfo.eu}
\author[1]{\fnm{Edward} \sur{Jiang}}
\author[1]{\fnm{Ruth} \sur{Mora-Soto}}
\author[1,3]{\fnm{Maciej} \sur{Lewenstein}}
\author[1,4]{\fnm{Marcin} \sur{P\l{}odzie\'{n}}}


\affil[1]{\orgdiv{ICFO-Institut de Ciencies Fotoniques}, \orgname{The Barcelona Institute of Science and Technology}, \orgaddress{\city{Castelldefels}, \postcode{08860}, \state{Barcelona}, \country{Spain}}}

\affil[2]{\orgname{Quside Technologies SL}, \orgaddress{\street{Carrer d’Esteve Terradas, 1}, \city{Castelldefels}, \postcode{08860}, \state{Barcelona}, \country{Spain}}}

\affil[3]{\orgname{ICREA}, \orgaddress{\street{Passeig Lluis Companys 23}, \postcode{08010}, \city{Barcelona}, \country{Spain}}}

\affil[4]{\orgname{Qilimanjaro Quantum Tech}, \orgaddress{\street{Carrer de Veneçuela 74}, \postcode{08019}, \city{Barcelona}, \country{Spain}}}


\abstract{Quantum machine learning (QML) seeks to exploit the intrinsic properties of quantum mechanical systems, including superposition, coherence, and quantum entanglement for classical data processing. However, due to the exponential growth of the Hilbert space, QML faces practical limits in classical simulations with the state-vector representation of quantum system. On the other hand, phase-space methods offer an alternative by encoding quantum states as quasi-probability functions.  Building on prior work in qubit phase-space and the Stratonovich-Weyl (SW) correspondence, we construct a closed, composable dynamical formalism for one- and multi-qubit systems in phase-space. This formalism replaces the operator algebra of the Pauli group with function dynamics on symplectic manifolds, and recasts the curse of dimensionality in terms of harmonic support on a domain whose dimension scales linearly with the number of qubits. It opens a new route for QML based on variational modelling over phase-space.}




\maketitle

\section{Introduction}

A central premise of Quantum Machine Learning (QML) is to combine classical machine learning and deep learning strategies with data encoded into quantum mechanical systems \cite{lewenstein1991quantum, lewenstein1994quantum, kak1995quantum, lagaris1997artificial, zak1998quantum, narayanan2000quantum, ventura2000quantum, schuld2014quest}.   It currently stands as a major research direction at the cross-section of Artificial Intelligence (AI) and Quantum Computing (QC) \cite{schuld2015introduction, biamonte2017quantum,benedetti2019parameterized,perez2020data,havlivcek2019supervised,schuld2019quantum,acampora2025quantum,Dawid2025,beer2020training, farhi2018classification,tacchino2019artificial}. This is due to the exponential growth of Hilbert space dimension opening the possibility of efficient embedding of classical data \cite{ lloyd2020quantum, naguleswaran2024quantum,schnabel2025quantum}. While this scaling is often presented as a resource, it also poses practical challenges. One such challenge, is that optimization over quantum states, particularly in variational models, becomes unwieldy. This is because describing a quantum system of many qubits with a state-vector wavefunction requires an exponentially large basis, and automatic differentiation (AD) over such objects is both memory- and computation-intensive \cite{ragone2024lie, larocca2024review}. Since modern AI is built on gradient-based optimization, this creates a tension between representation of quantum systems and classical training techniques. This motivates the search for alternative representations of quantum systems in which gradients, sampling, and composition are more tractable.

Historically, researchers have pursued a phase-space representation of quantum systems that render sampling, gradients, and composition more tractable.
Phase-space methods are a ubiquitous tool for continuous-variable quantum systems \cite{schleich2015quantum, klimov2009group}, with quasi-probability functions like the Wigner-, Husimi- \cite{husimi1940some}, and Glauber–Sudarshan- functions \cite{cohen1966generalized, sudarshan1963equivalence, glauber1963coherent} grounded in symplectic geometry. This representation has tools for measurement \cite{leonhardt1997measuring}, dissipation \cite{breuer2002theory}, classical limits \cite{schleich2015quantum}, sampling \cite{banaszek1997accuracy}, and simulation \cite{becca2017quantum}. Early work by Weidlich, Risken, and Haken \cite{haken1967quantum} developed phase-space methods for two-level atoms using characteristic functions, laying a foundation for master equation approaches to quantum optics. Haake and Lewenstein \cite{haake1983adiabatic, haake1982convolution} refined this via adiabatic expansions and convolution identities for bosonic quasiprobabilities, emphasizing ordering and diffusion structure. Later, Lewenstein \cite{lewenstein1994quantum} and collaborators \cite{gratsea2024storage, lewenstein2021storage,lewenstein1992storage} explored dynamical models of quantum neural networks using completely positive trace preserving maps and non-linear activation analogues to study storage and learning capacities of quantum systems. More recently, QML researchers studying continuous-variable quantum systems, like quantum optics, have returned to  the phase space formalism \cite{dugan2023q, hahmgenerative, rundle2021overview, huber2020phase}.  Yet, a general phase-space approach in qubit-based QML remains underdeveloped and fragmented, especially over finite-dimensional Hilbert spaces like qubits and qudits.

Several studies have developed phase-space representations for single-qubit systems \cite{brif1999phase, varilly1989moyal, klimov2002evolution, huber2020phase}. Brif et al. \cite{brif1999phase} developed an SW correspondence \cite{stratonovich1956jetp} for compact Lie groups, including $SU(2)$, yielding spin-$j$ quasidistributions and a Moyal star product. However, their approach applies only to one-body systems, and lacks mechanisms for composition, partial tracing, or dissipation. Similarly, Várilly et al. \cite{varilly1989moyal, varilly1989stratonovich} used $SU(2)$ Fourier analysis to define exact Moyal products via spherical harmonics, but without tools for correlations or subsystem dynamics. Other approaches use symmetric spin sums or bosonic mappings \cite{sanchez2025phase, klimov2017generalized}, or interpret qubit phase-space geometrically as a Kähler manifold \cite{klimov2017generalized, perelomov2002coherentstatesarbitrarylie, el1996supercoherent, onofri1975note, carosso2018geometricquantization, klimov2011qutrit} arising from Lie group cosets. While these reveal structural features, they offer no constructive dynamics beyond semi-classical approximations, and break down under entanglement or dissipation. Despite this long-standing foundation for single qubits, the extension to many-qubit systems has remained largely unexplored, with only recent efforts beginning to revisit the problem in earnest amid growing interest from quantum optics and quantum information communities \cite{navarette2025phase, rundle2021overview}.

\begin{table}[t!]
  \centering
  \small                              
  \setlength{\tabcolsep}{8pt}
  \renewcommand{\arraystretch}{2.2}

  \begin{tabular}{||c|c|c||}
    \hline
    \textbf{Object} &
    \textbf{Hilbert Space} &
    \textbf{Phase-Space} \\ \hline\hline

    Operators & $\hat A$ & $f^{(s)}_{\hat A}(\boldsymbol{\Omega})$ \\ \hline
    
    Commutators & $-i[\hat A,\hat B]$ & $[\![f^{(s)}_{\hat A}(\boldsymbol{\Omega}),f^{(s)}_{\hat B}(\boldsymbol{\Omega})]\!]$ \\ \hline

    Anti-commutators & $\{\hat A,\hat B\}$ & $\{\!\!\{f^{(s)}_{\hat A}(\boldsymbol{\Omega}),f^{(s)}_{\hat B}(\boldsymbol{\Omega})\}\!\!\}$ \\ \hline

    Dual & $\hat A^\dagger$ & $f^{(-s)}_{\hat A}(\boldsymbol{\Omega})$ \\ \hline

    Inner product & $\operatorname{Tr}\left[\hat A^\dagger \hat B\right]$ & $\displaystyle\int_{(S^2)^N}d\boldsymbol{\Omega}\,f^{(-s)}_{\hat A}(\boldsymbol{\Omega})f_{\hat B}^{(s)}(\boldsymbol{\Omega})$\\ \hline

    Tensor product & $\hat A^{(1)}\otimes \hat B^{(2)}$ & $f^{(s)}_{\hat A}(\boldsymbol{\Omega}^{(1)})f^{(s)}_{\hat B}(\boldsymbol{\Omega}^{(2)})$ \\ \hline

    Partial trace & $\operatorname{Tr}_2\left[\hat A^{(12)}\right]$ & $\displaystyle\int_{(S^2)^N}d\boldsymbol{\Omega}^{(2)}\,f_{\hat A}^{(s)}(\boldsymbol{\Omega}^{(1)}, \boldsymbol{\Omega}^{(2)})$ \\ \hline

    Unitary evolution & $\dfrac{\partial\hat\rho}{\partial t}=-i[\hat H,\hat\rho]$ & $\dfrac{\partial f^{(s)}_{\hat \rho}}{\partial t}=[\![f^{(s)}_{\hat H},f^{(s)}_{\hat \rho}]\!]$ \\ \hline
  \end{tabular}
    \medskip
  \caption{Comparison between the representation of various objects in the Hilbert space and phase-space formalisms. Operators in the Hilbert space are represented by quasi-probability distributions in phase-space, and mathematical operations (such as commutators or inner products) on Hilbert space operators have analogues in phase-space. The state of a many-qubit system, expressed as a density matrix $\hat\rho$, is a special case of such a quasi-probability distribution for which we can describe dynamics using the phase-space equivalent of the von-Neumann equation. For details, see main text.
  }\label{table:1}
\end{table}

In this work, we explore a compact phase-space formalism for many-qubit systems based on smooth product manifolds. Qubit states are represented as $s$-parametrised quasi-probability functions on $(S^2)^N$, extending techniques from quantum optics \cite{schleich2015quantum} to qubit systems. We develop the algebraic structure of this representation using the Stratonovich–Weyl (SW) correspondence \cite{stratonovich1956jetp}, Moyal star products \cite{Moyal1949}, and associated sine and cosine brackets. These operations generate a closed Jordan–Lie algebra on the image of the SW map and govern the dynamics of observables under unitary and dissipative evolution. Our construction highlights the symmetry properties of the Moyal bracket, its stability under unitary flows, and its relation to the classical Poisson bracket. We also show that the SW kernel can always be factorized into tensor products of local kernels. Even in open-system scenarios, by incorporating Stinespring's dilation theorem into the phase-space framework, this fact remains true. This enables a fully composable formalism that supports partial tracing, dilation, observable estimation, and the construction of many-body systems from local phase-space components. 

Furthermore, we emphasise the prospects of QML in phase-space as an alternative to Hilbert space in the context of deep learning. 
The contemporary approach to AI is based on deep neural networks, which in principle, are highly non-linear variational functions of their input parameters. Here, non-linearity is a foundation that underpins the expressiveness of a neural network architecture. In the typical formulation of QML, state vectors and operators live in a finite-dimensional Hilbert space. The description of a quantum system is given by a linear superposition of orthonormal basis vectors, with the dimension of the basis increasing exponentially in the number of qubits. As such, QML in the Hilbert space representation of a quantum systems lacks natural non-linearity of the wave function parameters, because these are coefficients of a linear superposition of the basis. Furthermore, we are tasked with handling an exponentially large domain per the curse of dimensionality. Phase-space, in contrast, allows a quantum system to be represented as a non-linear function whose input dimension \textit{scales linearly} in the number of qubits. This aligns more closely with neural-network models as non-linear function approximators over a domain with better scaling than Hilbert space. The nature of non-linearity of phase-space functions remains real-valued, smooth, and geometrically structured, making them compatible with gradient-based optimization. Observables can also be estimated by sampling or automatic differentiation, and variational models can be defined over function families rather than parametrised wavefunctions. We emphasise that this approach does not eliminate the curse of dimensionality, but reframes it. Complexity in phase-space is set by harmonic support, defined as the number of modes in a spherical harmonic expansion of the phase-space function describing a quantum system under consideration. This sits in contrast with Hilbert space, in which the curse manifests in the exponential dimension of the space itself. The overview of proposed formalism, and its comparison to the Hilbert space representation, is presented in Table \ref{table:1}. 

This paper is the first in a series of works developing a comprehensive phase-space approach to QML over finite-dimensional systems. Our goal is to establish the mathematical foundations and operational tools needed to apply phase-space methods to scalable, qubit-based QML models. Here we will primarily focus on the geometric and algebraic structure of the representation, laying the groundwork for future extensions to neural architectures, sampling techniques, and learning algorithms for both states and dynamics. While the formal ingredients of our approach such as the Stratonovich-Weyl correspondence~\cite{stratonovich1956jetp,varilly1989stratonovich,brif1999phase}, $s$-parametrised kernels~\cite{klimov2017generalized}, and Moyal brackets~\cite{varilly1989moyal}, are well known, our contributions lie in assembling these into a coherent, fully composable phase-space framework suitable for QML. We define a scalable representation over $(S^2)^N$ that supports dynamics, marginalisation, and dilation without recourse to Hilbert space. We recast quantum complexity in terms of harmonic support, and we identify quasi-probabilities as native function spaces for variational models and gradient-based learning~\cite{schuld2019quantum,rezende2020normalizing}. We also study its representation-theoretic properties in detail, showing that the $Q$-function realisation yields the most stable and differentiable structure for machine learning applications.

The rest of this work is structured as follows:
Sec.~\ref{sec:spin_phase_space} develops the phase-space representation of qubit states, including SW kernels, partial traces, and expectation algebra. Sec.~\ref{sec:spin_dynamics} introduces the dynamical theory via Moyal brackets and derives the von Neumann and Lindblad equations. Sec.~\ref{sec:dilation_unification} then offers a unified perspective on quantum dynamics in both representations, visualising the key concepts of this work through commuting diagrams. Here, we will prove that the SW kernel factorises for arbitrary dynamics, by realising Stinespring's theorem in phase-space through an extended SW kernel. We will then explain how to compute expectation values through moment generating functions (MGFs). Sec.~\ref{sec:discussion} then offers a discussion on the curse of dimensionality in this representation, and how its recasting could unlock new methods for QML in phase-space that are better-suited to neural approximation techniques. We also discuss the prospects for automatic differentiation in the phase-space picture in light of MGFs. Proofs of propositions and lemmas can be found in Appendix~\ref{app:proofs}, whilst instructive examples are given throughout the text. Readers unfamiliar with the phase-space picture of bosonic states are invited to consult Appendix~\ref{app:bosonic_phase_space}.

\section{Phase-Space of Qubits}
\label{sec:spin_phase_space}

In this section, we begin by reviewing the phase-space formulation of single qubit systems \cite{brif1999phase, varilly1989moyal, klimov2002evolution, huber2020phase, varilly1989stratonovich}, and then extend it to many-body settings. We first present the Stratonovich–Weyl (SW) correspondence \cite{stratonovich1956jetp}, which maps Hilbert space operators to functions on a smooth manifold, which for qubits is canonically the sphere $S^2$. The correspondence is indexed by a parameter $s \in [-1, 1]$, and different values of $s$ yield quasi-probability representations with distinct analytic and algebraic properties, most notably the spin-analogues of the $Q$-, $W$-, and $P$-functions (see Appendix~\ref{app:bosonic_phase_space}).

We then proceed to show how different quasi-probability functions are related, and extend them to the many-body case. Finally, we explore how algebraic operations, separability, and rank behave in phase-space, laying the foundation for dynamics in Sec.~\ref{sec:spin_dynamics}, including an $s$-parametrised definition of the $\star$-product.

\subsection{Single Qubit Stratonovich–Weyl (SW) Correspondence}
\label{sec:single_spin_sw}

The physical state space of a qubit is the complex projective line $\mathbb{CP}^1$, which arises from the Hilbert space $\mathcal{H}_2 \cong \mathbb{C}^2$ via quotienting by global phases and norm, $S^3/U(1) \cong \mathbb{CP}^1 \cong S^2$. Geometrically, $\mathbb{CP}^1$ is diffeomorphic to the two-sphere $S^2$, which thus serves as the canonical phase-space for a qubit.

To represent quantum states and observables directly on this manifold, we can employ the SW correspondence \cite{stratonovich1956jetp,varilly1989moyal}. This is an invertible, symmetry-respecting map between operators $\hat{A}$ on $\mathcal{H}_2$ and real-valued functions $f_{\hat{A}} : S^2 \rightarrow \mathbb{R}$, called phase-space functions or quasi-probability distributions.

To characterize this map, Stratonovich and Weyl proposed a set of axioms that encode the minimal structural requirements for a classical-like representation of quantum operators. They read as follows in the context of a qubit:

\subsubsection{Axioms}
Let us denote the manifold $\mathcal{M} = S^2$, equipped with its standard Haar measure $d\mu(\theta, \varphi) = \sin\theta\, d\theta\, d\varphi$. An SW correspondence is a map $\hat{A} \mapsto f_{\hat{A}}$ satisfying:

\begin{itemize}
    \item Linearity: The map is linear and invertible, \( f_{\alpha \hat{A} + \beta \hat{B}} = \alpha f_{\hat{A}} + \beta f_{\hat{B}}, \quad \forall \alpha, \beta \in \mathbb{C}.\)

    \item Reality: If $\hat{A}$ is Hermitian, then $f_{\hat{A}}$ is real-valued.

    \item Standardization: The trace of $\hat{A}$ is given by \(\mathrm{Tr}[\hat{A}] = \int_{S^2} f_{\hat{A}}(\theta, \varphi)\, d\mu(\theta, \varphi).\)
    \item Traciality: The Hilbert-Schmidt inner product is preserved, \(\mathrm{Tr}[\hat{A}^{\dagger} \hat{B}] = 4\pi\int_{S^2} f_{\hat{A}}(\theta, \varphi)\, f_{\hat{B}}(\theta, \varphi)\, d\mu(\theta, \varphi).\)
    \item Covariance: For any $g \in SU(2)$ and point $x \in S^2$, the symbol transforms as \(f_{\hat{U}_g \hat{A} \hat{U}_g^\dagger}(x) = f_{\hat{A}}(g^{-1} \cdot x),\)
    where $\hat{U}_g$ is the unitary representation of $g$ and $g \cdot x$ denotes the action of $g$ on $x \in S^2$.
\end{itemize}

As originally emphasized by Stratonovich~\cite{stratonovich1956jetp}, the linearity and traciality conditions carry a clear statistical role. Linearity means the mapping \( \hat{A} \mapsto W_{\hat{A}} \) respects superposition. Meanwhile traciality guarantees that expectation values become phase-space integrals, so that the quasi-probability symbols preserve the inner product structure of operator space.

The reality condition ensures that \( f_{\hat{A}}(x) \in \mathbb{R} \) when \( \hat{A} \) is self-adjoint, aligning measurable observables with real-valued functions. While the standardization condition enforces normalization by resolving the identity in phase-space. Finally, covariance preserves symmetry transformations of the quantum system, with compatible transformations of the phase-space. Together, these five axioms define the SW symbol of an operator, which is found using the so-called SW kernel whose formulation is described below.

\subsubsection{Kernel formulation}
The mapping of the SW correspondence is realized through an operator-valued kernel $\hat{\Delta}(\theta, \varphi)$, defined on $S^2$. Given such a kernel, the phase-space function of an operator $\hat{A}$, $f_{\hat{A}}(\theta, \varphi)$, is
\begin{align} \label{eq:SW_correspondence}
f_{\hat{A}}(\theta, \varphi) &= \mathrm{Tr}\left[\hat{A} \hat{\Delta}(\theta, \varphi)\right],
\end{align}
with the inverse given by
\begin{align}
\hat{A} &= 4\pi\int_{S^2} f_{\hat{A}}(\theta, \varphi)\, \hat{\Delta}(\theta, \varphi)\, d\mu(\theta, \varphi),
\end{align}
This duality holds provided the kernel satisfies the following properties:
\begin{itemize}
    \item Hermiticity: $\hat{\Delta}(\theta, \varphi)^\dagger = \hat{\Delta}(\theta, \varphi)$.
    
    \item Normalization: $\int_{S^2} \hat{\Delta}(\theta, \varphi)\, d\mu(\theta, \varphi) = \hat{\mathbb{I}}$.
    
    \item Reproducing Property: \(4\pi\int_{S^2} \mathrm{Tr}\left[\hat{\Delta}(x') \hat{\Delta}(x)\right]\, \hat{\Delta}(x)\, d\mu(x) = \hat{\Delta}(x').\)
    
    \item Covariance: $\hat{\Delta}(g \cdot x) = \hat{U}_g \hat{\Delta}(x) \hat{U}_g^\dagger$ for all $g \in SU(2)$.
\end{itemize}

The kernel $\hat{\Delta}(\theta, \varphi)$ therefore encodes the geometry of the phase-space, the symmetries of the quantum system, and the algebraic structure of observables. Through it, expectation values, state reconstruction, and even dynamics can be expressed entirely in terms of smooth functions on the sphere $S^2$.

Importantly, the Stratonovich-Weyl axioms do not uniquely fix the kernel. Instead, they admit a continuous family of solutions, labelled by a parameter $s \in [-1,1]$, which governs the ordering prescription used to map operators to phase-space functions. In quantum optics, this $s$-index interpolates between the Glauber-Sudarshan $P$ function ($s = +1$), the Wigner function ($s = 0$), and the Husimi $Q$ function ($s = -1$), each corresponding to a different quantization scheme \cite{bayen1978deformationI, woodhouse1992geometric}. This structure carries over to qubit systems, allowing us to define qubit-analogues of these quasi-probability distributions. We now construct these kernels explicitly.

\subsection{Qubit Quasi-probability}
\label{sec:single_spin_qwp}

Having seen $S^2$ as the canonical phase space in Sec.~\ref{sec:single_spin_sw}, we can parametrise its coordinates with respect to the computational basis via the coherent spin states,
\begin{align}
\ket{\theta,\varphi} = \cos\left(\tfrac{\theta}{2}\right)\ket{0} + e^{i\varphi}\sin\left(\tfrac{\theta}{2}\right)\ket{1},
\end{align}
with the usual spherical coordinates $(\theta,\varphi) \in S^2$. These coordinates form a natural basis for constructing kernel operators. Each $s$-parametrised SW kernel then takes the form 

\begin{align}
\hat{\Delta}^{(s)}=\frac{1}{2\pi}\left(\sqrt{3}^{1+s}\ketbra{\theta,\varphi}{\theta,\varphi}-\frac{\sqrt{3}^{1+s}-1}{2}\hat{\mathbb{I}}\right).
\end{align}

This defines three canonical representatives of the SW family, each corresponding to a different smoothing of operator content onto phase-space. Given a density matrix $\hat\rho \in \beta_{+}(\mathcal{H}_2)$, where $\beta_+(\mathcal{H}_2)$ is the space of positive semidefinite operators on $\mathcal{H}_2$, the corresponding $s$-parametrised quasi-probability function is defined by
\begin{align}
f^{(s)}_{\hat\rho}(\theta, \varphi) = \mathrm{Tr}[\hat\rho\, \hat{\Delta}^{(s)}(\theta,\varphi)].
\end{align}
These functions are real-valued, normalized on $S^2$ and provide a continuous family of phase-space representations, analogous to the bosonic setting (see App.~\ref{app:bosonic_phase_space}).
Among these, the Wigner function ($s = 0$) is unique in satisfying the full SW postulates, including the traciality (i.e. reproducing) condition,
\begin{align}
4\pi \int_{S^2} \mathrm{Tr}\left[ \hat{\Delta}^{(0)}(\theta, \varphi)\, \hat{\Delta}^{(0)}(\theta', \varphi') \right] \hat{\Delta}^{(0)}(\theta', \varphi')\, d\Omega' = \hat{\Delta}^{(0)}(\theta, \varphi).
\end{align}
In contrast, the $P$ and $Q$ kernels ($s = \pm 1$) only respect the traciality condition in duality with one another. That is,
\begin{align}
4\pi\int_{S^2}\operatorname{Tr}\left[\hat{\Delta}^{(s)}(\theta,\varphi)\hat{\Delta}^{(-s)}(\theta',\varphi')\right]\hat{\Delta}^{(s)}(\theta',\varphi')\,d\Omega'&=\hat{\Delta}^{(s)}(\theta,\varphi)
\end{align}

This is because the SW kernel is not uniquely fixed by symmetry and linearity alone; the choice of $s$ corresponds to a choice of quantization rule over $SU(2)$ kernels, analogous to normal, symmetric, or antinormal ordering in the bosonic setting (see App.~\ref{app:bosonic_phase_space}).

Though the phase-space picture is self-contained, such expressions allow direct comparison with standard density-matrix techniques and clarify how information is encoded in each representation. These forms are particularly helpful for those more familiar with the operator formalism, and will serve to ground our later generalizations. Beginning with $s = -1$,  we arrive at the $Q$ function,
\begin{align}
Q_{\hat\rho}(\theta,\varphi)=\frac{1}{2\pi}\bra{\theta,\varphi}\hat\rho\ket{\theta,\varphi},\label{eq:spin_q_func}
\end{align}
which we can express according to the following two propositions.

\begin{proposition}[Pauli expectation Q function expansion for single qubit]\label{prop:spin_q_func_exp}
The $Q$ function for a single qubit can be expressed equivalently as
\begin{align}
Q_{\hat\rho}(\theta,\varphi)=\frac{1}{4\pi} (\langle  \hat{\mathbb{I}}\rangle_{\hat\rho}+\langle\hat\sigma_x\rangle_{\hat\rho}\sin\theta\cos\varphi+\langle\hat\sigma_y\rangle_{\hat\rho}\sin\theta\sin\varphi+\langle\hat\sigma_z\rangle_{\hat\rho}\cos\theta),
\end{align}
where $\langle\hat{\mathbb{I}}\rangle_{\hat\rho},\langle\hat \sigma_x\rangle_{\hat\rho},\langle\hat \sigma_y\rangle_{\hat\rho},\langle\hat \sigma_z\rangle_{\hat\rho}$ denote the expectation values of the Pauli operators in state $\hat\rho$:
\begin{align}
\langle \hat{\mathbb{I}}\rangle_{\hat{\rho}}&=\operatorname{Tr}\left[\hat\rho\right],&\langle\hat \sigma_x\rangle_{\hat\rho}&=\operatorname{Tr}\left[\hat\rho\hat \sigma_x\right],&\langle\hat \sigma_y\rangle_{\hat\rho}&=\operatorname{Tr}\left[\hat\rho\hat \sigma_y\right],&\langle\hat \sigma_z\rangle_{\hat\rho}&=\operatorname{Tr}\left[\hat\rho\hat \sigma_z\right].
\end{align}
\end{proposition}

\begin{proposition}[Pauli expectation P function expansion for single spin]\label{prop:spin_p_func_exp}
The $P$ function of a qubit can be expressed equivalently as
\begin{align}
P_{\hat\rho}(\theta,\varphi)= \frac{1}{4\pi} (\langle \hat{\mathbb{I}}\rangle_{\hat\rho}+3\langle\hat \sigma_x\rangle_{\hat\rho}\sin\theta\cos\varphi+3\langle\hat \sigma_y\rangle_{\hat\rho}\sin\theta\sin\varphi+3\langle\hat \sigma_z\rangle_{\hat\rho}\cos\theta).
\end{align}
\end{proposition}

Lastly, the $W$ representation corresponds to the symmetric choice of SW kernel $(s=0)$, and was constructed explicitly by Várilly and Gracia-Bondía \cite{varilly1989moyal} as the unique solution satisfying full traciality. It yields the phase-space function
\begin{align}
W_{\hat\rho}(\theta,\varphi)=\frac{1}{4\pi}\left(\langle \hat{\mathbb{I}}\rangle_{\hat\rho}+\sqrt{3}\langle\hat \sigma_x\rangle_{\hat\rho}\sin\theta\cos\varphi+\sqrt{3}\langle\hat \sigma_y\rangle_{\hat\rho}\sin\theta\sin\varphi+\sqrt{3}\langle\hat \sigma_z\rangle_{\hat\rho}\cos\theta\right).
\end{align}

We observe that the $Q$, $P$, and $W$ representations differ only by the scaling factors of the non-constant terms. Finally, we also observe that  the phase-space picture can be related to the Bloch sphere in the following way. Expressing $\hat\rho = \tfrac{1}{2}(\hat{\mathbb{I}} + \vec{r}_{\hat\rho} \cdot \vec{\hat \sigma})$, one obtains
\begin{align}\label{eq:bloch_phase_function}
f^{(s)}_{\hat\rho(\theta, \varphi)} = \frac{1}{4\pi} \left(1 + \lambda(s)\, \vec{r}_{\hat\rho} \cdot \vec{n}(\theta,\varphi) \right) = \operatorname{Tr}(\hat{\rho} \hat{\Delta}^{(s)}),
\end{align}
where $\vec{n}(\theta, \varphi)$ is the unit vector on the Bloch sphere,
\begin{align}
\vec{n}(\theta, \varphi) = (\sin\theta\cos\varphi,\, \sin\theta\sin\varphi,\, \cos\theta),
\end{align}
and $\lambda(s)$ is a scalar weight determined by the $s$-index:
\begin{align} \label{eq:lambda_cases}
\lambda(s) = 
\begin{cases}
3, & s = +1 \quad\text{(P function)}, \\
\sqrt{3}, & s = 0 \quad\text{(Wigner function)}, \\
1, & s = -1 \quad\text{(Q function)}. \\
\end{cases}
\end{align}

As a first example, let us consider the Q-functions of the Pauli eigenstates:
\begin{example}[Pauli Eigenstates]
As an example, the $Q$ functions for the eigenstates of the Pauli operators are
\begin{align}
Q_{\ket{0}}(\theta,\varphi)&=\frac{1}{4\pi}\left(1+\cos\theta\right), & Q_{\ket{1}}(\theta,\varphi)&=\frac{1}{4\pi}\left(1-\cos\theta\right),\label{eq:q_func_z_states}\\
Q_{\ket{+}}(\theta,\varphi)&=\frac{1}{4\pi}\left(1+\sin\theta\cos\varphi\right), & Q_{\ket{-}}(\theta,\varphi)&=\frac{1}{4\pi}\left(1-\sin\theta\cos\varphi\right),\\
Q_{\ket{+i}}(\theta,\varphi)&=\frac{1}{4\pi}\left(1+\sin\theta\sin\varphi\right), & Q_{\ket{-i}}(\theta,\varphi)&=\frac{1}{4\pi}\left(1-\sin\theta\sin\varphi\right).
\end{align}
These are depicted in Fig.~\ref{fig:q_pauli_eigens}.
\begin{figure}
\captionsetup[subfigure]{aboveskip=0pt,belowskip=0pt}
\centering
\begin{subfigure}{0.49\textwidth}
    \centering
    \includegraphics[width=\textwidth]{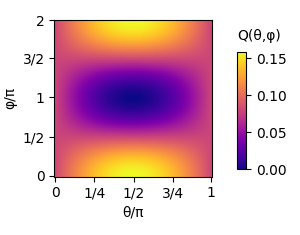}
    \caption{$\ket{+}$}
\end{subfigure}
\begin{subfigure}{0.49\textwidth}
    \centering
    \includegraphics[width=\textwidth]{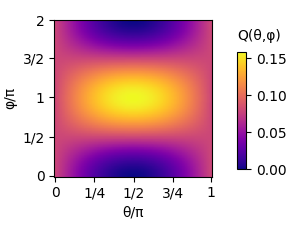}
    \caption{$\ket{-}$}
\end{subfigure}
\begin{subfigure}{0.49\textwidth}
    \centering
    \includegraphics[width=\textwidth]{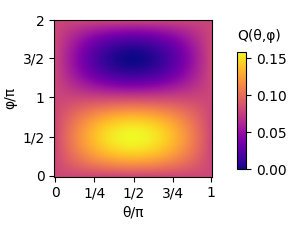}
    \caption{$\ket{{+}i}$}
\end{subfigure}
\begin{subfigure}{0.49\textwidth}
    \centering
    \includegraphics[width=\textwidth]{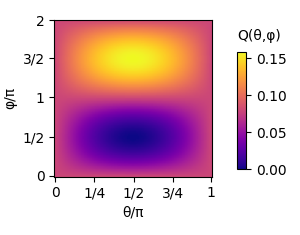}
    \caption{$\ket{{-}i}$}
\end{subfigure}
\begin{subfigure}{0.49\textwidth}
    \centering
    \includegraphics[width=\textwidth]{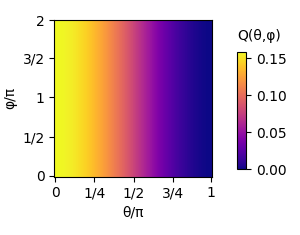}
    \caption{$\ket{0}$}
\end{subfigure}
\begin{subfigure}{0.49\textwidth}
    \centering
    \includegraphics[width=\textwidth]{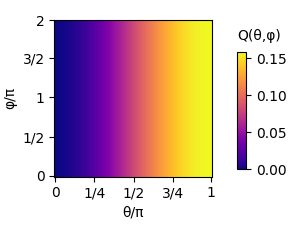}
    \caption{$\ket{1}$}
\end{subfigure}
\caption{$Q$ representations of the eigenstates of the Pauli operator $\hat\sigma_x$ (panels (a), (b)), $\hat\sigma_y$ (panels (c), (d)), and $\hat\sigma_z$ (panels (e), (f)).}
\label{fig:q_pauli_eigens}
\end{figure}

\end{example}

A common misconception holds that since the SW kernel uses the coherent basis, it must be insufficient to represent general mixed states. However, the $Q$-function construction does not rely on coherent-state expansion. It uses these states as probes \textit{through} the SW kernel, and this suffices to represent any density matrix. As shown in Examples~\ref{ex:classical_Q} and \ref{ex:q_func_thermal}, classical mixtures, and even thermal states admit smooth $Q$ functions over the sphere. 

\subsection{Relations between Phase-Space Representations}
\label{sec:qwp_relations}
Just like the bosonic case, qubit quasi-probabilities form a continuous family interpolating between different orderings of operators. In the bosonic case, this interpolation is governed by the complex Laplacian $\nabla^2_{\mathbb{C}}$, acting as a Weierstrass (Gaussian) convolution,
\begin{align}
f(\alpha,\alpha^*;s) &= \exp\left(-\tfrac{s}{2} \nabla^2_{\mathbb{C}}\right) W(\alpha,\alpha^*),
\end{align}
where $s = 1$, $0$, and $-1$ yield the $P$, Wigner, and $Q$ functions, respectively (see App.~\ref{app:bosonic_phase_space}).

\begin{figure}[h!]
\centering
\begin{tikzpicture}

\node (P) at (-4, 0) {$P(\theta,\varphi)$};
\node (W) at (0, 0) {$W(\theta,\varphi)$};
\node (Q) at (4, 0) {$Q(\theta,\varphi)$};

\draw [->] (W) to [out=150,in=30] node [above] {\small$\exp\left(-\frac{\log3}{4}\nabla^2_{S^2}\right)$} (P);
\draw [->] (P) to [out=330,in=210] node[below]{\small$\exp\left(\frac{\log3}{4}\nabla^2_{S^2}\right)$} (W);
\draw [->] (Q) to [out=150,in=30] node [above] {\small$\exp\left(-\frac{\log3}{4}\nabla^2_{S^2}\right)$} (W);
\draw [->] (W) to [out=330,in=210] node[below]{\small$\exp\left(\frac{\log3}{4}\nabla^2_{S^2}\right)$} (Q);

\end{tikzpicture}
\caption{Diffusion maps between quasi-probability representations for qubit systems.}
\label{fig:pwq_map_spin}
\end{figure}
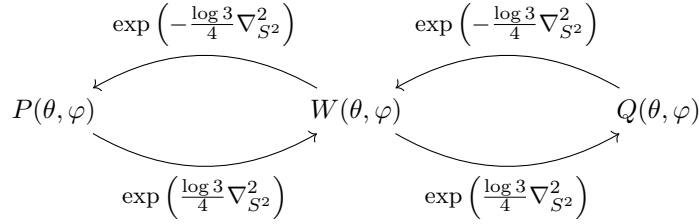

For qubit systems, an analogous structure holds, with the role of $\nabla^2_{\mathbb{C}}$ (see App.\ref{app:bosonic_phase_space} and Fig.~\ref{fig:pwq_map_cv}) played by the Laplace–Beltrami operator $\nabla^2_{S^2}$ on the two-sphere. The $s$-parametrised family of distributions is given by:
\begin{align}
f(\theta,\varphi; s) &= \exp\left(-\frac{s \log 3}{4} \nabla^2_{S^2} \right) W(\theta,\varphi),
\label{eq:change_rep_spins}
\end{align}
with $s = 1$, $0$, $-1$ corresponding again to the $P$, Wigner, and $Q$ functions. These flows are visualized in Fig.~\ref{fig:pwq_map_spin}.

Geometrically, this exponential operator acts as a heat kernel evolution on the sphere. As $s$ increases, the function is progressively smoothed and vice-versa. The parameter $s$ thus tunes the spectral content of the distribution in phase-space. This is made explicit by considering the eigenfunctions of $\nabla^2_{S^2}$, i.e the spherical harmonics $Y_{\ell m}$, which form an orthonormal basis for smooth functions on $S^2$:
\begin{align}
\nabla^2_{S^2} Y_{\ell m} = -\ell(\ell + 1) Y_{\ell m}.
\end{align}
The exponential operator then acts diagonally in this basis,
\begin{align}
\exp\left( -\frac{s \log 3}{4} \nabla^2_{S^2} \right) Y_{\ell m} = \exp\left( \frac{s \log 3}{4} \ell(\ell + 1) \right) Y_{\ell m}.
\end{align}
This shows that higher angular momentum modes are increasingly suppressed for $s < 0$ (e.g., $Q$ function) and increasingly enhanced for $s > 0$ (e.g., $P$ function). In this way, the $s$-parametrised family defines a spectral filter on operator content over $S^2$, and the transformations between different representations are merely a spectral flow generated by the heat kernel.

For every value of $s$, we define the $-s$ representation to be \emph{dual} to the $s$ representation. This duality emerges from the adjoint pairing of SW kernels from the traciality condition:
\begin{align}
\langle \hat A \rangle_{\hat{\rho}} = 4\pi\int_{S^2} d\Omega\, f_{\hat{\rho}}(\theta,\varphi; s)\, f_{\hat A}(\theta,\varphi; -s)
= 4\pi\int_{S^2} d\Omega\, f_{\hat{\rho}}(\theta,\varphi; -s)\, f_{\hat A}(\theta,\varphi; s).
\end{align}
In particular, the Wigner representation is self-dual:
\begin{align}
\langle \hat A \rangle_{\hat{\rho}} = 4\pi\int_{S^2} d\Omega\, W_{\hat{\rho}}(\theta,\varphi)\, W_{\hat A}(\theta,\varphi),
\end{align}
and the $P$ and $Q$ representations are dual to each other:
\begin{align}
\langle \hat A \rangle_{\hat{\rho}} 
= 4\pi\int_{S^2} d\Omega\, Q_{\hat{\rho}}(\theta,\varphi)\, P_{\hat A}(\theta,\varphi)
= 4\pi\int_{S^2} d\Omega\, P_{\hat{\rho}}(\theta,\varphi)\, Q_{\hat A}(\theta,\varphi).
\end{align}
Thus we have a mechanism to compute expectation values in this formalism. Let us now study how it scales in the many-body case.

\subsection{Generalization to Many Bodies}
\label{sec:many_body_phase_space}

The $Q$, $W$, and $P$ representations extend naturally to $N$-body systems. We can construct the many-body phase-space as the Cartesian product manifold $(S^2)^N:=S^2 = S^2 \times S^2\times \ldots \times S^2$, corresponding to the application of the single-body SW correspondence at each site. Each point on this manifold is specified by $(\boldsymbol{\theta}, \boldsymbol{\varphi})$, shorthand for the $N$-tuples $(\theta_1, \dots, \theta_N)$ and $(\varphi_1, \dots, \varphi_N)$, and defines the coherent product state
\begin{align}
\ket{\boldsymbol{\theta},\boldsymbol{\varphi}}=\ket{\theta_1,\varphi_1}\otimes\ket{\theta_2,\varphi_2}\otimes\cdots\otimes\ket{\theta_N,\varphi_N}  = : \ket{\boldsymbol{\Omega}},
\end{align}
where $\boldsymbol{\Omega}=(\theta_1, \dots, \theta_N, \varphi_1, \dots, \varphi_N)$.
\begin{remark}[$(S^2)^N \ncong S^{2N}$]
    We emphasise that the $N$-fold Cartesian product $(S^2)^N$ is not homeomorphic to the higher-dimensional sphere $S^{2N}$. The distinction is topological: for instance, their Betti numbers differ. Equivalently, by Morse theory, a simple height function on $S^{2N}$ has only two critical points (north and south poles), whereas a product height function on $(S^2)^N$ exhibits $2^N$ critical points, one for each choice of poles on the factors.
\end{remark}

On $(S^2)^N$ we may define a product kernel formed from $N$ copies of the single-body kernel:
\begin{align}
\hat{\Delta}^{(s)}(\boldsymbol{\theta}, \boldsymbol{\varphi}) := \bigotimes_{i=1}^N \hat{\Delta}^{(s)}(\theta_i, \varphi_i).
\end{align}
We note here that this is the unique separable extension of the SW map. We now focus on quasi-probabilities for many-qubit systems, choosing the $Q$-function ($s = -1$) as our representative of interest due to its dynamical stability properties (see Sec.~\ref{sec:spin_dynamics}). Given the many-qubit SW structure, the many-body $Q$ function for a state $\hat\rho$ is defined as
\begin{align}
Q_{\hat\rho}(\boldsymbol{\theta},\boldsymbol{\varphi}) = \frac{1}{(2\pi)^N} \bra{\boldsymbol{\theta},\boldsymbol{\varphi}} \hat\rho \ket{\boldsymbol{\theta},\boldsymbol{\varphi}},\label{eq:many_spin_q_func}
\end{align}
On the other hand, the $P$ representation can be found via resolution of the identity:
\begin{align}
\hat\rho = \int_{(S^2)^N} d\boldsymbol{\Omega} \, P_{\hat{\rho}}(\boldsymbol{\theta}, \boldsymbol{\varphi}) \ketbra{\boldsymbol{\theta},\boldsymbol{\varphi}}{\boldsymbol{\theta},\boldsymbol{\varphi}},
\end{align}
where $d \boldsymbol{\Omega}:= \prod_{j=1}^N d \Omega_j$ is a product-sum of standard Haar measures on each sphere.

\begin{proposition}[Many-body Pauli expectation expansion for Q and P functions]\label{prop:multi_spin_q_func_exp}
The many-body $Q$ and $P$ functions admit expansions over the product Pauli basis, with spherical harmonics encoding directional structure at each site:
\begin{align}
Q_{\hat\rho}(\boldsymbol{\theta},\boldsymbol{\varphi}) &= \frac{1}{(4\pi)^N} \sum_{\hat{P}_1,\dots,\hat{P}_N} \langle \hat{P}_1 \otimes \cdots \otimes \hat{P}_N \rangle_{\hat{\rho}} \prod_{i=1}^N Y_{\hat P_i}(\theta_i, \varphi_i), \label{eq:multi_spin_q_func_exp}\\
P_{\hat\rho}(\boldsymbol{\theta},\boldsymbol{\varphi}) &= \frac{1}{(4\pi)^N} \sum_{\hat{P}_1,\dots,\hat{P}_N} \langle \hat{P}_1 \otimes \cdots \otimes \hat{P}_N \rangle_{\hat{\rho}} \prod_{i=1}^N \tilde{Y}_{\hat P_i}(\theta_i, \varphi_i), \label{eq:multi_spin_p_func_exp}
\end{align}
where the sums run over all tensor products of $\hat{P}_i \in \{\hat{\mathbb{I}}, \hat{\sigma}_x, \hat{\sigma}_y, \hat{\sigma}_z\}$, and $A_{\hat\rho} = \operatorname{Tr}[\hat{A}\hat\rho]$. The functions $Y_{\hat P}$ and $\tilde{Y}_{\hat P}$ are unnormalized spherical harmonics associated with each Pauli operator:
\begin{align}
Y_{\hat{\mathbb{I}}}(\theta,\varphi) &= 1, & \tilde{Y}_{\hat{\mathbb{I}}}(\theta,\varphi) &= 1, \\
Y_{\hat\sigma_x}(\theta,\varphi) &= \sin\theta\cos\varphi, & \tilde{Y}_{\hat\sigma_x}(\theta,\varphi) &= 3\sin\theta\cos\varphi, \\
Y_{\hat\sigma_y}(\theta,\varphi) &= \sin\theta\sin\varphi, & \tilde{Y}_{\hat\sigma_y}(\theta,\varphi) &= 3\sin\theta\sin\varphi, \\
Y_{\hat\sigma_z}(\theta,\varphi) &= \cos\theta, & \tilde{Y}_{\hat\sigma_z}(\theta,\varphi) &= 3\cos\theta.
\end{align}
\end{proposition}

Expectation values are computed as in the single-body case, using duality of the $P$ and $Q$ representations:
\begin{align}\label{eq:expectation_phase_many_body}
\langle \hat A \rangle_{\hat\rho} = (4\pi)^N\int_{(S^2)^N} d \boldsymbol{\Omega} \, Q_{\hat\rho}(\boldsymbol{\theta},\boldsymbol{\varphi}) P_{\hat A}(\boldsymbol{\theta},\boldsymbol{\varphi}) = (4\pi)^N\int_{(S^2)^N} d \boldsymbol{\Omega} \, P_{\hat\rho}(\boldsymbol{\theta},\boldsymbol{\varphi}) Q_{\hat A}(\boldsymbol{\theta},\boldsymbol{\varphi}).
\end{align}

Although the phase-space \((S^2)^N\) grows only linearly in dimension with the number of qubits, the number of functionally independent components required to represent a generic quantum state grows exponentially as \(4^N\). This reflects the dimension of the full operator algebra \(\mathcal{H}_2^{\otimes N}\), and is already visible in Eq.~\eqref{eq:multi_spin_q_func_exp} as a sum over all tensor products of Pauli operators. Thus, the complexity of phase-space representations is not inherited from the geometry of \((S^2)^N\), but from the algebraic content of the Pauli group itself, see Sec.~\ref{sec:discussion} for further discussion.

A notable computational aspect of the phase-space formalism is that expectation values such as those of Eq.~(\ref{eq:expectation_phase_many_body}) reduce to high-dimensional integrals, which are particularly well-suited to Monte Carlo methods. Unlike vector-state representation and Exact Diagonalization, which involves costly matrix-vector contractions, or tensor network methods like MPS, which are optimized for 1D systems, phase-space techniques can be readily applied in arbitrary dimensions. In fact, the high dimensionality of $(S^2)^N$ becomes a virtue; Monte Carlo integration often improves in relative efficiency as the number of particles increases. To understand the way in which we can generate samples, we now consider the following example:

\begin{example}[Sampling outcomes via expectation values]
Suppose we want to obtain computational basis samples from an $N$-qubit state $\ket\psi$, given that we can only sample from the $Q$ function of $\ket\psi$. We can take advantage of the fact that the probability of obtaining outcome $b\in\{0,1\}$ after measuring a single qubit at index $i$ is equal to the expectation $\langle \hat{\Pi}^{(i)}_b\rangle_{\ket\psi}$, where
\begin{align}
\hat{\Pi}^{(i)}_0=\frac{\hat{\mathbb{I}}+\hat\sigma_z^{(i)}}{2},\qquad\qquad\qquad \hat{\Pi}^{(i)}_1=\frac{\hat{\mathbb{I}}-\hat\sigma_z^{(i)}}{2}.
\end{align}
These expectation values can be estimated via Eq.~(\ref{eq:expectation_phase_many_body}) using samples from $Q_{\ket\psi}$. Similarly, the conditional probability of obtaining outcome $b\in\{0,1\}$ after measuring the $i$-th qubit, conditioned on outcomes $c_{j_1},c_{j_2},\cdots, c_{j_k}$ on qubits $j_1,j_2,\cdots,j_k$ respectively, is given by the expectation
\begin{align}
p(b\,|\,c_{j_1},c_{j_2},\cdots, c_{j_k})=\frac{\langle \hat{\Pi}^{(j_1)}_{c_{j_1}}\otimes \hat{\Pi}^{(j_2)}_{c_{j_2}}\otimes\cdots\otimes\hat{\Pi}^{(j_k)}_{c_{j_k}}\otimes \hat{\Pi}^{(i)}_{b}\rangle_{\ket\psi}}{\langle \hat{\Pi}^{(j_1)}_{c_{j_1}}\otimes \hat{\Pi}^{(j_2)}_{c_{j_2}}\otimes\cdots\otimes\hat{\Pi}^{(j_k)}_{c_{j_k}}\rangle_{\ket\psi}}
\end{align}
This provides a straightforward algorithm to sample from the computational basis of all of the qubits: At the first step, sample from the computational basis of the first qubit, then in the $i$-th step for $1<i\leq N$, sample from the computational basis of the $i$-th qubit conditioned on all the samples from qubits $1,2,\cdots, i-1$.

\end{example}

Moreover, the marginal of the $Q$ function over a subset of sites yields the $Q$ function of the corresponding reduced state, as summarised by the following proposition:

\begin{proposition}[Partial tracing as marginalization] \label{prop:partial_trace_phase}
    Let $\hat\rho$ be an $N$-qubit state and $\hat\rho^{(i)}=\operatorname{Tr}_i[\hat\rho]$ be the reduced state after taking the partial trace of the qubit on site $i$, then
\begin{align}
Q_{\hat\rho^{(i)}}(\boldsymbol{\theta}\backslash\{\theta_i\},\boldsymbol{\varphi}\backslash\{\varphi_i\})=\int_{S^2}d\Omega_i\,Q_{\hat\rho}(\boldsymbol{\theta},\boldsymbol{\varphi}).
\end{align}
\end{proposition}
Thus the many-body phase-space representation is therefore compatible with partial tracing; marginalising over a subspace yields a new valid distribution over the subspace, for which 
Examples~\ref{ex:q_func_bell_states} and \ref{ex:q_func_ghz} are instructive.

\begin{example}[Bell States]\label{ex:q_func_bell_states}
The Q functions for the four 2-qubit Bell states are
\begin{align}
Q_{\ket{\Phi^+}}(\theta_1, \varphi_1,\theta_2,\varphi_2)&=\frac{1}{(4\pi)^2}\left(1+\cos\theta_1\cos\theta_2+\sin\theta_1\sin\theta_2\cos(\varphi_1+\varphi_2)\right), \\
Q_{\ket{\Phi^-}}(\theta_1, \varphi_1,\theta_2,\varphi_2)&=\frac{1}{(4\pi)^2}\left(1+\cos\theta_1\cos\theta_2-\sin\theta_1\sin\theta_2\cos(\varphi_1+\varphi_2)\right), \\
Q_{\ket{\Psi^+}}(\theta_1, \varphi_1,\theta_2,\varphi_2)&=\frac{1}{(4\pi)^2}\left(1-\cos\theta_1\cos\theta_2+\sin\theta_1\sin\theta_2\cos(\varphi_1-\varphi_2)\right), \\
Q_{\ket{\Psi^-}}(\theta_1, \varphi_1,\theta_2,\varphi_2)&=\frac{1}{(4\pi)^2}\left(1-\cos\theta_1\cos\theta_2-\sin\theta_1\sin\theta_2\cos(\varphi_1-\varphi_2)\right),
\end{align}
for which there is no factorisation into local functions. That is, $\nexists f_1(\theta_1,\varphi_1),f_2(\theta_2,\varphi_2)\;s.t. \;Q(\theta_1,\varphi_1,\theta_2,\varphi_2) = f_1 f_2$ for any of the four Bell states. Further, the marginalising operation over one of the two coordinates gives the maximally mixed phase-space function, in direct analogy to partial traces in  Hilbert space,
\begin{align}
\int_{S^2}d\Omega_1 \,Q_{\ket{\Psi^+}}(\theta_1,\varphi_1,\theta_2,\varphi_2)&= \\ \frac{1}{(4\pi)^2}\int_{S^2} d\theta_1 d \varphi_1 & \,\sin \theta_1 \left(1+\cos\theta_1\cos\theta_2+\sin\theta_1\sin\theta_2\cos(\varphi_1+\varphi_2)\right)\\
&=\frac{1}{4\pi}.
\end{align}
which is the $Q$ function of the maximally mixed state. See also Prop.~\ref{prop:partial_trace_phase} and Eq.~(\ref{eq:q_func_max_mixed_state}).
\end{example}

\subsection{Separability and Non-Classicality}
\label{sec:partial_trace_nonclassicality}
It is now worth studying how non-separability and other signs of non-classicality are encoded in this representation. Namely, generic quantum states yield $Q$ and $P$ functions that cannot be factorized into products over sites. Non-classical correlations are therefore encoded geometrically in the breakdown of separability across $(S^2)^N$, which we will now study in more detail. 

We will begin by characterising how separability manifests in the $Q$ function and show that this behaviour persists across the full $s$-family of representations. We then discuss non-classicality criteria and information-theoretic quantities such as the Wehrl entropy and matrix rank in phase-space. This will lead us to define a generalised Moyal product for qubits, making the phase-space representation ready for analysing quantum dynamics.

\begin{remark}[Q function for product state]
The Q function for a product state $\hat\rho = \hat\rho_1  \otimes \cdots \otimes \hat\rho_n$ is $Q_{\hat\rho}(\boldsymbol{\theta},\boldsymbol{\varphi}) = \prod\limits_{i = 1}^n Q_{\hat\rho_i}(\theta_i, \varphi_i)$. This follows directly from the definition of $Q$ as the overlap on coherent states. \label{rem: product state}
\end{remark}

Therefore, product states are mapped to factorizable $Q$ functions, in line with the probabilistic interpretation. In the same way, from the linearity of the correspondence, convex combinations of product states are mapped to the corresponding combination of products of local $Q$ functions. This is summarised by the following proposition:
\begin{proposition}[Separability and $Q$-function factorization]
\label{prop:sep}
Let $\hat\rho$ be a density operator on $\mathcal{H}_2^{\otimes N}$. Then the following are equivalent:
\begin{enumerate}
    \item Separability: $\hat\rho$ is separable, i.e., it admits a decomposition
    \begin{equation}
       \hat{\rho} = \sum_{k} p_k\, \hat\rho^{(k)}_1 \otimes \cdots \otimes \hat\rho^{(k)}_N,
    \end{equation}
    with $p_k \geq 0$, $\sum_k p_k = 1$;
    
    \item Convex factorization of the $Q$ function:
    \begin{equation}
        Q_{\hat\rho}(\boldsymbol\theta,\boldsymbol\varphi)
        = \sum_{k} p_k \prod_{i=1}^{N}
        Q_{\hat\rho^{(k)}_{i}}(\theta_i,\varphi_i).
    \end{equation}
\end{enumerate}
\end{proposition}

For example, we can see in phase-space that the so-called GHZ and W states belong to different entanglement classes:
\begin{example}[GHZ and W States]\label{ex:q_func_ghz}
The $Q$ function for the 3-qubit $|GHZ\rangle = \frac{1}{\sqrt{2}}(|000\rangle + |111\rangle)$ state is
\begin{align}
&Q_{\ket{GHZ}}(\theta_1,\varphi_1,\theta_2,\varphi_2,\theta_3,\varphi_3)\nonumber\\
&\qquad=\frac{1}{(4\pi)^3}\left(1+\cos\theta_1\cos\theta_2+\cos\theta_1\cos\theta_3+\cos\theta_2\cos\theta_3+\sin\theta_1\sin\theta_2\sin\theta_3\cos(\varphi_1+\varphi_2+\varphi_3)\right),
\end{align}
and the $Q$ function for the $|W\rangle = \frac{1}{\sqrt{3}}(|001\rangle + |010\rangle +|100\rangle)$ state is
\begin{align}
&Q_{\ket{W}}(\theta_1,\varphi_1,\theta_2,\varphi_2,\theta_3,\varphi_3)\nonumber\\
&\qquad\qquad=\frac{1}{3(4\pi)^3}\Big((1+\cos\theta_1)(1-\cos\theta_2\cos\theta_3+2\sin\theta_2\sin\theta_3\cos(\varphi_2-\varphi_3))\nonumber\\
&\qquad\qquad\qquad\qquad\qquad+(1+\cos\theta_2)(1-\cos\theta_1\cos\theta_3+2\sin\theta_1\sin\theta_3\cos(\varphi_1-\varphi_3))\nonumber\\
&\qquad\qquad\qquad\qquad\qquad+(1+\cos\theta_3)(1-\cos\theta_1\cos\theta_2+2\sin\theta_1\sin\theta_2\cos(\varphi_1-\varphi_2))\Big).
\end{align}
Notice that the separability and marginalisation in phase-space also reflects the fact that these two states belong to distinct entanglement classes; one is genuine tri-partite entanglement, whilst the other is three sets of bipartite entanglement. We can check this by marginalising via Prop.~\ref{prop:partial_trace_phase}.
\end{example}

Prop.~\ref{prop:sep} follows directly from the definition of the $Q$ function as an expectation over coherent product states (see Eq.~\ref{eq:many_spin_q_func}), which factorizes under tensor products of states and kernels. However, the first is a property of the system, while the $Q$ function is merely a representation in which this condition happens to be conveniently reproduced. A natural question is whether this criterion is $s$-index independent, or how it translates into the other representations, which we address with the following Remark and Corollary:

\begin{remark}[$s$-independence of separability]
\label{rem:s_indep}
The separability criterion of Prop.~\ref{prop:sep} is independent of the ordering parameter $s \in [-1,1]$. This is because the differential operator $\nabla^2_{S^2}$ used to relate $s$-parametrised representations acts independently and linearly on each factor of the product manifold $(S^2)^N$. Hence, factorizability in any representation implies factorizability in all others.
\end{remark}

\begin{corollary}[Separability across the $s$-family]
\label{cor:s_family_sep}
For all $s \in [-1,1]$, the following holds:
\begin{equation}
    \hat\rho \text{ separable} \;\Longleftrightarrow\;
f^{(s)}_{\hat\rho}(\boldsymbol{\theta}, \boldsymbol{\varphi}) = \sum_k p_k \prod_{i=1}^{N} f^{(s)}_{\hat\rho^{(k)}_i}(\theta_i, \varphi_i).
\end{equation}
\end{corollary}

Therefore the separability signature observed in the Q representation is not an artefact of choosing $s = -1$ but a robust feature of the full s-family phase-space. 

In the bosonic case, Wigner negativity is widely used as an indicator of non-classicality and quantum resources such as entanglement~\cite{kenfack2004negativity} (see App.~\ref{app:bosonic_phase_space}). This notion has been extended to spin-$j$ systems~\cite{davis2021wigner}, but for qubits, all pure states (including coherent ones) exhibit negativity in their Wigner functions. Thus, the classical–quantum boundary is already blurred at the single-qubit level.

However, the $Q$ function’s positivity allows us to define an information-theoretic entropy,
\begin{align}
 S_W[\hat\rho] := -\int_{(S^2)^N} Q_{\hat\rho}(\Omega) \log Q_{\hat\rho}(\Omega) \, d\mu(\Omega),
\end{align}
known as the Wehrl entropy~\cite{wehrl1979relation}. It upper-bounds the von-Neumann entropy and attains its minimum for coherent states. In the bosonic case, this was conjectured by Wehrl to be 1, whilst the Lieb conjecture gives $\frac{2j}{2j+1}$ for spin-$j$ systems~\cite{lieb1978proof, lieb2014proof, Schupp1999}. 

For a single qubit particle, the Wehrl entropy becomes trivial since all pure states are coherent. However, in the many-body context, the Wehrl entropy becomes non-trivial. We note here that the Wehrl entropy has not been studied in the context of $(S^2)^N$ phase-space. Its behaviour for entangled, thermal, or noisy many-qubit states could provide a new operational measure of non-classicality directly within phase-space, circumventing the need for Hilbert space constructs. In addition to the Wehrl entropy, another scalar diagnostic of non-classicality in phase-space is the purity, $\operatorname{Tr}[\hat\rho^2]$, which can also be directly expressed in phase-space, as summarised by the following proposition:

\begin{proposition}[Purity in phase-space]\label{prop:single_spin_purity}
Given the $Q$ function of a state $\hat\rho$, the purity of $\hat\rho$ can be computed as
\begin{align}
\operatorname{Tr}\left[\hat\rho^2\right]=6\pi\left[\int_{S^2}d\Omega\,Q_{\hat\rho}(\theta,\varphi)^2\right]-1,
\end{align}
\end{proposition}

For example, we can consider the classical state, and thermal state of a given Hamiltonian:
\begin{example}[Classical state]\label{ex:classical_Q}
Consider the ``classical'' state
\begin{align}
\hat\rho=p\ketbra{0}{0}+(1-p)\ketbra{1}{1},
\end{align}
where $0\leq p\leq 1$ is some probability. The $Q$ function corresponding to this state is
\begin{align}
Q_{\hat\rho}(\theta,\varphi)=\frac{1}{4\pi}(1+(2p-1)\cos\theta),\label{eq:q_func_classical_state}
\end{align}
In particular, the pure state cases $p=0$ and $p=1$ reduce to  Eq.~(\ref{eq:q_func_z_states}). The maximally mixed state with $p=\frac{1}{2}$ reduces to
\begin{align}
Q_{\frac{1}{2}\hat{\mathbb{I}}}(\theta,\varphi)=\frac{1}{4\pi},\label{eq:q_func_max_mixed_state}
\end{align}
which is the uniform distribution on $S^2$.

By Prop.~\ref{prop:single_spin_purity}, the purity of the classical state can be computed as
\begin{align}
\operatorname{Tr}\left[\hat\rho^2\right]&=6\pi\left[\int_{S^2}d\Omega\,\frac{1}{(4\pi)^2}\left(1+(2p-1)\cos\theta\right)^2\right]-1\\
&=p^2+(1-p)^2,
\end{align}
which is the expected result. The pure states produce the maximum purity of 1, whereas the maximally mixed state produces the minimum purity of $\frac{1}{2}$.

\end{example}

\begin{example}[Thermal state]\label{ex:q_func_thermal}
Consider the thermal state of $\hat{\sigma}_x$
\begin{equation}
    \hat\rho = \frac{e^{-\beta \hat{\sigma}_x}}{ \operatorname{Tr}[e^{-\beta \hat{\sigma}_x}]},
\end{equation}
where $\beta$ is the usual inverse temperature. The Q-function for this state reads
\begin{equation}
    Q_{\hat\rho}(\theta,\varphi) = \frac{1}{2}(1 - \tanh \beta \sin \theta \cos \varphi).
\end{equation}
As $\beta \rightarrow 0$ we recover the uniform distribution over $S^2$, that is, the infinite temperature quasi-probability. When $\beta \rightarrow \infty$ $Q_{\hat\rho}(\theta,\varphi)$ becomes peaked around the point $(\theta,\varphi)=(\frac{\pi}{2},0)$, giving a Pauli eigenstate. This is shown in Fig.~\ref{fig:q_thermal}.
\begin{figure}
\captionsetup[subfigure]{aboveskip=0pt,belowskip=0pt}
\centering
\begin{subfigure}{0.49\textwidth}
    \centering
    \includegraphics[width=\textwidth]{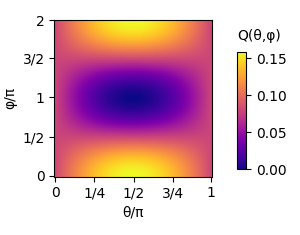}
    \caption{$\beta=\infty$}
\end{subfigure}
\begin{subfigure}{0.49\textwidth}
    \centering
    \includegraphics[width=\textwidth]{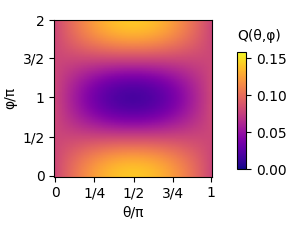}
    \caption{$\beta=1$}
\end{subfigure}
\begin{subfigure}{0.49\textwidth}
    \centering
    \includegraphics[width=\textwidth]{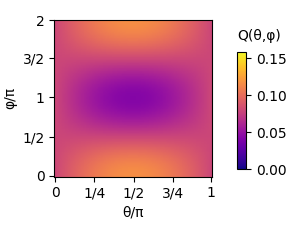}
    \caption{$\beta=0.5$}
\end{subfigure}
\begin{subfigure}{0.49\textwidth}
    \centering
    \includegraphics[width=\textwidth]{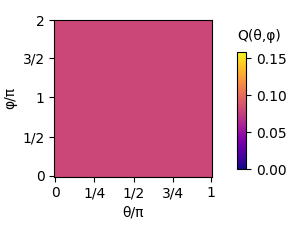}
    \caption{$\beta=0$}
\end{subfigure}
\caption{$Q$ representations of thermal states of the Hamiltonian $\hat H = \hat\sigma_x$, for different values of the inverse temperature $\beta$.}
\label{fig:q_thermal}
\end{figure}
\end{example}

\begin{example}[Thermal State of Two Qubits]
Consider the two-qubit Hamiltonian $\hat H=-\hat\sigma_z^{(1)}\hat\sigma_z^{(2)}$ and the thermal state $\hat\rho_\beta$ given by
\begin{align}
\hat\rho_{\beta}=\frac{e^{-\beta \hat H}}{\operatorname{Tr}\left[e^{-\beta \hat H}\right]},
\end{align}
The $Q$ function for this state is
\begin{align}
Q_{\hat\rho}(\theta_1,\varphi_1,\theta_2,\varphi_2)&=\frac{1}{(4\pi)^2}\left(1+\tanh\beta\cos\theta_1\cos\theta_2\right),
\end{align}
When $\beta\to\infty$, the $Q$ function approaches the uniform mixture over the two ground states $\ket{00}$ and $\ket{11}$. When $\beta=0$, the $Q$ function is simply the uniform distribution over $(S^2)^2$.
\end{example}

As a final consistency check, we recall a result of Ferrie and Emerson~\cite{ferrie2008frame}: any quasi-probability representation of quantum theory must either exhibit negativity or violate classical probability calculus. The phase-space framework manifests both. First, the Wigner function may be negative. Second, the $Q$ function, while positive, fails to obey the classical law of expectation: $\langle \hat{A} \rangle_{\hat\rho} \neq \int Q_{\hat{\rho}} Q_A$, and instead requires duality via the $P$ representation. This confirms that quantum structure cannot be fully hidden, even in smooth phase-space representations; the price of positivity is convolutional deformation, which we summarise by the following Remark:

\begin{remark}
Phase-space representations either expose quantum features (negativity) or encode them structurally (non-classical convolution). There is no classical limit that restores both positivity and classical probability theory simultaneously.
\end{remark}

\subsection{Phase-Space Rank, Algebraic Structure, and Physicality}
\label{sec:phase_space_rank}

The Stratonovich-Weyl correspondence provides a noncommutative algebra of functions on $(S^2)^N$ equipped with the Moyal $\star$-product \cite{Moyal1949}, defined as follows:

\begin{definition}[Moyal Star Product on $(S^2)^N$]
Let $\hat{A}, \hat{B}: \mathcal{H}_2^{\otimes N} \rightarrow \mathcal{H}_2^{\otimes N}$ be Hermitian operators, and let us define their phase-space images via the SW correspondence as
\begin{equation}
    f_{\hat{A}}^{(s)}(\boldsymbol{\Omega}) := \mathrm{Tr}[\hat{A} \hat{\Delta}^{(s)}(\boldsymbol{\Omega})], \qquad \boldsymbol{\Omega} \in (S^2)^N.
\end{equation}
Then the Moyal star product \cite{Moyal1949} is defined such that,
\begin{equation}
    f_{\hat{A}}^{(s)} \star f_{\hat{B}}^{(s)} := f_{\hat{A}\hat{B}}^{(s)}.
\end{equation}
\end{definition}

This $\star$-product turns the image of the SW map into a noncommutative algebra of real-valued functions on $(S^2)^N$. Although it is defined on the full space $L^2((S^2)^N)$, only a finite-dimensional subspace corresponds to physically meaningful functions. That is, phase-space functions that correspond to properly normalised, non-negative states of a Hilbert space. We denote this subspace by $\mathcal{F}_{\mathrm{phys}} \subset L^2((S^2)^N)$, which has dimension $4^N$ according to Prop.~\ref{prop:multi_spin_q_func_exp}. All operator symbols $f_{\hat{A}}$ lie in this subspace. Consequently, algebraic constructions such as the phase-space rank must be restricted to $\mathcal{F}_{\mathrm{phys}}$. 

This restriction enables us to check whether a given Q-function corresponds to a physical state under the SW correspondence. It suffices to verify that the following value of $\mathcal{L}$ is zero 
\begin{align}\label{eq:physical_subspace}
    \mathcal{L} =\int_{(S^2)^N} \left(\delta(\boldsymbol{\Omega}-\boldsymbol{\Omega}') - \sum_{l = 0}^1 Y_l(\boldsymbol{\Omega'}))Q(\boldsymbol{\Omega}') \right) d \boldsymbol{\Omega},
\end{align}
since the bracketed term acts as a projector out of the physical subspace. If the state is physical, it has zero amplitude with respect to such a projector. Its functional form comes from spherical harmonic decomposition of the delta function on the phase-space manifold,
\begin{equation}
    \sum_{l = 0}^{\infty} Y_{l}(\boldsymbol{\Omega})Y_{l}(\boldsymbol{\Omega}') = \delta(\boldsymbol{\Omega} - \boldsymbol{\Omega '}).
\end{equation}
This allows us to check general $s$-parametrised phase-space functions, by switching to the $Q$-representation via Fig.~\ref{fig:pwq_map_spin}.

Using the $\star$-product, we can also define a notion of matrix rank directly in phase-space, mirroring the operator rank of a quantum state but without reference to Hilbert space. This rank arises from the internal structure of the phase-space functions under the $\star$-algebra and signals the effective dimensionality of the support of the quantum state.
\begin{definition}[Phase-Space Rank]
Let $W_{\hat{\rho}}(\Omega)$ be the Wigner function of a density matrix $\hat\rho$ under the Stratonovich-Weyl correspondence. Then the phase-space rank of $W_{\hat{\rho}}$ is defined by
\begin{align}
    \mathrm{rank}_{\mathrm{ps}}(W_{\hat\rho}) := \dim \left\{ f_{\hat{A}} \mid \hat{A} \in \mathcal{B}(\mathcal{H}_2^{\otimes N}), \; f_{\hat{A}} \star W_{\hat\rho} = f_{\hat{A}} \right\}.
\end{align}
That is, the phase-space rank counts the number of linearly independent Wigner symbols that are left fixed under the $\star$-product with $W_{\hat{\rho}}$.
\end{definition}

\begin{proposition}[Equivalence with Operator Rank]
Let $\hat\rho$ be a density operator on $\mathcal{H}_2^{\otimes N}$, and let $W_{\hat{\rho}}(\Omega)$ denote its Wigner function. Then
\begin{align}
    \mathrm{rank}_{\mathrm{ps}}(W_{\hat\rho}) = \mathrm{rank}(\hat\rho).
\end{align}
\end{proposition}

We therefore have an \emph{intrinsic} definition of rank within phase-space, one that makes no reference to Hilbert space. The Wigner function carries an equivalent algebraic structure to rank in Hilbert space under the noncommutative convolution algebra defined by $\star$. We further note here that there is nothing special about choosing $s=0$ and the Wigner function. Deformation quantisation allows us to have a star product for the whole $s$-parametrised family \cite{bayen1978deformationI, bayen1978deformationII}. Given two functions $f_{\hat{A}}^{(s)}$ and $f_{\hat{B}}^{(s)}$ corresponding to operators $\hat{A}, \hat{B}$, the product
\begin{equation}
    f_{\hat{A}}^{(s)} \star^{(s)} f_{\hat{B}}^{(s)}(\boldsymbol{\Omega}) := \operatorname{Tr}\left[\hat{A}\hat{B} \, \hat{\Delta}^{(s)}(\boldsymbol{\Omega})\right],
\end{equation}
defines an associative, representation-dependent algebra on phase-space. This algebra smoothly interpolates between the highly non-local but positive $P$ and $Q$ representations and the symmetric but sign-indefinite Wigner case.

\section{Dynamics}
\label{sec:spin_dynamics}

Having defined the algebraic image of quantum states in phase-space, we now turn to dynamics. We derive a time evolution equations for phase-space function both for unitary and non-unitary dynamics, i.e. equivalents of von-Neumann and Lindblad equations for density matrices, as well as imaginary time evolution equation after a Wick rotation. This requires introducing a bracket structure on phase-space functions, serving as the analogue of operator commutators and anti-commutators. We begin by discussing these brackets for a single-qubit system, highlighting a subtlety that makes the $Q$-function most appropriate for modelling dynamics. We will often use the compact notation $Q_{\hat{A}}(\boldsymbol{\Omega}):= f_{\hat{A}}^{(-1)}(\boldsymbol{\Omega})$ for this case. We then show how they extend intuitively to the many-body case.

\subsection{One-Qubit Dynamics}
\label{sec:one_body_dynamics}

To elevate qubit phase-space to a fully autonomous formulation of quantum dynamics, we must endow it with internal algebraic structure sufficient to represent operator products and evolution laws. Specifically, we aim to define analogues of the commutator and anti-commutator directly on the image of the Stratonovich-Weyl (SW) correspondence.

Let $\hat{\Delta}^{(s)}(\Omega)$ be the SW kernel on $S^2$ for a qubit with fixed index $s$, and define the symbol of an operator $\hat{A}$ as
\begin{align}
    f^{(s)}_{\hat{A}}(\Omega) := \operatorname{Tr}[\hat{A} \hat{\Delta}^{(s)}(\Omega)], \quad \Omega \in S^2.
\end{align}
We now define bilinear operations on $f^{(s)}_{\hat A}$ and $f^{(s)}_{\hat B}$ that reproduce operator commutators and anti-commutators:
\begin{align}
    [\![f^{(s)}_{\hat A}, f^{(s)}_{\hat B}]\!] &:= f^{(s)}_{-i[\hat{A}, \hat{B}]},\\
    \{\!\!\{f^{(s)}_{\hat A}, f^{(s)}_{\hat B}\}\!\!\} &:= f^{(s)}_{\{\hat{A}, \hat{B}\}}.
\end{align}
These define the \emph{sine} and \emph{cosine} brackets on $L^2(S^2)$.     The overall operator composition structure on the image of the SW correspondence is therefore
\begin{equation}
    f^{(s)}_{\hat{A}} \star f^{(s)}_{\hat{B}} 
= f^{(s)}_{\hat{A}\hat{B}} 
= \frac{1}{2} \left( \{\!\!\{f^{(s)}_{\hat{A}}, f^{(s)}_{\hat{B}}\}\!\!\} + i[\![f^{(s)}_{\hat{A}}, f^{(s)}_{\hat{B}}]\!] \right).
\end{equation}
In particular, the sine bracket is independent of $s$ thanks to the following lemma:

\begin{lemma}[Covariance induces Lie action]
If the Stratonovich-Weyl kernel satisfies the covariance property under $SU(2)$ conjugation,
\begin{equation}
    \hat U \hat{\Delta}^{(s)}(\Omega) \hat U^\dagger = \hat{\Delta}^{(s)}(R_U \cdot \Omega),
\end{equation}
where $R_U \cdot \Omega$ denotes the usual rotation action of $\hat U \in SU(2)$ on $\Omega \in S^2$, then the sine bracket implements a representation of the Lie algebra \( \mathfrak{su}(2) \) as first-order differential operators (Killing vector fields) on the phase-space \( S^2 \).
\end{lemma}

\begin{proposition}[Sine bracket in coordinates]
Let $\hat P \in \{ \hat \sigma_x, \hat \sigma_y, \hat \sigma_z \}$ be a Pauli operator and $\hat\rho$ a single-qubit state. Then, for all $s \in [-1,1]$, we have
\begin{align}
    [\![f^{(s)}_{\hat{P}}, f^{(s)}_{\hat{\rho}}]\!] = 2\mathcal{J}_{\hat{P}} f^{(s)}_{\hat{\rho}},
\end{align}
where $\mathcal{J}_P$ is a first-order differential operator on $S^2$, explicitly given by:
\begin{align}
\mathcal{J}_{\hat{\mathbb{I}}}&=0, \\
\mathcal{J}_{\hat \sigma_x}&=\sin\varphi\frac{\partial}{\partial\theta}+\cot\theta\cos\varphi\frac{\partial}{\partial\varphi}, \\
\mathcal{J}_{\hat \sigma_y}&=-\cos\varphi\frac{\partial}{\partial\theta}+\cot\theta\sin\varphi\frac{\partial}{\partial\varphi}, \\
\mathcal{J}_{\hat \sigma_z}&=-\frac{\partial} {\partial\varphi} .
\end{align}
These form a representation of $\mathfrak{su}(2)$ as Killing vector fields on $S^2$. 
\label{prop: sine_bracket}
\end{proposition}

While the sine bracket yields an $s$-independent geometric action corresponding to the Lie algebra $\mathfrak{su}(2)$, the cosine bracket behaves differently. Its differential form depends explicitly on the representation index $s$, and this dependence encodes the failure of certain phase-space algebras to close under simple geometric action. More concretely, if we define
\begin{align}
    \{\!\!\{f^{(s)}_{\hat{P}}, f^{(s)}_{\hat{\rho}}\}\!\!\}^{(s)} := 2\mathcal{K}_P^{(s)} f^{(s)}_{\hat{\rho}},
\end{align}
the operators $\mathcal{K}_P^{(s)}$ are no longer independent of $s$, and the cosine bracket comes equipped with an index $\{\!\{\cdot,\cdot\}\!\}^{(s)}$. Unlike the $\mathcal{J}_P$, they do not commute with the Laplace-Beltrami operator $\nabla^2_{S^2}$, and therefore do not commute with the diffusion semigroup $e^{\alpha \nabla^2_{S^2}}$ which relates different $s$-parametrised representations (see Fig.~\ref{fig:pwq_map_spin}). In particular, the cosine bracket in the $s = -1$ representation (i.e., the $Q$-function) corresponds to second-order differential operators with smooth coefficients. But in representations with $s \to 0$ or $s \to 1$, these coefficients diverge as the spherical harmonic decomposition becomes increasingly singular. This is  a failure of the differential representation to capture the non-local structure of the star product, rather than a geometric singularity. In this sense, the $Q$-representation is the only one in which the algebra closes neatly under differential operators. For other $s$, the algebra remains formally well-defined, but the bracket's realisation via differential operators becomes unbounded. For this reason, we will proceed by constructing a formulation of dynamics only for the $s=-1$ index, i.e. for $Q$-functions.

\begin{proposition}[Cosine bracket in coordinates for Q-function]\label{prop:cosine_bracket}
Let $\hat P \in \{\hat{\mathbb{I}}, \hat \sigma_x,\hat  \sigma_y, \hat \sigma_z \}$ be a Pauli operator and $\hat\rho$ a single-qubit state. Then as differential operators on the $Q$-function, we have
\begin{align}
\mathcal{K}_{\hat{\mathbb{I}}}&=1,\\
\mathcal{K}_{\hat \sigma_x}&=\sin\theta\cos\varphi+\cos\theta\cos\varphi\frac{\partial}{\partial\theta}-\csc\theta\sin\varphi\frac{\partial}{\partial\varphi},\\
\mathcal{K}_{\hat \sigma_y}&=\sin\theta\sin\varphi+\cos\theta\sin\varphi\frac{\partial}{\partial\theta}+\csc\theta\cos\varphi\frac{\partial}{\partial\varphi},\\
\mathcal{K}_{\hat \sigma_z}&=\cos\theta-\sin\theta\frac{\partial}{\partial\theta}.
\end{align}
\end{proposition}

We now summarize the differential structure of the bracket algebra in the Q-representation:
\begin{corollary}
Let $\hat P \in \{\hat{\mathbb{I}}, \hat \sigma_x, \hat \sigma_y, \hat \sigma_z\}$, and let $\hat\rho$ be a single-qubit state. Then:
\begin{align}
[\![Q_{\hat P}, Q_{\hat\rho}]\!] &= 2 \mathcal{J}_{\hat P} Q_{\hat\rho},\\
\{\!\!\{Q_{\hat P}, Q_{\hat\rho}\}\!\!\} &= 2 \mathcal{K}_{\hat P} Q_{\hat\rho},
\end{align}
where the differential operators $\mathcal{J}_{\hat P}$ and $\mathcal{K}_{\hat P}$ are defined above. 
\end{corollary}

This result endows the Q-representation with a Lie–Jordan algebraic structure; the sine bracket defines a Lie algebra (commutator), and the cosine bracket defines a symmetric bilinear product (anti-commutator). Together, they reconstruct operator composition geometrically on $S^2$, which shows how the dynamics of a qubit system can be fully formulated within the geometry of the sphere. The operator formalism is no longer required to define time evolution, expectation values, or composition, and the Q-function algebra suffices.

In addition to their differential realisations, the sine and cosine brackets admit integral definitions derived from the kernel structure of the SW transform. To derive the integral forms of the sine and cosine brackets, we begin again from the general structure of the Stratonovich-Weyl (SW) correspondence, recalling Eq.~\ref{eq:SW_correspondence}.
This allows us to define a noncommutative $\star$-product on functions by requiring
\begin{equation}
    f^{(s)}_{\hat A\hat B}(\Omega) := \operatorname{Tr}[\hat{A}\hat{B} \, \hat{\Delta}^{(s)}(\Omega)] = (f^{(s)}_{\hat A} \star f^{(s)}_{\hat B})(\Omega),
\end{equation}
per our Remarks in Sec.\ref{sec:phase_space_rank}. Substituting the inverse maps, we obtain an explicit kernel representation:
\begin{align}
    f^{(s)}_{\hat A} \star f^{(s)}_{\hat B}(\Omega)
    = \int_{S^2} d\Omega' \int_{S^2} d\Omega'' \,
    f^{(s)}_{\hat A}(\Omega') f^{(s)}_{\hat B}(\Omega'') \, K^{(s)}(\Omega, \Omega', \Omega''),
\end{align}
where the \emph{triple kernel} is defined by
\begin{align}
    K^{(s)}(\Omega, \Omega', \Omega'') := 4\pi \operatorname{Tr}\left[\hat{\Delta}^{(-s)}(\Omega') \hat{\Delta}^{(-s)}(\Omega'') \hat{\Delta}^{(s)}(\Omega)\right].
\end{align}
This gives a generic integral representation for the star product. The sine and cosine brackets are then recovered from the antisymmetric and symmetric parts of the star product \cite{Moyal1949, bayen1978deformationII},
\begin{align}
    \{\!\!\{f, g\}\!\!\} &:= f \star g + g \star f, \\
    [\![f, g]\!] &:= -i (f \star g - g \star f),
\end{align}

This mirrors the bosonic case, where the Moyal bracket admits both a Fourier-based integral form and a differential expansion (see App.~\ref{app:bosonic_phase_space}). Here, the coherent-state kernel plays the analogous role on \( S^2 \).

Phase-space also supports direct geometric operations. The sine bracket reduces to a Poisson bracket on each sphere:
\begin{equation}
    [\![f, g]\!] = \frac{1}{\sin\theta} \left( \partial_\theta f \, \partial_\varphi g - \partial_\varphi f \, \partial_\theta g \right),
\end{equation}
while he cosine bracket in the Q representation takes the usual form of the (inverse) metric,
\begin{align}
\{\!\!\{f,g\}\!\!\}=fg+\partial_\theta f\,\partial_\theta g+\frac{1}{\sin^2\theta}\partial_\varphi f\,\partial_\varphi g.
\end{align}
This is the only representation (the Q-representation) where the cosine bracket has this form, as discussed above. Such a differential and integral algebra structure on \( S^2 \) constitutes a deformation quantization of the qubit phase-space, with the Q-representation realising a particularly regular model.  In the following, we provide a many-body generalisation of this formalism.

\subsection{Many-Qubit Dynamics}

To extend the dynamics to many-body systems, the bracket operations must respect the tensor-product compositional structure of many-body Hilbert space. The following identity guarantees that the Q-brackets are indeed compatible with bipartite operator embeddings.

\begin{proposition}[Tensor Compatibility]\label{prop:bracket_tensor}
Let $\hat{A} \in \beta_+(\mathcal{H}_i)$, $\hat{B} \in \beta_+(\mathcal{H}_j)$, and $\hat{C} \in \beta_+(\mathcal{H}_i \otimes \mathcal{H}_j)$. Then the brackets satisfy:
\begin{align}
[\![Q_{\hat A \otimes \hat B}, Q_{\hat C}]\!] &= \tfrac{1}{2} \left([\![Q_{\hat A}, \{\!\!\{Q_{\hat B}, Q_{\hat C}\}\!\!\}]\!] + \{\!\!\{Q_{\hat A}, [\![Q_{\hat B}, Q_{\hat C}]\!]\}\!\!\} \right), \label{eq:sine_bracket_tensor}\\
\{\!\!\{Q_{\hat A \otimes \hat B}, Q_{\hat C}\}\!\!\} &= \tfrac{1}{2} \left(-[\![Q_{\hat A}, [\![Q_{\hat B}, Q_{\hat C}]\!]]\!] + \{\!\!\{Q_{\hat A}, \{\!\!\{Q_{\hat B}, Q_{\hat C}\}\!\!\}\}\!\!\} \right). \label{eq:cosine_bracket_tensor}
\end{align}
\end{proposition}

Thus tensor embeddings preserve the bracket relations, extending the algebra to many-body systems. This has profound implications as it means that the bracket structure respects locality; operators acting on distinct subsystems induce bracket operations on distinct coordinate sets. As a result, the entire machinery of quantum dynamics can be lifted into phase-space while preserving its compositional structure. This allows us to recast all types of standard quantum dynamics, from unitary and dissipative, to imaginary-time, entirely in terms of bracket operations on functions over $(S^2)^N$. We now walk through each case in turn, each one corresponding to a distinct bracket flow.

\begin{figure}[t!]
\includegraphics[width=\textwidth]{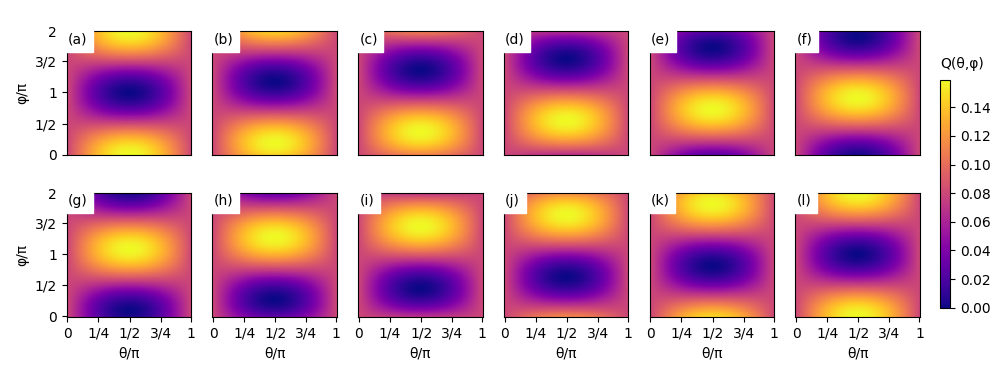}
\caption{$Q$ representation of the time evolution of the $\ket{+}$ state under the Hamiltonian $\hat H=\hat \sigma_z$, for total time $T=2\pi$. Panels (a) -- (l) represent the $Q$ function of the state at times $t \in \{\frac{k \pi}{11}\}_{k = 0}^{11}$ respectively. Observe that the $Q$ function appears to translate in the $\varphi$ direction, corresponding to a rotation around the $z$-axis.}
\label{fig:z_evol}
\end{figure}

\begin{proposition}[Unitary Evolution]\label{prop:unitary_evol_eq}
The von Neumann equation is represented in phase-space as:
\begin{align}
\frac{\partial Q_{\hat\rho}}{\partial t} = [\![Q_{\hat H}, Q_{\hat\rho}]\!],    
\end{align}
where $Q_{\hat{H}}$ is the phase-space representation of the Hamiltonian, and $Q_{\hat{\rho}}$ that of the state. 
\end{proposition}
Unitary evolution therefore corresponds to a sine-bracket flow, which we can see more explicitly in the following two examples:
\begin{example}[Single-Qubit evolution]\label{ex:one_qubit_relaxation}
Let the initial state be
\begin{equation}
    \hat\rho = \frac{1}{2} \left( \hat{\mathbb{I}} + \hat \sigma_x \right),
\end{equation}
and the Hamiltonian be $\hat{H} = \hat{\sigma}_z$. In the $Q$-representation, the state becomes,
\begin{equation}
    Q_{\hat\rho}(\theta, \varphi) = \frac{1}{2} \left( 1 +  \sin\theta \cos\varphi \right),
\end{equation}
and the Hamiltonian is,
\begin{equation}
    Q_{\hat H}(\theta, \varphi) = \cos\theta.
\end{equation}
The unitary evolution is then governed by the sine bracket:
\begin{equation}
    \frac{\partial Q_{\hat\rho}}{\partial t} = [\![Q_{\hat H}, Q_{\hat\rho}]\!] = \frac{1}{\sin\theta} \left( \partial_\theta Q_{\hat H} \cdot \partial_\varphi Q_{\hat\rho} - \partial_\varphi Q_{\hat H} \cdot \partial_\theta Q_{\hat\rho} \right).
\end{equation}
Since $Q_H$ generates rotations about the $z$-axis of $S^2$, the state precesses azimuthally,
\begin{equation}
    Q_{\hat\rho}(\theta, \varphi, t) = Q_{\hat\rho}(\theta, \varphi - 2t),
\end{equation}
for which the dynamics are shown in Fig.~\ref{fig:z_evol}.
\end{example}

\begin{figure}[h!]
\begin{subfigure}{\textwidth}
\includegraphics[width=\textwidth]{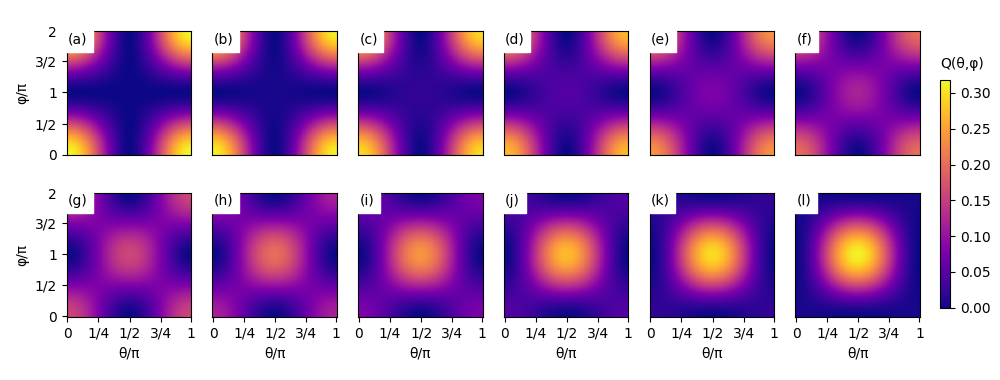}
\caption{}
\end{subfigure}
\begin{subfigure}{\textwidth}
\includegraphics[width=\textwidth]{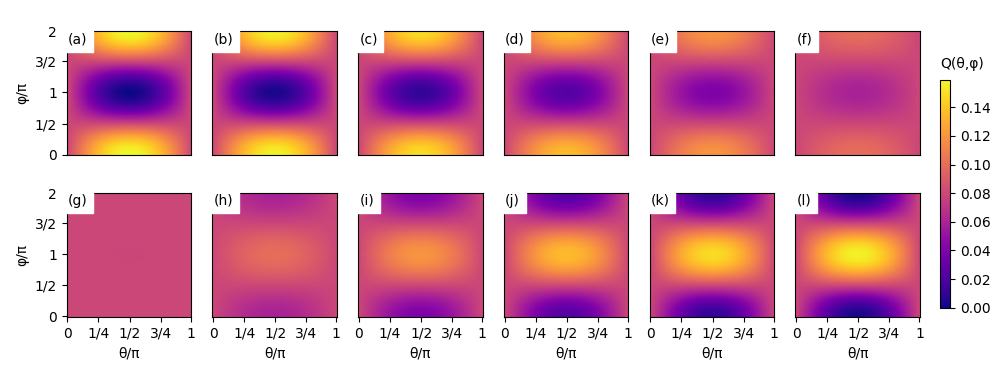}
\caption{}
\end{subfigure}
\caption{$Q$ representation of the time evolution of the $\ket{{+}{+}}$ state under the Hamiltonian $\hat H=\hat\sigma_z\otimes\hat\sigma_z$, for total time $T=\pi$, for (a) the $\theta_1=\theta_2=\frac{\pi}{2}$ slice and (b) marginalizing over the second qubit. Panels (a) -- (l) in both of the plots represent the $Q$ function of the state at times $t \in \{\frac{k \pi}{12}\}_{k = 0}^{12}$
respectively. Non-separability implies entanglement, which means that the partial trace is mixed. This is visible in (b) (lower), where we have visualised only the single-qubit marginal. Since the Hamiltonian is entangling, it will change the degree to which this qubit is mixed. We see this in the snap-shots since the panels of (b) show how the second qubit oscillates between being pure and maximally mixed. 
Specifically, at $t=\pi/2$ (the first plot in the second row of (b)), we see a snapshot of the maximally mixed state (i.e. uniform density), and at the end-points we recognise the heat map of a Pauli eigenstate, which is pure (see Fig.~\ref{fig:q_pauli_eigens}).}
\label{fig:zz_evol}
\end{figure}

\begin{example}[Two-qubit evolution]
Suppose that $\hat H=\hat\sigma^{(1)}_z\hat\sigma^{(2)}_z$. By the tensor compatibility of the sine bracket,
\begin{equation}
    [\![Q_{\hat H}, Q_{\hat\rho}]\!] = \tfrac{1}{2} \left( [\![Q^{(1)}_{\hat \sigma_z}, \{\!\!\{Q^{(2)}_{\hat \sigma_z}, Q_{\hat\rho}\}\!\!\}] \!]+ \{\!\!\{Q^{(1)}_{\hat \sigma_z}, [\![Q^{(2)}_{\hat \sigma_z}, Q_{\hat\rho}]\!] \}\!\!\} \right),
\end{equation}
where \( Q_{\hat H} = Q_{\hat \sigma_z^{(1)} \hat \sigma_z^{(2)}} \). Applying the differential forms of the brackets then yields
\begin{equation}
    \frac{\partial Q}{\partial t} = 2\left( \mathcal{K}^{(1)}_{\hat\sigma_z} \mathcal{J}^{(2)}_{\hat \sigma_z} + \mathcal{J}^{(1)}_{\hat\sigma_z} \mathcal{K}^{(2)}_{\hat \sigma_z} \right) Q.
\end{equation}
Using the definitions for the rotation and boost operators, we find
\begin{align}
\frac{\partial Q}{\partial t}&=
2\left[\left(-\cos\theta_1+\sin\theta_1\frac{\partial}{\partial\theta_1}\right)\frac{\partial}{\partial\varphi_2}+\frac{\partial}{\partial\varphi_1}\left(-\cos\theta_2+\sin\theta_2\frac{\partial}{\partial\theta_2}\right)\right]Q.
\end{align}
This unitary evolution entangles the qubits through azimuthally coupled dynamics across both spheres. The dynamics for this system cannot be visualised directly in two spatial dimensions. As such we visualise one subspace's marginal and a heat-map of one variable per system in Fig.~\ref{fig:zz_evol}.
\end{example}

\begin{figure}[t!]
\includegraphics[width=\textwidth]{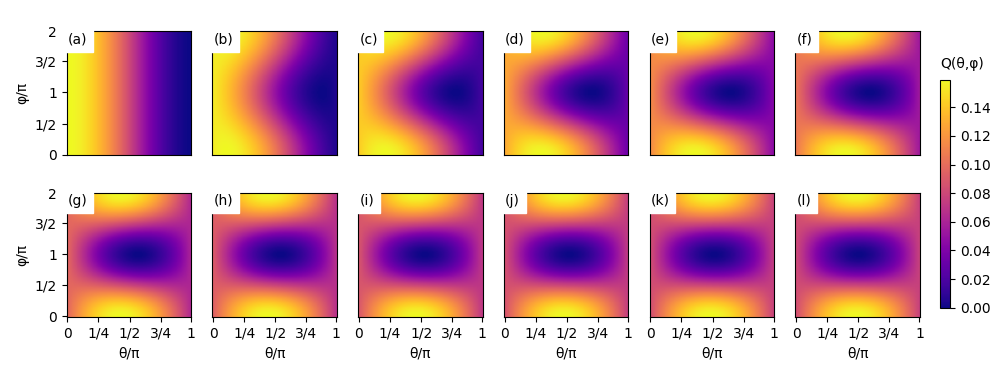}
\caption{$Q$ representation of the imaginary time evolution of the $\ket{0}$ state under the Hamiltonian $\hat H=-\hat\sigma_x$, for total imaginary time $\tau=2$, normalized so that it represents a valid density. Panels (a) -- (l) represent the $Q$ function of the state at imaginary times $\tau \in \{\frac{k}{6}\}_{k=0}^{11}$
respectively. The $Q$ function in the infinite-time limit is exactly the $Q$ function of the $\ket{+}$ state, which is the ground state of $\hat H$.}
\label{fig:x_thermal}
\end{figure}

On the other hand, imaginary-time evolution, often used in ground-state preparation and statistical mechanics, corresponds to a gradient flow generated by the cosine bracket. This is summarised in the following proposition:
\begin{proposition}[Imaginary-Time Evolution]\label{prop:imag_time_evol_eq}
The gradient flow associated to imaginary time evolution is:
\begin{align}
    \frac{\partial Q_{\hat{\rho}}}{\partial t} = -\{\!\!\{Q_{\hat H}, Q_{\hat\rho}\}\!\!\}.
\end{align}
\end{proposition}
We supplement this proposition with the following example:
\begin{example}[Imaginary time evolution]
Suppose we want to find the ground state of the Hamiltonian $\hat H=-\hat\sigma_x$. We may start in any state that is not an eigenstate of $\hat H$, say $\ket{0}$, then perform imaginary time evolution for some (large) time $T$. By Prop.~\ref{prop:imag_time_evol_eq}, the equation of motion for the $Q$ function is given by
\begin{align}
\frac{\partial Q}{\partial t}&=-\{\!\!\{Q_{-\hat\sigma_x},Q\}\!\!\}=2\mathcal{K}_{\hat\sigma_x}Q,
\end{align}
where from Prop.~\ref{prop:cosine_bracket}, we know that
\begin{equation}
\mathcal{K}_{\hat\sigma_x}=\sin\theta\cos\varphi+\cos\theta\cos\varphi\frac{\partial}{\partial\theta}-\csc\theta\sin\varphi\frac{\partial}{\partial\varphi}.
\end{equation}
Hence, we have a PDE for the quasi-probability, for which the dynamics are shown in Fig.~\ref{fig:x_thermal}.
\end{example}

\begin{figure}
\includegraphics[width=\textwidth]{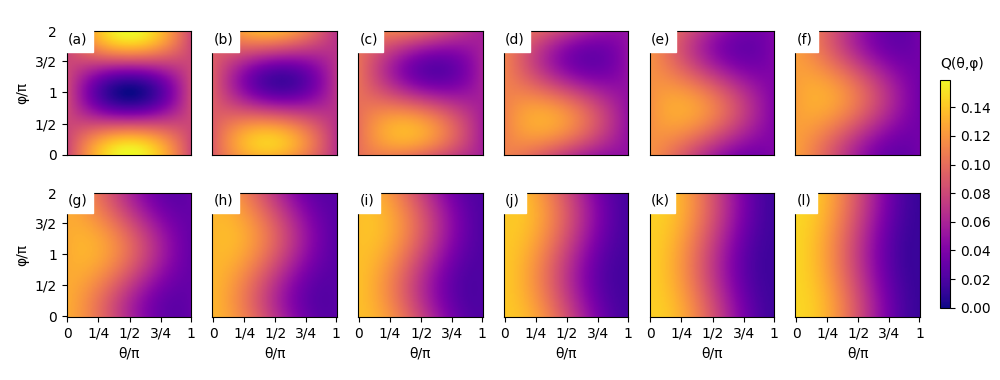}
\caption{$Q$ representation of the Lindblad evolution of the $\ket{+}$ state under the Hamiltonian $\hat H=\hat\sigma_z$ for total time $T=2\pi$, and subject to dephasing and amplitude damping noise with Lindblad operators $\hat\sigma_z$ and $\hat\sigma_-$ respectively. The noise strength is set to $\gamma=0.3$. Panels (a) -- (l) represent the $Q$ function of the state at times $t \in \{\frac{2k \pi}{11}\}_{k = 0}^{11}$
respectively. Observe that, over time, the peak of the distribution is gradually smeared out in the $\varphi$ direction and drifts towards the $-\theta$ direction. In the infinite-time limit, the $Q$ function approaches that of the $\ket{0}$ state.}
\label{fig:dephase}
\end{figure}

Next, we turn our attention to the Lindblad equation governing the dynamics of open quantum systems:

\begin{proposition}[Lindblad Evolution]\label{prop:lindblad_evol_eq}
Let $\{\hat L_i\}$ be Lindblad jump operators and $\gamma$ a decoherence rate. In Hilbert space, the Lindblad master equation takes the usual form,
\begin{equation}
    \frac{d\hat\rho}{dt} 
    = -i[\hat{H}, \hat\rho] 
    + \gamma \sum_i \left( \hat{L}_i \hat\rho \hat{L}_i^\dagger 
    - \frac{1}{2} \left\{ \hat{L}_i^\dagger \hat{L}_i, \hat{\rho} \right\} \right),
\end{equation}
whilst in phase-space, its form is given by
\begin{align}
\frac{\partial Q_{\hat\rho}}{\partial t} &= [\![Q_{\hat H}, Q_{\hat\rho}]\!] + \frac{\gamma}{4} \sum_i \Big([\![Q_{\hat L_i}, [\![Q_{\hat L_i^\dagger}, Q_{\hat\rho}]\!]]\!] + [\![Q_{\hat L_i^\dagger}, [\![Q_{\hat L_i}, Q_{\hat\rho}]\!]]\!] \nonumber\\
&\quad + i[\![Q_{\hat L_i}, \{\!\!\{Q_{\hat L_i^\dagger}, Q_{\hat\rho}\}\!\!\}]\!] - i[\![Q_{\hat L_i^\dagger}, \{\!\!\{Q_{\hat L_i}, Q_{\hat\rho}\}\!\!\}]\!] \Big).
\end{align}
\end{proposition}
Here, the first term $[\![Q_{\hat H}, Q_{\hat\rho}]\!]$ governs unitary evolution, while the remaining terms define the dissipative flow on the manifold. The Lindblad generators become a combination of nested sine and cosine brackets, preserving complete positivity within the phase-space formalism. Curiously, the imaginary unit now enters through the dissipators, not the Hamiltonian; a reversal of the standard Hilbert space formulation. Let us now consider the following two examples:
\begin{example}[Lindblad evolution]
Suppose we have a system in initial state $\ket{+}$ evolving under the Hamiltonian $\hat H=\hat\sigma_z$, and also subject to dephasing and amplitude damping with strength $\gamma$. The jump operators corresponding to these processes are respectively $\hat\sigma_z$ and $\hat\sigma_-=\frac{1}{2}\left(\hat\sigma_x+i\hat\sigma_y\right)$. By Prop.~\ref{prop:lindblad_evol_eq}, the equation of motion for the $Q$ function is given by
\begin{align}
\frac{\partial Q}{\partial t} &= 2\mathcal{J}_{\hat\sigma_z}Q+\frac{\gamma}{2}\left(\mathcal{J}_{\hat\sigma_x}^2+\mathcal{J}_{\hat\sigma_y}^2+\mathcal{J}_{\hat\sigma_x}\mathcal{K}_{\hat\sigma_y}-\mathcal{J}_{\hat\sigma_y}\mathcal{K}_{\hat\sigma_x}+2\mathcal{J}_{\hat\sigma_z}^2\right)Q,
\end{align}
where the operators $\mathcal{J}_{\hat P}$ and $\mathcal{K}_{\hat P}$, $\hat{P}\in\{\hat\sigma_x,\hat\sigma_y,\hat\sigma_z\}$ are respectively given by Prop.~\ref{prop: sine_bracket} and Prop.~\ref{prop:cosine_bracket}. This again yields a PDE for the quasi-probability, for which the dynamics are shown in Fig.~\ref{fig:dephase}.
\end{example}

\begin{example}[1D Ising model]
Consider a 1D transverse-field Ising model, for which the Hamiltonian is given by
\begin{align}
\hat H=\sum_{i=1}^Nh_i\hat\sigma_x^{(i)}+\sum_{i=1}^{N-1}J_i\hat\sigma_z^{(i)}\hat\sigma_z^{(i+1)}
\end{align}
Consider also that each particle is subject to depolarization noise with strength $\gamma$, where the jump operators are
\begin{align}
\hat L_{i,1}&=\hat\sigma_x \qquad\qquad \hat L_{i,2}=\hat\sigma_y \qquad\qquad \hat L_{i,3}=\hat\sigma_z
\end{align}
for all $1\leq i\leq N$.
By Prop.~\ref{prop:bracket_tensor} and \ref{prop:lindblad_evol_eq}, the time evolution of the $Q$ function under open system dynamics is given by
\begin{equation}
    \begin{split}
        \frac{\partial Q}{\partial t}=2\bigg[\sum_{i=1}^Nh_i\mathcal{J}_{\hat\sigma_x}^{(i)}+&\sum_{i=1}^{N-1}J_i\bigg(\mathcal{J}_{\hat\sigma_z}^{(i)}\mathcal{K}_{\hat\sigma_z}^{(i+1)}+\mathcal{K}_{\hat\sigma_z}^{(i)}\mathcal{J}_{\hat\sigma_z}^{(i+1)}\bigg) \\ +\frac{\gamma}{2}&\sum_{i=1}^N\bigg(\mathcal{J}_{\hat\sigma_z}^{(i)}{}^2+\mathcal{J}_{\hat\sigma_z}^{(i)}{}^2+\mathcal{J}_{\hat\sigma_z}^{(i)}{}^2\bigg)\bigg]Q\label{eq:tfim_evol}
    \end{split}
\end{equation}
The first sum corresponds to the transverse field and is a first-order differential operator. The second sum corresponds to the Pauli $ZZ$ couplings, and is a second-order differential operator that is hyperbolic or wave-like in nature. The last sum corresponds to the depolarization noise, and is also a second-order differential operator but is parabolic or diffusion-like in nature.

The Hamiltonian can be augmented with other higher-order interactions, in which case the equation of motion would contain additional terms taking the form from Prop.~\ref{prop:bracket_tensor}. For example, with a next-nearest-neighbour interaction of the form
\begin{align}
\hat H_{nnn}=\sum_{i=1}^{N-2}K_i\hat\sigma_z^{(i)}\hat\sigma_z^{(i+1)}\hat\sigma_z^{(i+2)}
\end{align}
the dynamical equation for $Q$ would contain third-order differential operators of the form
\begin{equation}
\begin{split}
    \frac{\partial Q}{\partial t}=2\bigg[\sum_{i=1}^{N-1}K_i\bigg(\mathcal{J}_{\hat\sigma_z}^{(i)}\mathcal{K}_{\hat\sigma_z}^{(i+1)}\mathcal{K}_{\hat\sigma_z}^{(i+2)}&+\mathcal{K}_{\hat\sigma_z}^{(i)}\mathcal{J}_{\hat\sigma_z}^{(i+1)}\mathcal{K}_{\hat\sigma_z}^{(i+2)} \\
    &+\mathcal{K}_{\hat\sigma_z}^{(i)}\mathcal{K}_{\hat\sigma_z}^{(i+1)}\mathcal{J}_{\hat\sigma_z}^{(i+2)}-\mathcal{J}_{\hat\sigma_z}^{(i)}\mathcal{J}_{\hat\sigma_z}^{(i+1)}\mathcal{J}_{\hat\sigma_z}^{(i+2)}\bigg)\bigg]Q
\end{split}
\end{equation}
Note that the order of the resulting PDE is exactly the coupling order of the Hamiltonian.

\end{example}

\subsection{Dynamics using Feynman-Style Propagators}

Finally, we conclude this section by describing time-evolution through Green's functions, akin to the path-integral formulation. Let $\mathcal{A}=\mathcal{B}(\mathcal{H}_2^{\otimes N})$ be the set of bounded-operators on an $N$-qubit Hilbert space, and its SW image $SW_s: \mathcal{A}\rightarrow\mathcal{F}_s$, where $\mathcal{F}_s$ denotes the image of $\mathcal{A}$ under $SW_s$, that is $\mathcal{F}_s=\mathrm{span}\!\left\{\prod_{j=1}^N Y_{\ell_j m_j}(\Omega_j):\ \ell_j\in\{0,1\},\ m_j=-\ell_j,\ldots,\ell_j\right\}$ (i.e., $\{1,n_{jx},n_{jy},n_{jz}\}$ per site).

\begin{proposition}[Green's functions for Linear Superoperators]
    Given any linear superoperator $\hat \Lambda_t:\mathcal{A}\to\mathcal{A}$,  its SW-induced action in phase space,
    \begin{align}
    \Phi_t:=SW_s \circ\,\hat \Lambda_t\, \circ SW_s^{-1}:\mathcal{F}_s\to\mathcal{F}_s,
    \end{align} 
    admits a propagator,
    \begin{align}
    f_{\hat\rho(t)}^{(s)}(\boldsymbol{\Omega}) =(\Phi_t f^{(s)}_{\hat \rho(0)})(\boldsymbol{\Omega})=\int_{(S^2)^N} G^{(s)}_t(\boldsymbol{\Omega},\boldsymbol{\Omega}')\,f^{(s)}_{\hat \rho(0)}(\boldsymbol{\Omega}')\,d\boldsymbol{\Omega}',
    \end{align}
    where the propagator is defined as
    \begin{align}
    G^{(s)}_t(\boldsymbol{\Omega},\boldsymbol{\Omega}')=(4\pi)^N\text{Tr}\big[\hat \Lambda_t(\hat\Delta^{(-s)}(\boldsymbol{\Omega}'))\hat\Delta^{(s)}(\boldsymbol{\Omega})\big].
\end{align}
\end{proposition}

Hence, we can describe both real and imaginary time evolution propagators by choosing $\hat \Lambda_t(\hat \rho) = \hat{U}(t) \hat \rho \hat{U}(t)^{\dagger}$, where $\hat U(t) = e^{-i \hat H t}$ for real time, and $\hat{U} = e^{-\tau \hat H}$ for imaginary time. Explicitly, this means that the propagator for real-time evolution reads
\begin{align}
G^{(s)}_t(\boldsymbol{\Omega},\boldsymbol{\Omega}')&=(4\pi)^N\,\mathrm{Tr}\left[e^{i\hat Ht}\hat\Delta^{(s)}(\boldsymbol{\Omega})e^{-i\hat Ht}\hat\Delta^{(-s)}(\boldsymbol{\Omega}')\right],
\end{align}
whereas imaginary time reads
\begin{align}
G^{(s)}_{\tau}(\boldsymbol{\Omega},\boldsymbol{\Omega}')&=(4\pi)^N\,\mathrm{Tr}\left[e^{-\hat H\tau}\hat\Delta^{(s)}(\boldsymbol{\Omega})e^{-\hat H\tau}\hat\Delta^{(-s)}(\boldsymbol{\Omega}')\right].
\end{align}

Next, we highlight two important properties of propagators. First, that the composition of two linear maps in Hilbert space $\hat \Lambda_1 \circ \hat \Lambda_2$ is obtained by convolution in phase space:

\begin{proposition}[Composition by convolution]
    If $G_1$ is the propagator of $\hat \Lambda_1$ and $G_2$ is the propagator of $\hat \Lambda_2$, then composition $\Phi_1 \circ \Phi_2$ in phase space is obtained by convolution,
    \begin{align}
    (G_1*G_2)(\boldsymbol{\Omega},\boldsymbol{\Omega}''):=\int_{(S^2)^N} G_1(\boldsymbol{\Omega},\boldsymbol{\Omega}')\,G_2(\boldsymbol{\Omega}',\boldsymbol{\Omega}'')\,d\boldsymbol{\Omega}'.
    \end{align}
\end{proposition}
This endows the propagators with a group or semi-group structure as summarised by the following propositions:
\begin{proposition}[Group structure]
    When linear maps, $\hat{\Lambda}_t$ are one-parameter groups on on $\mathcal{A}$, then the kernels $\{G^{(s)}_t\}_{t\in\mathbb{R}}$ form a group under convolution.
\end{proposition}

\begin{proposition}[Semi-group structure]
    Let $\mathcal{\hat L}: \mathcal{A} \rightarrow\mathcal{A}$ be a Lindblad generator and set $\hat \Lambda_t=e^{t\mathcal{\hat L}}:\mathcal{A}\rightarrow\mathcal{A}$ for $t\ge 0$. Then the induced convolution algebra defines a semigroup in phase space.
\end{proposition}

\section{Phase-Space Dilation and Unification}
\label{sec:dilation_unification}

In this section, we unify quantum dynamics and expectation algebras in the Hilbert and phase-space representations through a series of commuting diagrams. This allows us to see a unified view of Hilbert space and the phase-space construction from Secs.~\ref{sec:spin_phase_space} and \ref{sec:spin_dynamics}, treating them as dual representations of the same underlying dynamics. It also enables us to prove that there is always a phase-space representation for qubit dynamics for which the SW kernel factorises. Without such a guarantee, the kernel itself would inherit an exponential scaling, making the phase-space representation no more tractable than Hilbert space. Our proof is based on realising Stinespring's Theorem in phase-space through the SW kernel. Finally we will derive moment generating functions (MGFs) for qubit phase-space, and briefly discuss their role in the prospect for automatic differentiation (AD) in the phase-space picture, see Sec.~\ref{sec:QML_prospects_discussion} for further discussion.

\subsection{The SW Kernel Always Factorises}

We begin by briefly reviewing Stinespring's dilation theorem in the context of states and dynamics \cite{stinespring1955positive}. Let $\mathcal{H}_A$ be a Hilbert space of dimension $d$, and let $\mathcal{H}_{AB} \supseteq \mathcal{H}_A$ be an extended space with dimension $D \geq d^2$.  
Then, any mixed state $\hat\rho$ on $\mathcal{H}_A$ can be realized as the partial trace of a pure state $\ket{\psi} \in \mathcal{H}_{AB}$, of the extended space $\mathcal{H_{AB}}$,
\begin{equation}
    \hat\rho = \operatorname{Tr}_{B} \left[ \ket{\psi}\bra{\psi} \right],
\end{equation}
Moreover, Stinespring's theorem also applies to operators, allowing us to dilate Lindbladian dynamics into unitary dynamics on the extended space \cite{stinespring1955positive, choi1975completely}. That is, we can take the CPTP map $\hat{\mathcal{E}}(t) = \exp(t \hat{L})$, for a given Lindbladian $\hat{L}$, and dilate it into a unitary $\hat{U}_{\mathcal{E}}(t)$. Hence, any Lindblad evolution on $\mathcal{H}_A$ can be promoted to unitary dynamics on $\mathcal{H}_{AB}$ followed by partial trace,
\begin{equation}
    \hat\rho(t) = \operatorname{Tr}_B \left[ \hat{U}_{\mathcal{E}}(t) \ket{\psi_0}\bra{\psi_0} \hat{U}^\dagger_{\mathcal{E}}(t) \right],
\end{equation}
where $\ket{\psi_0} \in \mathcal{H}_{AB}$ is the purification of $\hat\rho \in \beta_+(\mathcal{H}_A)$ at $t = 0$, and $\hat{U}_{\mathcal{E}}(t)$ is the dilation of the CPTP map $\hat{\mathcal{E}}(t) = \exp(t \hat{\mathcal{L}})$ for a given Lindbladian $\hat{\mathcal{L}}$. We note here that while we can dilate $\hat{\mathcal{E}}(t)$, we cannot dilate the Lindbladian directly while keeping the environment size fixed and the system-environment Hamiltonian time-independent \cite{breuer2002theory, watrous2018theory}. Stinespring's Theorem in Hilbert space can be understood as the following commuting diagram:
\begin{center}
\begin{tikzpicture}[>=Latex, scale=1.5]

\node (A) at (0, 1) {$\hat\rho(t = 0)$};
\node (B) at (3, 1) {$\hat\rho(t)$};
\node (C) at (0, 0) {$\ket{\psi(t = 0)}$};
\node (D) at (3, 0) {$\ket{\psi(t)}$};

\draw[->] (A) -- (B) node[midway, above] {$\hat{\mathcal{E}}(t)$};
\draw[->] (C) -- (D) node[midway, below] {$\hat{U}_{\mathcal{E}}(t)$};

\draw[->] ([xshift=-2pt]A.south) -- ([xshift=-2pt]C.north) node[midway, left] {\small S.D.};
\draw[->] ([xshift=+2pt]C.north) -- ([xshift=+2pt]A.south) node[midway, right] {\small Tr$_B$};

\draw[->] ([xshift=-2pt]B.south) -- ([xshift=-2pt]D.north) node[midway, left] {\small S.D.};
\draw[->] ([xshift=+2pt]D.north) -- ([xshift=+2pt]B.south) node[midway, right] {\small Tr$_B$};

\end{tikzpicture}
\end{center}

Here $\hat{\mathcal{E}}(t):\mathcal{H}_S \times \mathbb{R}\rightarrow\mathcal{H}_S$ is any CPTP operator on the system's Hilbert space, $\hat{\mathcal{E}}(t) = \exp(\hat{L}t)$ where $\hat{L}$ is some Lindbladian operator or Hamiltonian. The symbol $\text{S.D}$ refers to a purification by Stinespring's dilation theorem, with $\hat{U}_{\mathcal{E}}(t): \mathcal{H}_A \otimes \mathcal{H}_B \times\mathbb{R} \rightarrow \mathcal{H}_A \otimes \mathcal{H}_B $ the corresponding unitary operator in the image of the purification, and $\dim(\mathcal{H}_A \otimes \mathcal{H}_B) \geq \dim(\mathcal{H}_A)^2$. Finally,
$\mathrm{Tr}_{B}: \mathcal{H}_{AB} \rightarrow\mathcal{H}_A$ refers to tracing out the auxiliary degrees of freedom.

Next, let us consider how the SW correspondence works on this commuting diagram. Let $SW_A^{(s)}$ denote the (invertible) SW map for the $A$ system only, with its usual $s$-index. Further, let $SW_{AB}^{(s)}$ denote the SW map for the composite $AB$ Hilbert space. Then we may construct the following commuting diagram,

\begin{center}
\begin{tikzpicture}[>=Latex, scale=1.5]

\node (A) at (0, 1) {$\hat\rho(t = 0)$};
\node (B) at (3, 1) {$\hat\rho(t)$};
\node (C) at (0, 0) {$\ket{\psi(t = 0)}$};
\node (D) at (3, 0) {$\ket{\psi(t)}$};

\node (A') at (-1, 2) {$f^{(s)}_{\hat\rho(0)}$};
\node (B') at (4, 2) {$f^{(s)}_{\hat\rho(t)}$};
\node (C') at (-1, -1) {$f^{(s)}_{\ket{\psi(0)}}$};
\node (D') at (4, -1) {$f^{(s)}_{\ket{\psi(t)}}$};

\draw[->] (A) -- (B) node[midway, above] {$\hat{\mathcal{E}}(t)$};
\draw[->] (C) -- (D) node[midway, below] {$\hat{U}_{\mathcal{E}}(t)$};

\draw[->] ([xshift=-2pt]A.south) -- ([xshift=-2pt]C.north) node[midway, left] {\small S.D.};
\draw[->] ([xshift=+2pt]C.north) -- ([xshift=+2pt]A.south) node[midway, right] {\small Tr$_B$};

\draw[->] ([xshift=-2pt]B.south) -- ([xshift=-2pt]D.north) node[midway, left] {\small S.D.};
\draw[->] ([xshift=+2pt]D.north) -- ([xshift=+2pt]B.south) node[midway, right] {\small Tr$_B$};

\draw[->] (A') -- (B') node[midway, above] {$\mathfrak{E}(t)$};
\draw[->, dashed] (A') -- (C') node[midway, left] {$\mathcal{S.D.}$};
\draw[->] (C') -- (D') node[midway, below] {$\mathfrak{U}(t)$};
\draw[->] (D') -- (B') node[midway, right] {$\mathcal{T}_B$};

\draw[<->] (A') -- (A) node[midway, above,xshift=10pt] {$SW_A^{(s)}$};
\draw[<->] (B') -- (B) node[midway, above,xshift=-10pt] {$SW_A^{(s)}$};
\draw[<->] (C') -- (C) node[midway, below,xshift=10pt] {$SW_{AB}^{(s)}$};
\draw[<->] (D') -- (D)node[midway, below,xshift=-10pt] {$SW_{AB}^{(s)}$};

\end{tikzpicture}
\end{center}
Here $\mathfrak{E}(t)$ is the phase-space correspondence of $\hat{\mathcal{E}}(t)$, for example arising from the application of the exponential map of some Hamiltonian or Lindbladian on the $A$-system. Next, $\mathcal{T}_B$ refers to marginalising over the dilation coordinates in phase-space via Prop~\ref{prop:partial_trace_phase}, and $\mathfrak{U}(t)$ is the phase-space correspondence of $\hat{U}_{\mathcal{E}}(t)$ in phase-space. Note that $\mathfrak{U}(t)$ is strictly unitary as it is the image of the SW map for a purified CPTP map.

The implication of this diagram is two-fold. First, notice that since the phase-space construction is consistent with partial tracing; we can therefore choose whether to employ the SW correspondence on a mixed state, or its purification and get consistent results when taking into account general dynamics.  

Second, notice the dashed line can be found by function composition, $\mathcal{S.D} = SW_{AB}^{(s)} \circ S.D \circ (SW_{A}^{(s)})^{-1}$, where $(\cdot)^{-1}$ denotes the inverse mapping per Eq.~\ref{eq:SW_correspondence}. We therefore arrive at the following proposition:
\begin{proposition}[Dilation in Phase-Space]\label{prop:dilation_in_phase_space}
    Consider a mixed state $\hat\rho \in \beta(\mathcal{H}_A)$ and its dilation into a composite system $\ket{\psi} \in \mathcal{H}_A \otimes \mathcal{H}_B$, where $\dim (\mathcal{H}_A \otimes \mathcal{H}_B) \geq \dim(\mathcal{H}_A)^2$. Then defining $f^{(s)}_{\hat\rho} = SW_A(\hat\rho) = \operatorname{Tr}_A(\hat\rho \hat{\Delta}_A)$ and $f^{(s)}_{\ket{\psi_{}}} = SW_{AB}(\ket{\psi_{}})=\operatorname{Tr}_{AB}(\ket{\psi_{}}\!\bra{\psi_{}} \hat{\Delta}_{AB})$. Then we may write
    \begin{equation}
        f^{(s)}_{\hat\rho} = SW_A(\hat\rho_{})=\int_{\mathcal{M}_B}d\mu_B\;SW_{AB}(\ket{\psi_{}}),
    \end{equation}
    where $\mathcal{M}_B = \bigtimes_{N(N-1)} S^2$ is the minimal dimension of the $B$-system's manifold.
\end{proposition}
Prop.~\ref{prop:dilation_in_phase_space} closes the above diagram, and more compactly gives the following commuting diagram,
\begin{center}
\begin{tikzpicture}[>=Latex, scale=1.5]

\node (A) at (0, 1) {$f^{(s)}_{\hat\rho(0)}$};
\node (B) at (3, 1) {$f^{(s)}_{\hat\rho(t)}$};
\node (C) at (0, 0) {$f^{(s)}_{\ket{\psi(0)}}$};
\node (D) at (3, 0) {$f^{(s)}_{\ket{\psi(t)}}$};

\draw[->] (A) -- (B) node[midway, above] {$\hat{\mathfrak{E}}(t)$};
\draw[->] (C) -- (D) node[midway, below] {$\mathfrak{U}(t)$};

\draw[->] ([xshift=-2pt]A.south) -- ([xshift=-2pt]C.north) node[midway, left] {\small $\mathcal{S.D}$};
\draw[->] ([xshift=+2pt]C.north) -- ([xshift=+2pt]A.south) node[midway, right] {\small$\mathcal{T}_B$};

\draw[->] ([xshift=-2pt]B.south) -- ([xshift=-2pt]D.north) node[midway, left] {\small S.D.};
\draw[->] ([xshift=+2pt]D.north) -- ([xshift=+2pt]B.south) node[midway, right] {\small Tr$_B$};
\end{tikzpicture}
\end{center}
which formally establishes a direct analogue of Stinespring’s theorem within phase-space via an extended kernel. Note that this diagram is just the outer layer of the first one. Now that we have a mechanism for purifying states and dilating operators, we need not make further references to Hilbert space beyond the SW kernel; applying the SW correspondence to a reduced state $\hat\rho$ yields the same result as marginalising over the ancillary coordinates in the joint quasi-probability distribution of its purification, as guaranteed by Prop.~\ref{prop:partial_trace_phase}.

\begin{remark}
    We identify the kernel in the dilated space:  $\hat{\Delta}_{AB} = \hat{\Delta}^{(s)}_A \otimes \frac{1}{d_B} \hat{\Delta}_B$ and the relation with the open system kernel: $\hat{\Delta}_A^{(s)} = Tr_B(\hat{\Delta}_{AB})$. In particular, since in the purified setting the kernel factorizes, we see that it is always possible to find a representation of open-system dynamics for which the kernel factorises by dilating in phase-space.
\end{remark}

This observation is especially powerful in the context of dynamics. As shown in \cite{navarette2025phase}, simulating open-system evolution directly in phase-space requires evolving the phase-space kernel itself; an operation that becomes computationally expensive unless the kernel factorises. However, by dilating the dynamics into a higher-dimensional phase-space $\mathcal{M}_{AB}$, where the SW kernel always factorises by construction, we can circumvent this problem. Even if the kernel fails to factorise on $\mathcal{M}_A$, there always exists a dilation to $\mathcal{M}_{AB}$
where it does, allowing dynamics to be lifted into a representation with more tractable structure. We emphasise that such a higher-dimensional space has a quadratically increased domain, per the Hilbert space picture. However, since the domain in phase-space scales linearly in the number of qubits, this is only a quadratic cost with respect to system size, rather than an exponential one.

Furthermore, since Stinespring's Theorem in phase-space is $s$-independent, we also have the following commuting diagram
\begin{center}
\begin{tikzpicture}[>=Latex, scale=1.5]

\node (A) at (0, 1) {$f^{(s)}_{\hat\rho_A}$};
\node (B) at (3, 1) {$f^{(s')}_{\hat\rho_A}$};
\node (C) at (0, 0) {$f^{(s)}_{\ket{\psi_{AB}}}$};
\node (D) at (3, 0) {$f^{(s')}_{\ket{\psi_{AB}}}$};

\draw[<->] (A) -- (B) node[midway, above] {$\mathcal{C.R}$};
\draw[<->] (C) -- (D) node[midway, below] {$\mathcal{C.R}$};

\draw[->] ([xshift=-2pt]A.south) -- ([xshift=-2pt]C.north) node[midway, left] {\small $\mathcal{S.D}$};
\draw[->] ([xshift=+2pt]C.north) -- ([xshift=+2pt]A.south) node[midway, right] {\small$\mathcal{T}_B$};

\draw[->] ([xshift=-2pt]B.south) -- ([xshift=-2pt]D.north) node[midway, left] {\small $\mathcal{S.D}$};
\draw[->] ([xshift=+2pt]D.north) -- ([xshift=+2pt]B.south) node[midway, right] {\small Tr$_B$};
\end{tikzpicture}
\end{center}
where $\mathcal{C.R}$ refers to a change in phase-space representation according to Fig.~\ref{fig:pwq_map_spin} and Eq.~(\ref{eq:change_rep_spins}). In practise, this means we are free to choose a representation in which dilation is most practical. Let us now see dilation in practise,

Having unified the phase-space and Hilbert space representations of quantum dynamics, we now turn our attention to a unified view of expectation algebra in phase-space.

\subsection{Moment Generating Functions}
Let us now discuss expectation algebra in this unified picture. Following Eq.~\ref{eq:expectation_phase_many_body} expectation values can be computed by phase-space integration. As such, numerical integration can be effectively perform with the Monte-Carlo integration techniques \cite{becca2017quantum, landau2021guide}.  Additionally, expectation values of the correlation functions can be obtained via differentiation of Moment Generating Functions (MGFs), opening an avenue for using automatic differentiation (AD) for observable estimation \cite{griewank2008evaluating, baydin2018automatic} (see Sec.~\ref{sec:QML_prospects_discussion} for further discussion on AD).

\begin{proposition}[One Qubit Moment Generating Function]
    The one-qubit MGF, $\chi_{{\hat\rho}}^{(s)}: \mathbb{R}^3 \times [-1,1] \rightarrow \mathbb{R}$ is given by
    \begin{equation}
    \chi_{\hat{\rho}}^{(s)}(\omega) := \int_{S^2} d\Omega \; f^{(s)}_{{\hat\rho}}(\Omega) \, e^{\omega \cdot n(\Omega)},
    \end{equation}
    where $\omega = (\omega_x, \omega_y,\omega_z)\in\mathbb{R}^3$, $s \in [-1,1]$ is the usual phase-space representation index, and the dual vector field $n: S^2 \hookrightarrow \mathbb{R}^3$ is the Bloch vector,
    \begin{equation}
    n = (\sin \theta \cos \varphi, \sin \theta\sin \varphi,\cos \theta).
    \end{equation}
    From this generating function, Pauli moments are obtained via
    \begin{equation}
    \left\langle \hat{\sigma}_{\mu} \right\rangle_{\hat\rho}
    = \frac{3}{\lambda(s)}\left. \frac{\partial \chi_{{\hat\rho}}^{(s)}}{\partial \omega_{\mu}} (\omega)\right|_{\omega = 0}
    \end{equation}
    where $\lambda(s)$ is given by Eq.~(\ref{eq:lambda_cases}).
\end{proposition}
To see how this works in practise, we turn to the following example:
\begin{example}[One Qubit $Q$-function moments]
    we can see the Q-function at $s = -1$ is given by 
\begin{equation}
    \chi_{\hat{\rho}}^{(-1)}(\omega) = \int_{S^2} d \Omega \; Q_{{\hat\rho}}(\Omega) \, e^{\omega \cdot n},
\end{equation}
Then $\langle \hat{\sigma}_x \rangle$ is obtained as
\begin{equation}
    \langle \hat{\sigma}_x \rangle = \left. 3 \cdot\frac{\partial \chi_{{\hat\rho}}^{(-1)} (\omega)}{\partial \omega_x} \right|_{\omega = 0},
\end{equation}
where the prefactor \( 3/\lambda(s) = 3 \) since $\lambda(-1) = 1$.
\end{example}

Owing to the structure of the Pauli algebra, only first-order moments in each local subspace matter. Higher powers of Pauli operators reduce via $\hat{\sigma}_\mu^2 = \hat{\mathbb{I}}$ and the $\mathfrak{su}(2)$ commutation relations. We now turn to many-qubit MGFs, for which we arrive at the following proposition:

\begin{proposition}[Many-qubit MGFs]
    The many-qubit MGF, $\chi_{\hat{\rho}}^{{(s)}}: \mathbb{R}^{3N} \times [-1,1] \rightarrow \mathbb{R}$ is given by
    \begin{equation}\label{eq:MGF_many_body_dfn}
        \chi_{\hat{\rho}}^{(s)}(\boldsymbol{\omega}) = \int_{(S^2)^N} d\boldsymbol{\Omega} \; f^{(s)}_{\hat{\rho}}(\boldsymbol{\Omega}) \, \exp\left( \boldsymbol{\omega} \cdot \boldsymbol{n} \right),
    \end{equation}
    where $\boldsymbol{\omega} \in \mathbb{R}^{3N}$ is such that $\boldsymbol{\omega} = (\omega_x^{(1)},\omega_y^{(1)},\omega_z^{(1)},\ldots,\omega_z^{(N)})$, and $\boldsymbol{n} = (n^{(1)},\ldots,n^{(N)}): (S^2)^N \hookrightarrow \mathbb{R}^{3N}$ is such that
    \( n^{(i)}(\Omega_i) = (\sin \theta_i \cos \varphi_i, \sin \theta_i \sin \varphi_i, \cos \theta_i)\).
    Then for any Pauli string of full weight $N$, we have
    \begin{equation} \label{eq:many_body_MGF}
        \left. \frac{\partial^N \chi_{\hat{\rho}}^{(s)}}{\partial \omega_{\mu_1}^{(1)} \cdots \partial \omega_{\mu_N}^{(N)}} (\boldsymbol{\omega})\right|_{\boldsymbol{\omega} = 0} 
        = \left(\frac{3}{\lambda(s)}\right)^N \left\langle \hat  \sigma_{\mu_1}^{(1)} \otimes \cdots \otimes \hat \sigma_{\mu_N}^{(N)} \right\rangle_{\hat\rho},
    \end{equation}
    where $\lambda(s)$ is given by Eq.~(\ref{eq:lambda_cases})
\end{proposition}

In direct analogy to the one-qubit MGF, only moments which are first-order in each subspace matter. This is because, as before, $\hat{\sigma}_{\mu_j}^n = \hat{\mathbb{I}}$ for even $n$ and $\hat{\sigma}_{\mu_j}^n = \hat{\sigma}_{\mu_j}$ for odd $n$. Further, the structure of the many-qubit Pauli algebra means products of local operators always reduce to first order in each subspace. The following example highlights how to use MGFs for two-point correlations:
\begin{example}[Many-qubit two-point correlation of $Q$-function]
    to compute a two-point correlation function $\langle \hat{\sigma}_{\mu_i} \hat{\sigma}_{\mu_j}\rangle$ for qubits $i \neq j$, we evaluate
\begin{equation}
    \left\langle \hat{\sigma}_{\mu_i} \hat{\sigma}_{\mu_j} \right\rangle_{\hat{\rho}}
= \left.\left( \frac{3}{\lambda(s)} \right) ^2 \;\frac{\partial^2 \chi_{\hat{\rho}}^{(s)}}{\partial \omega_{\mu_i}^{(i)} \partial \omega_{\mu_j}^{(j)}} (\boldsymbol{\omega}) \right|_{\boldsymbol{\omega} = 0}.
\end{equation}
\end{example}
Although we chose the $Q$-functions as examples, we may also work in other representations, changing representations directly at the level of the MGF. This is because For different $s$-representations, MGFs are related by a rescaling,
\begin{equation} \label{eq:rep_change_chi}
    \chi_{\hat{\rho}}^{(s')}(\omega) = \chi_{\hat{\rho}}^{(s)}\left( \tfrac{\lambda(s')}{\lambda(s)} \cdot \omega \right), \quad \text{for } s,s' \ne 0,
\end{equation}
due to the pre-factor $\lambda(s)$ in Eq.~(\ref{eq:bloch_phase_function}). This remains true for many-bodies where we have \( Y^{(s)}_{\sigma_{\mu}}(\Omega_j) = \lambda(s) n_{\mu}(\Omega_j) \), where $\mu \in \{x,y,z\}$.

The moment generating function $\chi_{\hat{\rho}}^{(s)}(\omega)$ encodes the full content of the phase-space function $f^{(s)}(\Omega)$ through its derivatives. In fact, for systems of qubits, the inverse problem is trivial; the first-order derivatives of $\chi_{\hat{\rho}}^{(s)}$ at $\omega = 0$ suffice to reconstruct the coefficients of $f^{(s)}$ in the dual basis. More generally, we may formally regard $f^{(s)}$ as a Laplace-type inverse of $\chi_{\hat{\rho}}^{(s)}$ pulled back along the dual vector field $n^{(-s)}(\Omega)$,
\begin{equation}\label{eq:chi_to_phase_space}
    f^{(s)}_{\hat{\rho}}(\Omega) = \frac{1}{(2\pi)^3} \int_{\mathbb{R}^3} d^3\omega \; \chi_{\hat{\rho}}^{(s)}(\omega) \, e^{-\omega \cdot n(\Omega)}, \quad \Omega \in S^2
\end{equation}
in direct analogy to the Fourier transform that relates MGFs to phase-space function in quantum optics (see App.\ref{app:bosonic_phase_space}).

We now have two equivalent mechanisms for computing expectation values over the Pauli algebra: either via direct phase-space integration, or via derivatives of MGFs. These constructions are connected by the following commuting diagram:
\begin{center}
\begin{tikzpicture}[>=Latex, scale=1.4]

\node (A) at (0, 2) {$f^{(s)}_{\hat{\rho}}$};
\node (B) at (4, 2) {$f^{(s')}_{\hat{\rho}}$};
\node (C) at (0, 0) {$\chi_{\hat{\rho}}^{(s)}$};
\node (D) at (4, 0) {$\chi_{\hat{\rho}}^{(s')}$};

\node (E) at (2, 1) {$\left\langle \hat{\sigma}_{\mu_1} \cdots \hat{\sigma}_{\mu_n} \right\rangle_{\hat{\rho}}$};

\draw[<->] (A) -- (B) node[midway, above] {$\mathcal{C.R}$};

\draw[<->] ([xshift=-2pt]A.south) -- ([xshift=-2pt]C.north) node[midway, left] {MGF};
\draw[<->] ([xshift=+2pt]B.south) -- ([xshift=+2pt]D.north) node[midway, right] {MGF};

\draw[->] (A) -- (E) node[midway, above] {$\int$};
\draw[->] (B) -- (E) node[midway, above] {$\int$};

\draw[->] (C) -- (E) node[midway, below] {$\partial$};
\draw[->] (D) -- (E) node[midway, below] {$\partial$};
\draw[<->](C) -- (D) node[midway, below] {$C.R$};

\end{tikzpicture}
\end{center}
Here $\mathcal{C.R}$ refers to changing representation in phase-space, whilst $C.R$ refers to changing representation at the level of the MGF via Eq.~(\ref{eq:rep_change_chi}), and the vertical arrows apply the MGF transformation from Eq.~(\ref{eq:MGF_many_body_dfn}), with the inverse given by Eq.~(\ref{eq:chi_to_phase_space}). Lastly, the symbols $\int$ and $\partial$ are placeholders for phase-space integration, Eq.~(\ref{eq:expectation_phase_many_body}), and MGF differentiation at the origin, Eq.~(\ref{eq:many_body_MGF}), respectively.

\section{Discussion}\label{sec:discussion}

What is striking here is the uniformity of the framework. The same two bracket operations suffice to encode every kind of evolution. In Hilbert space, unitarity, dissipation, and imaginary time involve qualitatively distinct mechanisms. Here, they are all unified in a single algebraic language acting on a geometric domain. This allows us to think differently about quantum dynamics. Instead of operators evolving vectors in a linear space, we now have functions evolving under bracket flows on a curved manifold. Let us now discuss the curse of dimensionality in this representations, and its prospects for QML.

\subsection{On the Curse of Dimensionality}\label{sec:curse_of_dim}

The Stratonovich–Weyl correspondence is a linear and injective map, so Hilbert-space and phase-space representations ultimately encode the same exponential number of degrees of freedom. What changes is the way these degrees of freedom appear: in phase space they manifest as the spherical-harmonic content of a real function. In this view, the complexity of a many-body state is measured not by the number of basis amplitudes, but by the harmonic resolution required to express the function. A key observation is that the domain of this function, $(S^2)^N$, has dimension $2N$ growing only linearly with system size. While this does not eliminate the exponential overhead, it recasts it as bandwidth rather than coordinate dimension.


The broader benefits of a qubit phase-space construction is structural unification with infinite-dimensional Hilbert spaces like those from quantum optics. In both cases, phase-space representations assign quasi-probability functions to quantum states, and dynamics proceeds via deformation of the pointwise product. The sine and cosine brackets are the qubit analogues of the Moyal and symmetric products. The underlying geometry changes (from \(\mathbb{R}^{2N}\) to \((S^2)^N\)), but the algebraic framework persists. This allows a single language for the analysis of finite and infinite-dimensional quantum systems. Both can now be analysed in the same representational terms. Hybrid systems like cavity QED and spin-boson models then fit naturally into this structure.

\subsection{On the prospects for QML}\label{sec:QML_prospects_discussion}

A many-qubit phase space could be a promising avenue for neural-network-based inference and control because its structure is more amenable to neural network approximation. The reason for this is two-fold. First, as discussed in Sec.~\ref{sec:curse_of_dim}, phase-space contains the curse of dimensionality in a fundamentally different way than Hilbert space. Second, the wave-function is a smooth, non-linear function of the domain with continuous-valued inputs. As such, learning or representing quantum states in the phase-space picture may be a more natural language for deep learning, with non-linear function fitting being once of its central ideas \cite{prince2023understanding}.

For example, we can consider how NQS \cite{carleo2017solving,lange2024architectures} could work differently in this picture.  In Hilbert space, as originally done by \cite{carleo2017solving}, a NQS learns a map from a binary input to the complex numbers, or reals if we use the POVM representation (and change the domain from $\{0,1\}^N$ to $\{0,1,2,3\}^N$). In contrast, a phase space NQS would learn a map $f: (S^2)^N \rightarrow\mathbb{R}$. This distinction is conceptually useful: nonlinear function fitting over continuous coordinates is an area where machine learning has shown scalable success. Unlike neural quantum states (NQS) on discrete inputs, the SW framework frames learning as approximation of a continuous function on a linearly growing input space, while keeping the curse of dimensionality in the harmonic resolution. Moreover, we have a direct mechanism for any loss function to be physics-informed in the physical subspace of $L^2\big((S^2)^N\big)$ via Eq.~(\ref{eq:physical_subspace}) as a Lagrange multiplier that should converge to zero once an architecture is trained.

A second prospect for QML in phase-space is that the MGF from Eq.~(\ref{eq:MGF_many_body_dfn}) provides a natural route to extracting Pauli expectation values via automatic differentiation (AD). Since these moments correspond to derivatives of $\chi_{\hat{\rho}}^{(s)}$ evaluated at $\omega = 0$, one may compute observables directly by differentiating through the exponential kernel in the definition of the MGF. When $f^{(s)}$ is known, this process may be accelerated using fast-Fourier transform techniques \cite{nussbaumer1982fast, kunis2003fast, mohlenkamp1999fast, hecht2018fast}.

However, executing this forward transform requires approximating integration over $(S^2)^N$, which becomes increasingly costly with system size and non-separability \cite{kunis2003fast, hecht2018fast}. For this reason, it is natural to consider modelling $\chi_{\hat{\rho}}^{(s)}$ directly, bypassing the need to explicitly construct $f^{(s)}$ at all. In this setting, a learned representation of the MGF acts as a differentiable surrogate for the quantum state itself. More ambitiously, one may model the logarithm, $\log \chi_{\hat{\rho}}^{(s)}$, where the combinatorics of differentiation simplify: products of moments become sums of cumulants, and partial derivatives may be more numerically stable \cite{hardy2006combinatorics}.

Finally, let us consider how the prospects of QML in dynamical settings within the phase-space picture. Phase-space methods evolve real-valued functions over a symplectic manifold, with some early works already breaking ground on how this can be integrated with neural-network-based numerical methods \cite{dugan2023q, lin2024real, hahmgenerative}. Dynamics could be implemented by time-stepping the $Q$-function via bracket flows, using recursive updates. Truncating the spherical harmonic expansion offers a natural approximation hierarchy, analogous to bond dimension in TNs or truncation depth in variational circuits. Score-matching techniques may also be applied, enabling the learning of vector fields or Hamiltonians that drive bracket evolution, in direct analogy with how continuous normalising flows learn in generative modelling \cite{chen2018neural, kidger2022neural, heightman2024solving}. Because the $Q$-functions are smooth and real-valued, gradient-based learning over $L^2\big((S^2)^N\big)$ becomes tractable, and physical constraints can be enforced via projection onto the admissible subspace through Eq.~(\ref{eq:physical_subspace}). This enables a new class of neural architectures for quantum dynamics, built on geometrically structured flows of functions over curved space, rather than circuit ans\"atze or tensor contractions.

\section{Conclusions and Outlook}
In this work, we presented a self-contained formulation of many-qubit states and dynamics in their phase-space representation. We first defined the SW kernel, extended it to many bodies, and analysed the signatures of non-classicality in phase-space, defining a Moyal $\star$-product in the process. We then derived the phase-space equivalents of the von-Neumann, imaginary time, and Lindblad equations via the sine and cosine brackets. Next we constructed a unified picture of quantum dynamcis in Hilbert and phase-space, proving that for qubits the Stinespring's theorem can be realised in phase-space via an extended kernel. As a consequence, it is always possible to choose an phase-space over which the SW kernel factorises, even when the original dynamics are from an open system. Finally, we showed an efficient way of calculating many-body correlators with the help derivatives of Moment Generating Functions, opening avenues for automatic differentiation techniques to estimate observables. This formalism represents quantum systems with smooth functions over a domain that has linear scaling in the number of qubits, lifting the curse of dimensionality from the domain (as we saw in Hilbert space) to harmonic support.

With bracket dynamics defined geometrically, new variational principles, control strategies, and approximation schemes can be formulated directly in function space. This perspective opens a computational direction that remains largely unexplored, with strong prospects for deep-learning-based methods as discussed in Sec.~\ref{sec:QML_prospects_discussion}. The structure of $\chi_{\hat{\rho}}^{(s)}$, particularly its logarithm, invites modelling strategies that align naturally with gradient-based learning and variational estimation. These methods offer an alternative to sampling-based expectation value estimation such as Monte-Carlo based integration.

As the first in a series of papers, this work develops a phase-space formalism for many-qubit quantum machine learning. Future work will extend this foundation toward practical modelling, including neural architectures over spin phase-space, normalising-flow-based variational modelling, sampling-based inference, and variational learning schemes in dynamics. In particular, we aim to explore how function-based models can be trained directly using bracket flows, and how geometric priors—such as symmetry, locality, and smoothness—can be embedded into learning over $(S^2)^N$ without reverting to Hilbert space representations. This will lead to applications in tomography, Hamiltonian learning, open-system characterisation, and optimal control. 

\section*{Acknowledgments}

TH acknowledges support from the Government of Spain (Severo Ochoa CEX2019-000910-S, Quantum in Spain, FUNQIP and European Union NextGenerationEU PRTR-C17.I1), the European Union (PASQuanS2.1, 101113690 and Quantera Veriqtas), Fundació Cellex, Fundació Mir-Puig, Generalitat de Catalunya (CERCA program), the ERC AdG CERQUTE and the AXA Chair in Quantum Information Science. 

EJ acknowledges support from the predoctoral program AGAUR-FI ajuts (2025
FI-1 00298) Joan Oró, which is backed by the Secretariat of Universities and
Research of the Department of Research and Universities of the Generalitat
of Catalonia, as well as the European Social Plus Fund.

MP and ML acknowledge support from: European Research Council AdG NOQIA;
MCIN/AEI (PGC2018-0910.13039/501100011033, CEX2019-000910-S/10.13039/501100011033, Plan National FIDEUA PID2019-106901GB-I00, Plan National STAMEENA PID2022-139099NB, I00, project funded by MCIN/AEI/10.13039/501100011033 and by the “European Union NextGenerationEU/PRTR" (PRTR-C17.I1), FPI); QUANTERA DYNAMITE PCI2022-132919, QuantERA II Programme co-funded by European Union’s Horizon 2020 program under Grant Agreement No 101017733; Ministry for Digital Transformation and of Civil Service of the Spanish Government through the QUANTUM ENIA project call - Quantum Spain project, and by the European Union through the Recovery, Transformation and Resilience Plan - NextGenerationEU within the framework of the Digital Spain 2026 Agenda;  Fundació Cellex; Fundació Mir-Puig; Generalitat de Catalunya (European Social Fund FEDER and CERCA program; Barcelona Supercomputing Center MareNostrum (FI-2023-3-0024); Funded by the European Union. Views and opinions expressed are however those of the author(s) only and do not necessarily reflect those of the European Union, European Commission, European Climate, Infrastructure and Environment Executive Agency (CINEA), or any other granting authority. Neither the European Union nor any granting authority can be held responsible for them (HORIZON-CL4-2022-QUANTUM-02-SGA PASQuanS2.1, 101113690, EU Horizon 2020 FET-OPEN OPTOlogic, Grant No 899794, QU-ATTO, 101168628), EU Horizon Europe Program (This project has received funding from the European Union’s Horizon Europe research and innovation program under grant agreement No 101080086 NeQSTGrant Agreement 101080086 — NeQST);
ICFO Internal “QuantumGaudi” project.

\section*{Statements and Declarations}

The authors declare no competing or financial interests.

\begin{itemize}
 
\item Ethics approval and consent to participate: \emph{Not applicable}
\item Consent for publication: \emph{Not applicable}
\item Data availability: \emph{Not applicable}
\item Materials availability: \emph{Not applicable}
\item Code availability: \emph{Not applicable}
\item Author contribution: \emph{Not applicable}
\end{itemize}
 
\bibliography{sn-bibliography}

\begin{appendices}

\section{Quantum Optics in Phase-Space}\label{app:bosonic_phase_space}

In this section, we offer a brief exposition on the representation of bosonic states in phase-space, and their dynamics. This is in order to draw direct analogies to our qubit phase-space construction in later sections. 
\subsection{Bosonic States in Phase-Space}

At the heart of state representation theory in quantum optics is the Stratonovich-Weyl (SW) correspondence. Let \( \mathcal{H} \) be a Hilbert space and \( \Gamma \) a classical phase-space with measure \( d\mu(x) \). The SW correspondence defines a map
\begin{equation}
    \hat{A} \mapsto W_A(x) := \operatorname{Tr}[\hat{A} \, \hat{\Delta}(x)],
\end{equation}
where \( \hat{\Delta}(x) \in \mathcal{L}(\mathcal{H}) \) is the SW kernel. The inverse map is given by
\begin{equation}
    \hat{A} = \int_\Gamma W_A(x) \, \hat{\Delta}(x) \, d\mu(x).
\end{equation}
The kernel must satisfy the following axioms,
\begin{enumerate}
    \item \textbf{Covariance:} \( \hat{\Delta}(g \cdot x) = U(g) \hat{\Delta}(x) U(g)^\dagger \), for a unitary representation \( U \) of a group \( G \).
    \item \textbf{Reality:} \( \hat{\Delta}(x)^\dagger = \hat{\Delta}(x) \), hence \( W_A(x)^* = W_{A^\dagger}(x) \).
    \item \textbf{Traciality:} \( \operatorname{Tr}[\hat{\Delta}(x)] = 1 \), \quad \( \operatorname{Tr}[\hat{\Delta}(x) \hat{\Delta}(y)] = \delta(x - y) \)
    \item \textbf{Resolution of the Indentity} \( \int_\Gamma \hat{\Delta}(x) \, d\mu(x) = \hat{\mathbb{I}} \).
\end{enumerate}
From this perspective, bosonic systems' phase-space representation are merely a consequence of this correspondence over the Heisenberg-Weyl group \cite{klimov2009group}. We may now consider how it is manifest in the Hilbert space of the HW group, $\mathcal{F} = \text{span}(\ket{n_1,\ldots n_N})$. This is simply the Hilbert space of a system of $n$ bosonic modes with creation and annihilation operators, $\hat{a}$, $\hat{a}^{\dagger}$ satisfying
\[
[\hat{a}_j, \hat{a}_k^{\dagger}] = \delta_{jk}, \quad [\hat{a}_j, \hat{a}_k]  = 0 = [\hat{a}_j^{\dagger}, \hat{a}_k^{\dagger}].
\]
The SW correspondence of this algebra establishes
an invertible mapping between states and operators described by these operators, and density functionals on a symplectic manifold, $M = (\mathbb{R}^{2n}, \omega)$ with $\omega = \sum_i dq^i \wedge dp_i$, being the usual symplectic two-form in the basis of quadrature operators,
\begin{equation}
    \hat{q}_j = \frac{1}{\sqrt{2}}(\hat{a}_j + \hat{a}^{\dagger}),\quad \hat{p}_j = \frac{1}{\sqrt{2}i}(\hat{a}_j - \hat{a}_j^{\dagger}),
\end{equation}
satisfying
\begin{equation}
    [\hat{q}_j, \hat{p}_k] = i \delta_{jk},\quad [\hat{q}_j, \hat{q}_k] = 0 = [\hat{p}_j, \hat{p}_k].
\end{equation}
In the HW algebra, the forward-direction of the SW transform can be executed via a trace against a kernel operator \( \hat{\Delta}^{(s)}(x) \):
\begin{equation}
    f^{(s)}_{\hat{O}}(x) = \operatorname{Tr}[\hat{O} \, \hat{\Delta}^{(s)}(x)],
\end{equation}
with $-1 \leq s \leq 1$. The choice of \( \hat{\Delta}^{(s)} \) is determined by the ordering of the field operators constructing $\hat{O}(\hat{a},\hat{a}^{\dagger})$, and thereby the nature of the corresponding phase-space function. 
This is because constructing density functionals over commutative variables means we must pre-specify an ordering of quantum operators if a phase-space map is to be invertible. A canonical choice is the Weyl symbol (symmetric ordering), but the full family of $s$-ordered representations arises by varying \( s \in [-1,1] \), interpolating between normal, symmetric, and anti-normal ordering. This family unifies the Glauber–Sudarshan \( P \)-function, the Wigner function, and the Husimi \( Q \)-function under a common framework known as $s$-\textit{parametrised quasi-probability distributions}.

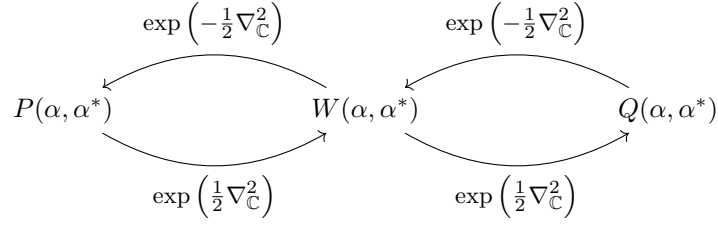
\begin{figure}
\centering
\begin{tikzpicture}

\node (P) at (-4, 0) {$P(\alpha, 
\alpha^*)$};
\node (W) at (0, 0) {$W(\alpha,\alpha^*)$};
\node (Q) at (4, 0) {$Q(\alpha, \alpha^*)$};

\draw [->] (W) to [out=150,in=30] node [above] {\small$\exp\left(-\frac{1}{2}\nabla^2_\mathbb{C}\right)$} (P);
\draw [->] (P) to [out=330,in=210] node[below]{\small$\exp\left(\frac{1}{2}\nabla^2_\mathbb{C}\right)$} (W);
\draw [->] (Q) to [out=150,in=30] node [above] {\small$\exp\left(-\frac{1}{2}\nabla^2_\mathbb{C}\right)$} (W);
\draw [->] (W) to [out=330,in=210] node[below]{\small$\exp\left(\frac{1}{2}\nabla^2_\mathbb{C}\right)$} (Q);

\end{tikzpicture}
\caption{The mapping between $Q$, $W$, and $P$ representations for bosonic systems.}
\label{fig:pwq_map_cv}
\end{figure}

To that end, let \( f^{(s)}_{\hat{\rho}}(x) \) be a quasi-probability distribution arising from an operator ordering determined by Wick's theorem \cite{scully1997quantum}. The index \( s \in [-1,1] \) specifies the ordering:
\begin{itemize}
    \item \( s = +1 \): normal ordering \( \Rightarrow \) Glauber–Sudarshan \( P \)-function, defined as a coherent basis expansion in Fock space, $\mathcal{F}$. Let \( \hat\rho \) be a quantum state on our Fock space. Then the \( P \)-function is defined as
    \begin{equation}
        \hat\rho = \int_{\mathbb{C}^n} P(\bm{\alpha}) \, |\bm{\alpha}\rangle\langle \bm{\alpha}| \, \frac{d^2\bm{\alpha}}{\pi},
    \end{equation}
    where \( \{ |\bm{\alpha}\rangle = \ket{\alpha_1,\ldots,\alpha_N} \} \) are the coherent eigenstates of the annihilation operator satisfying the resolution of the identity:
    \begin{equation}
        \int_{\mathbb{C}^n} |\alpha_j\rangle\langle \alpha_j| \, \frac{d^2\alpha}{\pi} = \hat{\mathbb{I}}_j,
    \end{equation}
    on each individual mode.
    \item \( s = 0 \): symmetric ordering \( \Rightarrow \) Wigner function defined as,
    \begin{equation}
    W(\bm{\alpha}, \bm{\alpha}^*) = \frac{1}{(2\pi)^n} \int_{\mathbb{R}^n} d^n \bm{\alpha} \;
    e^{-i \bm{p} \cdot \bm{\alpha}} \;
    \left\langle \bm{q} + \frac{\bm{\alpha}}{2} \middle| \hat{\hat\rho} \middle| \bm{q} - \frac{\bm{\alpha}}{2} \right\rangle,
    \end{equation}
    where $\bm{q} = \ket{q_1,\ldots,q_N}$ are the eigenstates of the position operators. 
    \item \( s = -1 \): antinormal ordering \( \Rightarrow \) Husimi \( Q \)-function, defined as
    \begin{align}
    Q_{\hat\rho}(\bm{\alpha},\bm{\alpha}^*)=\frac{1}{\pi^N}\bra{\bm{\alpha}}\hat\rho\ket{\bm{\alpha}}
    \end{align}
    for a state $\hat\rho \in\mathcal{F}$ 
\end{itemize}
Each distribution is related by a Gaussian convolution:
\begin{equation}
    f^{(s)}(x) = \frac{2}{\pi(1 - s)} \int_{\mathbb{R}^{2n}} e^{-\frac{2}{1 - s} \|x - y\|^2} f^{(1)}(y)\, dy \quad \text{for } s < 1,
\end{equation}
which is summarised graphically in Fig.\ref{fig:pwq_map_cv}. We note here that the Gaussian convolution formula used to relate the \( f^{(s)} \) representations is well-defined only for \( s < 1 \). As \( s \to 1^- \), the Gaussian kernel becomes sharply peaked and tends toward a delta distribution. Therefore in the limit \( s \to 1 \), this expression does not yield the Glauber–Sudarshan \( P \)-function. Instead, the $P$-function is constructed directly from a coherent basis expansion as defined above (i.e. by resolving the identity).

Expectation values of observables can be computed directly in phase-space via weighted integrals against the corresponding quasi-probability distributions. Let \( \hat{A}(\hat{a},\hat{a}^{\dagger}) \) be an operator with associated $s$-symbol \( f^{(s)}_{\hat{A}}(x) \), and let \( f^{(-s)}_{\hat{\rho}}(x) \) be the quasi-probability representation of the state in the dual ordering. Then the expectation value is given by
\begin{equation}
    \langle \hat{A} \rangle = \operatorname{Tr}[\hat\rho \hat{A}] = \int_{\mathbb{R}^{2n}} f^{(-s)}_{\hat{\rho}}(x) \, f^{(s)}_{\hat{A}}(x) \, d^{2n}x.
\end{equation}
This dual pairing means that observable averages are invariant under the choice of ordering parameter \( s \), provided the state and operator functions are defined with opposite orderings. In the special case where both are expressed in symmetric ordering, the formula becomes
\begin{equation}
    \langle \hat{A} \rangle = \int_{\mathbb{R}^{2n}} W_{\hat\rho}(x) \, W_{\hat{A}}(x) \, d^{2n}x,
\end{equation}
where \( W_{\hat\rho} \) and \( W_{\hat{A}} \) are the Wigner representations of the state and observable, respectively. This formulation mirrors the Hilbert-Schmidt inner product in operator space, but transplanted entirely to the phase-space setting.

Having established how states, observables, and expectation algebra are represented within this framework, we now turn to the question of representing dynamics on the phase-space manifold $\mathbb{R}^{2n}$.

\subsection{Dynamics of Bosonic States in Phase-Space}
We begin with unitary dynamics. Let \( \hat{H}: \mathcal{F} \rightarrow\mathcal{F} \) be a Fock-space Hamiltonian, and let \( \hat\rho(t) \) be the density operator of a quantum state at time \( t \). The time evolution in the Schrödinger picture is governed by the usual Von-Neumann equation,
\begin{equation}
    \frac{d\hat\rho}{dt} = -i [\hat{H}, \hat\rho].
\end{equation}
Under symmetric ordering, the resultant Wigner function \( W_{\hat\rho}(x, t) \), evolves according to the so-called quantum Liouville equation,
\begin{equation}
    \frac{\partial W_{\hat\rho}}{\partial t} = \{\!\!\{\hat H, W_{\hat\rho}\}\!\!\},
\end{equation}
where the Moyal bracket,
\begin{equation}
    \{\!\!\{f, g\}\!\!\} := \frac{1}{i} (f \star g - g \star f),
\end{equation}
is defined via the \(\star\)-product:
\begin{equation}
    (f \star g)(\bm{p},\bm{q}) := f(\bm{p},\bm{q}) \exp\left( \frac{i}{2} \overleftarrow{\partial_i} \omega^{ij} \overrightarrow{\partial_j} \right) g(\bm{p},\bm{q}),
\end{equation}
where 
\begin{equation}
    \omega = \bigoplus_{k = 1}^n \begin{pmatrix}
        0 & 1 \\
        -1 & 0 
    \end{pmatrix}.
\end{equation}
means this quadratic form can be expressed as
\begin{equation}
    \overleftarrow{\partial}_i \, \omega^{ij} \, \overrightarrow{\partial}_j = \sum_k \left( \overleftarrow{\partial}_{q_k} \, \overrightarrow{\partial}_{p_k} - \overleftarrow{\partial}_{p_k} \, \overrightarrow{\partial}_{q_k} \right).
\end{equation}
Here, the left arrow \( \overleftarrow{\partial_i} \) indicates that derivatives act to the left (on \( f \)), and the right arrow \( \overrightarrow{\partial_j} \) means derivatives act to the right (on \( g \)). And the summed (repeated) indices sum over the entire phase-space $\mathbb{R}^{2n}$, i.e. covering both $\bm{p}$ and $\bm{q}$. The bracket reduces to the Poisson bracket in the classical limit.

Dynamics of states can be expressed for any $s$-index representation. Notably, the $Q$-function evolves according to
\begin{equation}
    \frac{\partial Q}{\partial t} (\bm{\alpha}, \bm{\alpha}^*)
= \{\!\!\{H, Q\}\!\!\}(\bm{\alpha}, \bm{\alpha}^*) =\frac{1}{i} \left( H \star_Q Q - Q \star_Q H \right) (\bm{\alpha}, \bm{\alpha}^*),
\end{equation}
\begin{equation}
    \text{with} \quad
(f \star_Q g)(\bm{\alpha}, \bm{\alpha}^*) := f(\bm{\alpha}, \bm{\alpha}^*) \exp\left( \frac{1}{2} \overleftarrow{\partial}_{\alpha_j}  \omega^{ij} \, \overrightarrow{\partial}_{\alpha_j^*} \right) g(\bm{\alpha}, \bm{\alpha}^*),
\end{equation}
with 
\begin{equation}
    \omega = \bigoplus_{k = 1}^n \begin{pmatrix}
        0 & i \\
        -i & 0 
    \end{pmatrix}.
\end{equation}
for this representation. Similarly, the $P$-function evolves as
\begin{equation}
    \frac{\partial P}{\partial t} (\bm{\alpha}, \bm{\alpha}^*) = \{\!\!\{P, H\}\!\!\} (\bm{\alpha}, \bm{\alpha}^*) = 
    = \frac{1}{i} \left( H \star_P P - P \star_P H\right) (\bm{\alpha}, \bm{\alpha}^*),
\end{equation}
\begin{equation}
    \text{with} \quad
(f \star_P g)(\bm{\alpha}, \bm{\alpha}^*) := f(\bm{\alpha}, \bm{\alpha}^*) \exp\left( -\frac{1}{2} \overleftarrow{\partial}_i \, \omega^{ij} \, \overrightarrow{\partial}_j \right) g(\bm{\alpha}, \bm{\alpha}^*),
\end{equation}

with $\omega$ the same as the $Q$-function. This is to be expected since $Q$- and $P$-functions are dual. 

\subsection{Open System Dynamics in Phase-Space}
The phase-space Lindblad equation mirrors the structure of its Hilbert-space counterpart. For example for $s=0$ (Wigner function), we have
\begin{equation}
\begin{split}
    \frac{\partial W(x)}{\partial t}
    &= \{\!\!\{ H_W(x), W(x) \}\!\!\} \\
    &+ \sum_k \bigg(
    L_k(x) \star W(x) \star L_k^*(x) \\
    &- \frac{1}{2} \left[
    L_k^*(x) \star L_k(x) \star W(x)
    + W(x) \star L_k^*(x) \star L_k(x)
    \right]
    \bigg),
\end{split}
\end{equation}

Notice that the sine bracket encodes Hamiltonian flow, while dissipation appears through cyclic compositions of the $\star$-product. Each term preserves the kernel-induced structure of the phase-space algebra. No operator traces remain, there is only only function algebra on phase-space. For $P$ and $Q$ functions, the Lindblad equation takes the same form, but with $\star$ replaced by its $s = \pm 1$ variant — corresponding to normal and anti-normal ordering.

\subsection{Non-classicality indicators in bosonic W and Q functions}
In bosonic quantum systems, quasi-probability distributions such as the Wigner and Husimi $Q$-functions provide valuable phase-space representations that reveal classical or non-classical features of quantum states. 

A central indicator of non-classicality is Wigner function negativity \cite{kenfack2004negativity}. According to Hudson’s theorem \cite{hudson1974wigner}, a pure quantum state has a non-negative Wigner function if and only if it is Gaussian. The negative regions in the $W$ can be interpreted as signature of superposition and wave function interference, phenomena with no classical counterpart.  However, a positive Wigner function does not guarantee classicality in an absolute sense (e.g. highly entangled mixed states can have $W\ge0$ everywhere). So negativity in the Wigner function is a sufficient condition for non classicality as it implies the impossibility of interpreting the state as a classical statistical mixture of pure phases

Within this context, Gaussian states, including coherent, squeezed, and thermal states, are taken as the classical reference for continuous variable systems. Another reason for this choice is that non-gaussianity serves itself as a quantum resource, since Gaussian states and transformations on them generated by Hamiltonians that are inhomogeneous quadratics in the canonical operators are efficient to simulate classically \cite{bartlett2002efficient}.

The Husimi $Q$-function, being strictly positive, lacks negativity as a witness but enables an alternative indicator through the Wehrl entropy, defined by
\begin{equation}
    S_W[\hat\rho] = -\int Q_{\hat{\rho}}(\alpha) \log Q_{\hat{\rho}}(\alpha)\, d^2\alpha.
\end{equation}
Coherent states minimize the Wehrl entropy, and deviations from this minimum reflect non-classical features~\cite{wehrl1979relation, lieb1978proof}. The Wehrl entropy also upper bounds the von Neumann entropy, capturing coarse-grained quantum correlations. The theoretical significance of Wehrl entropy is that it measure how spread out or delocalized a quantum state is in phase-space, relative to the minimum-uncertainty reference of a coherent state.

In terms of the Glauber–Sudarshan $P$-representation, a quantum state is defined to be classical if its $P$ function is positive and well-behaved. This corresponds to a statistical mixture of coherent states and implies that the Wigner or $Q$ function is a convex sum of localized blobs, with no interference fringes or negative regions. 

Beyond negativity, structural properties such as zeros in the $Q$ function or non-factorizability indicate non-classicality thanks to the notion of non-classical depth. The non-classical depth, introduced by Lee~\cite{lee1991measure}, quantifies how much Gaussian smoothing is required to convert a state’s quasi-probability distribution into a well behaved probability distribution. 

\section{Proofs of Propositions} \label{app:proofs}

\begin{proofprop}[Pauli expectation Q function expansion for single qubit]
Substituting the identity
\begin{align}
\ketbra{\theta,\varphi}{\theta,\varphi}=\frac{1}{2}\left[  \hat{\mathbb{I}}+\hat\sigma_x\sin\theta\cos\varphi+\hat\sigma_y\sin\theta\sin\varphi+\hat\sigma_z\cos\theta\right]\label{eq:coherent_ketbra}
\end{align}
into  Eq.~(\ref{eq:spin_q_func}) gives the desired result.
\end{proofprop}

\begin{proofprop}[Pauli expectation P function expansion for single qubit]
The result follows by considering the resolution of the identity
\begin{align}
    \hat{\rho} = \frac{1}{4 \pi}\int_{S^2} \ket{\theta,\varphi}\!\bra{\theta,\varphi} \hat{\rho} \; d \mu(\theta, \varphi),
\end{align}
where $d \mu(\theta, \varphi)$ is the usual Haar measure. If we trace both sides, and invoke the orthonormality of spherical harmonics
\begin{align}
Y_{0,0}=\frac{1}{\sqrt{4\pi}},\quad Y_{1,1}=\sqrt{\frac{3}{4\pi}}\sin\theta\cos\varphi,\quad Y_{1,-1}=\sqrt{\frac{3}{4\pi}}\sin\theta\sin\varphi,\quad Y_{1,0}=\sqrt{\frac{3}{4\pi}}\cos\theta
\end{align}
\end{proofprop}

\begin{proofprop}[Many-body Pauli expectation expansion for Q and P functions]
Similar to the proofs of Prop.~\ref{prop:spin_q_func_exp} and \ref{prop:spin_p_func_exp}, except using tensor products of the coherent states.
\end{proofprop}

\begin{proofprop}[Partial tracing as marginalization]
First, note that since $\langle\hat A\rangle_{\hat\rho^A}=\langle(\hat{\mathbb{I}}\otimes\hat A)\rangle_{\hat\rho^{1A}}$ for all observables $\hat A$ on qubits $2,\cdots,N$, we have by Prop.~\ref{prop:multi_spin_q_func_exp} that
\begin{align}
Q_{\hat\rho^A}(\boldsymbol{\theta}\backslash\{\theta_1\},\boldsymbol{\varphi}\backslash\{\varphi_1\})&=\frac{1}{(4\pi)^{N-1}}\sum_{\hat P_2,\cdots,\hat P_N}\langle \hat P_2\otimes\cdots\otimes \hat P_N\rangle_{\hat\rho^A}\prod_{i=2}^NY_{\hat P_i}(\theta_i,\varphi_i)\nonumber\\
&=\frac{1}{(4\pi)^{N-1}}\sum_{\hat P_2,\cdots,\hat P_N}\langle \hat{\mathbb{I}}\otimes \hat P_2\otimes\cdots\otimes \hat P_N\rangle_{\hat\rho^{1A}}\prod_{i=2}^NY_{\hat P_i}(\theta_i,\varphi_i)
\end{align}
On the other hand, the marginal of $Q_{\hat\rho^{1A}}$ is
\begin{align}
\int_{S^2}d\Omega_1&\,Q_{\hat\rho^{1A}}(\boldsymbol{\theta},\boldsymbol{\varphi})=\frac{1}{(4\pi)^N}\int_{S^2}d\Omega_1 \sum_{\hat P_1,\hat P_2,\cdots,\hat P_N}\langle \hat P_1\otimes \hat P_2\otimes\cdots\otimes \hat P_N\rangle_{\hat\rho^{1A}}\prod_{i=1}^NY_{\hat P_i}(\theta_i,\varphi_i)\nonumber\\
&=\frac{1}{(4\pi)^N} \sum_{\hat P_1,\hat P_2,\cdots,\hat P_N}\langle \hat P_1\otimes \hat P_2\otimes\cdots\otimes \hat P_N\rangle_{\hat\rho^{1A}}\prod_{i=2}^NY_{\hat P_i}(\theta_i,\varphi_i)\int_{S^2}d\Omega_1Y_{\hat P_1}(\theta_1,\varphi_1)\nonumber\\
&=\frac{1}{(4\pi)^{N-1}} \sum_{\hat P_2,\cdots,\hat P_N}\langle \hat{\mathbb{I}}\otimes \hat P_2\otimes\cdots\otimes \hat P_N\rangle_{\hat\rho^{1A}}\prod_{i=2}^NY_{\hat P_i}(\theta_i,\varphi_i)
\end{align}
\end{proofprop}

\begin{proofprop}[Separability and Q-function factorization]
    Forward direction follows from the previous Remark~\ref{rem: product state} and linearity.\\
\end{proofprop}

\begin{proofprop}[Purity in phase-space]

For a single qubit use the single qubit relation $P_{\hat{\rho}}(\Omega) = 3 Q_{\hat{\rho}}(\Omega) - 2$ from Prop.~\ref{prop:spin_q_func_exp} and \ref{prop:spin_p_func_exp} in the phase-space expression of the trace: $\mathrm{tr}[\hat\rho^2] = \int Q_{\hat{\rho}}(\Omega) P_{\hat{\rho}}(\Omega) \, d\mu(\Omega)$. The result follows immediately.
For many-bodies subsystem purity takes the form
\begin{align}
\operatorname{Tr}\left[\hat\rho^2\right]&=\frac{1}{2}\int_{S^2} d\boldsymbol{\Omega}\,\{\!\!\{Q_{\hat{\rho}}(\boldsymbol{\Omega}),Q_{\hat{\rho}}(\boldsymbol{\Omega})\}\!\!\},
\end{align}
since trivially $\hat\rho^2=\frac{1}{2}\{\hat\rho,\hat\rho\}$ maps to the cosine bracket (see Sec.\ref{sec:spin_dynamics} for full discussion on brackets and  $\star$-products).

\end{proofprop}
\begin{proofprop}[Equivalence with operator rank]
The Stratonovich-Weyl correspondence is a monomorphism between bounded operators on $\mathcal{H}_2^{\otimes N}$ and a finite-dimensional space of real-valued functions on $(S^2)^N$. The star product satisfies
\begin{equation}
    f_{\hat{A}} \star W_{\hat{\rho}} = f_{\hat{A} \hat\rho}.
\end{equation}
Thus, the condition $f_{\hat{A}} \star W_{\hat{\rho}} = f_{\hat{A}}$ implies $\hat{A} \hat\rho = \hat{A}$. That is, $\hat{A}$ is a left fixed point of right-multiplication by $\hat\rho$.

This space of operators is isomorphic to the left module of $\hat\rho$, whose dimension equals the rank of $\hat\rho$. Hence the result follows.
\end{proofprop}

\begin{proofprop}[Sine bracket in coordinates]
It suffices to verify that the sine bracket reproduces the $\mathfrak{su}(2)$ commutation relations under the SW correspondence. In particular, the sine bracket $[\![f^{(s)}_{\hat P}, f^{(s)}_{\hat Q}]\!]^{(s)}$ must map to $2f^{(s)}_{[\hat P,\hat Q]}$ for Pauli operators $\hat P, \hat Q \in \{\hat \sigma_x, \hat \sigma_y, \hat \sigma_z\}$. Since the $s$-parametrised SW transform maps each Pauli operator $\hat \sigma_i$ to a smooth function $Q_{\hat \sigma_i}$ on $S^2$, the phase-space version of the commutator is encoded in the action of differential operators $\mathcal{J}_i$ on these functions. 

The operators $\mathcal{J}_{\hat P}$ act as Killing vector fields generating rotations on the sphere, and satisfy the same Lie algebra as the Pauli matrices under commutation: 
\begin{equation}
    [\mathcal{J}_{\hat \sigma_x}, \mathcal{J}_{\hat \sigma_y}] = 2\mathcal{J}_{\hat \sigma_z}, \quad
[\mathcal{J}_{\hat \sigma_y}, \mathcal{J}_{\hat \sigma_z}] = 2\mathcal{J}_{\hat \sigma_x}, \quad
[\mathcal{J}_{\hat \sigma_z}, \mathcal{J}_{\hat \sigma_x}] = 2\mathcal{J}_{\hat \sigma_y}.
\end{equation}
Therefore, verifying the action of each $\mathcal{J}_i$ on the basis functions $Q_{\sigma_j}$ confirms that the bracket correctly implements the $\mathfrak{su}(2)$ structure at the level of quasi-probability functions.

Indeed, we see that
\begin{align}
\mathcal{J}_{\hat{\mathbb{I}}}Q_{\hat{\mathbb{I}}}&=0&\mathcal{J}_{\hat{\mathbb{I}}}Q_{\hat\sigma_x}&=0&\mathcal{J}_{\hat{\mathbb{I}}}Q_{\hat\sigma_y}&=0&\mathcal{J}_{\hat{\mathbb{I}}}Q_{\hat\sigma_z}&=0\\
\mathcal{J}_{\hat\sigma_x}Q_{\hat{\mathbb{I}}}&=0&\mathcal{J}_{\hat\sigma_x}Q_{\hat\sigma_x}&=0&\mathcal{J}_{\hat\sigma_x}Q_{\hat\sigma_y}&=Q_{\hat\sigma_z}&\mathcal{J}_{\hat\sigma_x}Q_{\hat\sigma_z}&=-Q_{\hat\sigma_y}\\
\mathcal{J}_{\hat\sigma_y}Q_{\hat{\mathbb{I}}}&=0&\mathcal{J}_{\hat\sigma_y}Q_{\hat\sigma_x}&=-Q_{\hat\sigma_z}&\mathcal{J}_{\hat\sigma_y}Q_{\hat\sigma_y}&=0&\mathcal{J}_{\hat\sigma_y}Q_{\hat\sigma_z}&=Q_{\hat\sigma_x}\\
\mathcal{J}_{\hat\sigma_z}Q_{\hat{\mathbb{I}}}&=0&\mathcal{J}_{\hat\sigma_z}Q_{\hat\sigma_x}&=Q_{\hat\sigma_y}&\mathcal{J}_{\hat\sigma_z}Q_{\hat\sigma_y}&=-Q_{\hat\sigma_x}&\mathcal{J}_{\hat\sigma_z}Q_{\hat\sigma_z}&=0
\end{align}
which matches the expected algebra:
\begin{equation}
    [\hat \sigma_i, \hat \sigma_j] = 2i\epsilon_{ijk}\hat \sigma_k \quad \Rightarrow \quad [\![Q_{\hat \sigma_i}, Q_{\hat \sigma_j}]\!] = 2\mathcal{J}_{\hat\sigma_i} Q_{\hat \sigma_j} = 2 Q_{\hat \sigma_k}.
\end{equation}
Hence, the proposition follows.
\end{proofprop}

\begin{proofprop}[Cosine bracket in coordinates for Q-function]
Analogously to Prop.~\ref{prop: sine_bracket}, we verify the action of the cosine bracket $\{\!\!\{f^{(s)}_{\hat P}, f^{(s)}_{\hat{\rho}}\}\!\!\}^{(s)}$ on the basis functions $Q_{\hat \sigma_i}$. For $s = -1$, the cosine bracket corresponds to the pointwise Jordan product on the operator level, which maps to the $\mathcal{K}_{\hat P}$ differential operators in phase-space. These act on functions $Q_{\hat \sigma_i}$ to encode the symmetric part of the Moyal product.

Evaluating their action on the basis $\{Q_{\hat{\mathbb{I}}}, Q_{\hat\sigma_x}, Q_{\hat\sigma_y}, Q_{\hat\sigma_z}\}$ gives:
\begin{align}
\mathcal{K}_{\hat{\mathbb{I}}}Q_{\hat{\mathbb{I}}}&=Q_{\hat{\mathbb{I}}}&\mathcal{K}_{\hat{\mathbb{I}}}Q_{\hat\sigma_x}&=Q_{\hat\sigma_x}&\mathcal{K}_{\hat{\mathbb{I}}}Q_{\hat\sigma_y}&=Q_{\hat\sigma_y}&\mathcal{K}_{\hat{\mathbb{I}}}Q_{\hat\sigma_z}&=Q_{\hat\sigma_z}\\
\mathcal{K}_{\hat\sigma_x}Q_{\hat{\mathbb{I}}}&=Q_{\hat\sigma_x}&\mathcal{K}_{\hat\sigma_x}Q_{\hat\sigma_x}&=Q_{\hat{\mathbb{I}}}&\mathcal{K}_{\hat\sigma_x}Q_{\hat\sigma_y}&=0&\mathcal{K}_{\hat\sigma_x}Q_{\hat\sigma_z}&=0\\
\mathcal{K}_{\hat\sigma_y}Q_{\hat{\mathbb{I}}}&=Q_{\hat\sigma_y}&\mathcal{K}_{\hat\sigma_y}Q_{\hat\sigma_x}&=0&\mathcal{K}_{\hat\sigma_y}Q_{\hat\sigma_y}&=Q_{\hat{\mathbb{I}}}&\mathcal{K}_{\hat\sigma_y}Q_{\hat\sigma_z}&=0\\
\mathcal{K}_{\hat\sigma_z}Q_{\hat{\mathbb{I}}}&=Q_{\hat\sigma_z}&\mathcal{K}_{\hat\sigma_z}Q_{\hat\sigma_x}&=0&\mathcal{K}_{\hat\sigma_z}Q_{\hat\sigma_y}&=0&\mathcal{K}_{\hat\sigma_z}Q_{\hat\sigma_z}&=Q_{\hat{\mathbb{I}}}
\end{align}
which is a representation the symmetric part of the Pauli algebra.
\end{proofprop}

\begin{proofprop}[Tensor Compatibility]
This is simply a consequence of the SW kernel factorising locally over the product manifold $(S^2)^N$. Let $\hat{A} \in \mathcal{B}(\mathcal{H}_i)$ and $\hat{B} \in \mathcal{B}(\mathcal{H}_j)$, and assume the SW kernel factorizes as
\begin{equation}
    \hat{\Delta}^{(s)}(\Omega_i, \Omega_j) = \hat{\Delta}^{(s)}(\Omega_i) \otimes \hat{\Delta}^{(s)}(\Omega_j).
\end{equation}
Then, for any product operator $\hat{A} \otimes \hat{B}$, the phase-space symbol factorizes:
\begin{equation}
    Q_{A \otimes B}(\Omega_i, \Omega_j) = \operatorname{Tr}[(\hat{A} \otimes \hat{B}) (\hat{\Delta}^{(s)}(\Omega_i) \otimes \hat{\Delta}^{(s)}(\Omega_j))] = Q_A(\Omega_i) Q_B(\Omega_j).
\end{equation}
Now let $\hat{C}$ be an operator on $\mathcal{H}_i \otimes \mathcal{H}_j$. The operator identity
\begin{equation}
    -i[\hat{A} \otimes \hat{B}, \hat{C}] = \tfrac{1}{2} \left(-i[\hat{A}, \{\hat{B}, \hat{C}\}] + \{\hat{A}, -i[\hat{B}, \hat{C}]\}\right)
\end{equation}
follows from repeated application of the Leibniz rule and the (anti)commutator identities.

Applying the SW map to both sides and using bilinearity of the sine and cosine brackets then yields the claimed identities for $[\![\cdot,\cdot]\!]$ and $\{\!\!\{\cdot,\cdot\}\!\!\}$.
\end{proofprop}

\begin{proofprop}[Unitary-Time Evolution]
Starting from $\dot{\hat\rho} = -i[H,\hat\rho]$, apply the SW map:
\begin{equation}
    \frac{\partial f^{(s)}_{\hat{\rho}}}{\partial t}(\Omega) = \mathrm{Tr}\bigl[\dot{\hat\rho}\, \hat{\Delta}(\Omega)\bigr]
= -i\,\mathrm{Tr}\bigl[[H,\hat\rho]\, \hat{\Delta}(\Omega)\bigr]
= f^{(s)}_{-i[H,\hat\rho]}.
\end{equation}
The time derivative passes through the trace trivially since the Hilbert space is finite-dimensional. By definition of the sine bracket, $[\![f^{(s)}_H, f^{(s)}_{\hat{\rho}}]\!] := f^{(s)}_{-i[H,\hat\rho]}$, we obtain
\begin{equation}
    \frac{\partial f^{(s)}_{\hat{\rho}}}{\partial t} = [\![f^{(s)}_{\hat{H}}, f^{(s)}_{\hat{\rho}}]\!].
\end{equation}
This holds for arbitrary $s$ and extends to many-body systems with product kernels. This equation is equivalent to the operator one because the SW map is injective.
\end{proofprop}
We note that the original proposition was in terms of the $Q$-function (i.e. $s = -1$), but in fact this remains true independently of $s$. We emphasise the $s = -1$ case as this has sine and cosine brackets which are bounded. See discussion in the main text.

\begin{proofprop}[Imaginary-Time Evolution]
Analogous to previous proof starting with $\partial_\tau \hat\rho = -\{\hat{H}, \hat\rho\}$, and using the definition of the cosine bracket.
\end{proofprop}

\begin{proofprop}[Lindblad Evolution]
Starting from 
$\dot{\hat\rho} = -i[\hat{H}, \hat\rho] + \gamma \sum_i \left( \hat L_i \hat\rho \hat L_i^\dagger - \tfrac{1}{2} \{\hat L_i^\dagger \hat L_i, \hat\rho\} \right)
$, the dissipative term can be rewritten as:
\begin{align}
\hat L_i \hat\rho \hat  L_i^\dagger - \tfrac{1}{2} \{\hat L_i^\dagger \hat  L_i, \hat\rho\}  = \frac{1}{4} ([\hat L, \{\hat L^\dagger, \hat\rho\}]] - [\hat L, [\hat L^\dagger, \hat\rho]] - [\hat L^\dagger, \{\hat L, \hat\rho\}] - [\hat L^\dagger, [\hat L, \hat\rho]])
\end{align}

We first check that the proposed identity holds. We start by expanding the right-hand side:
\begin{equation}
    \begin{split}
        [\hat L, \{\hat L^\dagger, \hat\rho\}] &= \hat L \hat L^\dagger \hat\rho + \hat L \hat\rho \hat L^\dagger - \hat L^\dagger \hat\rho \hat L - \hat\rho \hat L^\dagger \hat L, \\
        [\hat L, [\hat L^\dagger, \hat\rho]] &= \hat L \hat L^\dagger \hat\rho - \hat L \hat\rho \hat L^\dagger - \hat L^\dagger \hat\rho \hat L + \hat\rho \hat L^\dagger \hat L, \\
        [\hat L^\dagger, \{\hat L, \hat\rho\}] &= \hat L^\dagger \hat L \hat\rho + \hat L^\dagger \hat\rho \hat L - \hat L \hat\rho \hat L^\dagger - \hat\rho \hat L \hat L^\dagger, \\
        [\hat L^\dagger, [\hat L, \hat\rho]] &= \hat L^\dagger \hat L \hat\rho - \hat L^\dagger \hat\rho \hat L - \hat L \hat\rho \hat L^\dagger + \hat\rho \hat L \hat L^\dagger.
    \end{split}
\end{equation}

Now summing the terms with the correct signs:
\begin{equation}
    \begin{split}
        &\quad\ [\hat L, \{\hat L^\dagger, \hat\rho\}]
- [\hat L, [\hat L^\dagger, \hat\rho]]
- [\hat L^\dagger, \{\hat L, \hat\rho\}]
- [\hat L^\dagger, [\hat L, \hat\rho]] \\
&= (\hat L \hat L^\dagger \hat\rho + \hat L \hat\rho \hat L^\dagger - \hat L^\dagger \hat\rho \hat L - \hat\rho \hat L^\dagger \hat L) \\
&\quad - (\hat L \hat L^\dagger \hat\rho -\hat  L \hat\rho\hat  L^\dagger - \hat L^\dagger \hat\rho \hat L + \hat\rho \hat L^\dagger \hat L) \\
&\quad - (\hat L^\dagger \hat L \hat\rho + \hat L^\dagger \hat\rho\hat  L - \hat L \hat\rho\hat  L^\dagger - \hat\rho \hat L \hat L^\dagger) \\
&\quad - (\hat L^\dagger \hat L \hat\hat \rho - \hat L^\dagger \hat\rho \hat L - \hat L \hat\rho \hat L^\dagger + \hat\rho \hat L \hat L^\dagger) \\
&= 4 \hat L \hat\rho \hat L^\dagger - 2\hat  L^\dagger\hat  L \hat\rho - 2 \hat\rho \hat L^\dagger \hat L.
    \end{split}
\end{equation}
Mapping the commutators, with the corresponding $i$ factor, and anti-commutators to the corresponding sine and cosine brackets yields the result. This expression holds in any finite-dimensional Hilbert space, with equivalence to the operator evolution guaranteed by injectivity of the SW correspondence.
    
\end{proofprop}

\begin{proofprop}[Propagators]
    The definition of the propagator was constructive,
    \begin{align}
    G^{(s)}_t(\boldsymbol{\Omega},\boldsymbol{\Omega}')=(4\pi)^N\text{Tr}\big[\hat \Lambda_t(\hat\Delta^{(-s)}(\boldsymbol{\Omega}'))\hat\Delta^{(s)}(\boldsymbol{\Omega})\big],
    \end{align}
    which we can use to show,
    \begin{align*}
    (\Phi_t f)(\boldsymbol{\Omega})
    &:= f^{(s)}_{\hat\Lambda_t[\,SW_s^{-1}f\,]}(\boldsymbol{\Omega})
    = (4\pi)^N\,\mathrm{Tr}\!\Big[\hat\Lambda_t\!\big(SW_s^{-1}f\big)\,\hat\Delta^{(s)}(\boldsymbol{\Omega})\Big] \\
    &= (4\pi)^N\,\mathrm{Tr}\!\Big[\hat\Lambda_t\!\Big(\int f(\boldsymbol{\Omega}')\,\hat\Delta^{(-s)}(\boldsymbol{\Omega}')\,d\boldsymbol{\Omega}'\Big)\,\hat\Delta^{(s)}(\boldsymbol{\Omega})\Big] \\
    &= \int (4\pi)^N\,\mathrm{Tr}\!\big[\hat\Lambda_t(\hat\Delta^{(-s)}(\boldsymbol{\Omega}'))\,\hat\Delta^{(s)}(\boldsymbol{\Omega})\big]\;f(\boldsymbol{\Omega}')\,d\boldsymbol{\Omega}' = \int G_t^{(s)}(\boldsymbol{\Omega},\boldsymbol{\Omega}')\,f(\boldsymbol{\Omega}')\,d\boldsymbol{\Omega}',
    \end{align*}

\end{proofprop}

\begin{proofprop}[Composition by convolution]
    Let $\Phi_i:=SW_s\circ\hat\Lambda_i\circ SW_s^{-1}$ and let $G_i$ be its kernel,
    \begin{align}
    G_i(\boldsymbol{\Omega},\boldsymbol{\Omega}') = (4\pi)^N\text{Tr}\big[\hat\Lambda_i(\hat\Delta^{(-s)}(\boldsymbol{\Omega}'))\,\hat\Delta^{(s)}(\boldsymbol{\Omega})\big].
    \end{align}
    The kernel of $\Phi_1\circ\Phi_2$ is
    \begin{align}
        G_{1\circ2}(\boldsymbol{\Omega},\boldsymbol{\Omega}'')=(4 \pi)^N\text{Tr}\big[(\hat\Lambda_1\!\circ\!\hat\Lambda_2)\big(\hat\Delta^{(-s)}(\boldsymbol{\Omega}'')\big)\,\hat\Delta^{(s)}(\boldsymbol{\Omega})\big].
    \end{align}
    Using the SW inverse for the operator $\hat\Lambda_2\big(\hat\Delta^{(-s)}(\boldsymbol{\Omega}'')\big)$,
    \begin{align}
        \hat\Lambda_2\big(\hat\Delta^{(-s)}(\boldsymbol{\Omega}'')\big)
        =\int_{(S^2)^N} G_2(\boldsymbol{\Omega}',\boldsymbol{\Omega}'')\,\hat\Delta^{(-s)}(\boldsymbol{\Omega}')\,d\boldsymbol{\Omega}',
    \end{align}
    we obtain
    \begin{align}
    G_{1\circ2}(\boldsymbol{\Omega},\boldsymbol{\Omega}'')
    &=(4 \pi)^N \int G_2(\boldsymbol{\Omega}',\boldsymbol{\Omega}'')
    \,\text{Tr} \big[\hat\Lambda_1(\hat\Delta^{(-s)}(\boldsymbol{\Omega}'))\,\hat\Delta^{(s)}(\boldsymbol{\Omega})\big]\,d\boldsymbol{\Omega}'\\
    &=\int G_1(\boldsymbol{\Omega},\boldsymbol{\Omega}')\,G_2(\boldsymbol{\Omega}',\boldsymbol{\Omega}'')\,d\boldsymbol{\Omega}'
    =(G_1*G_2)(\boldsymbol{\Omega},\boldsymbol{\Omega}'').
    \end{align}

\end{proofprop}

\begin{proofprop}[Kernels' group structure]
    Let $\{\hat \Lambda_t\}_{t\in\mathbb{R}}$ be a one-parameter group on $\mathcal{A}$ with $\hat \Lambda_0=\mathrm{id}$ and $\hat \Lambda_{t}\circ\hat \Lambda_{s}=\hat \Lambda_{t+s}$. Transporting to phase space via $SW_s$ yields $\Phi_t=SW_s\circ\hat \Lambda_t\circ SW_s^{-1}: \mathcal{F}_s \rightarrow\mathcal{F}_s$, with kernels $G^{(s)}_t$ as above, and a direct substitution shows 
    \begin{align}
    (\Phi_t\circ\Phi_s)f
    &=\int G^{(s)}_t(\boldsymbol{\Omega},\boldsymbol{\Omega}')\!\left(\int G^{(s)}_s(\boldsymbol{\Omega}',\boldsymbol{\Omega}'')\,f(\boldsymbol{\Omega}'')\,d\boldsymbol{\Omega}''\right)d\boldsymbol{\Omega}' \nonumber\\
    &=\int \big(G^{(s)}_t*G^{(s)}_s\big)(\boldsymbol{\Omega},\boldsymbol{\Omega}'')\,f(\boldsymbol{\Omega}'')\,d\boldsymbol{\Omega}''.
    \end{align}
    Hence $\Phi_t\circ\Phi_s=\Phi_{t+s}$ (follows from $\{\hat \Lambda_t\}_{t\in\mathbb{R}}$ being a one parameter group and definition of $\Phi_t$) implies
    \begin{align}
    G^{(s)}_{t+q}(\boldsymbol{\Omega},\boldsymbol{\Omega}'')
    =\int_{(S^2)^N} G^{(s)}_{t}(\boldsymbol{\Omega},\boldsymbol{\Omega}')\,G^{(s)}_{q}(\boldsymbol{\Omega}',\boldsymbol{\Omega}'')\,d\boldsymbol{\Omega}'
    \;=\; (G^{(s)}_{t}*G^{(s)}_{q})(\boldsymbol{\Omega},\boldsymbol{\Omega}'').
    \end{align}
    Identity follows from SW orthogonality,
    \begin{align}
    G^{(s)}_{0}(\boldsymbol{\Omega},\boldsymbol{\Omega}')=\delta(\boldsymbol{\Omega},\boldsymbol{\Omega}'),
    \end{align}
    associativity from Fubini's theorem, and inverses from $\hat \Lambda_t^{-1}=\hat \Lambda_{-t}$:
    \begin{align}
    G^{(s)}_{t}*G^{(s)}_{-t}=G^{(s)}_{0}=G^{(s)}_{-t}*G^{(s)}_{t}.
    \end{align}
    Thus $\{G^{(s)}_t\}_{t\in\mathbb{R}}$ forms a group under convolution.
\end{proofprop}

\begin{proofprop}[Kernels' semi-group structure]
    Pushing forward, $\Phi_t = SW_s \circ \hat \Lambda_t \circ SW_s^{-1}$, then the kernels again satisfy the convolution property
    \begin{align}
    G^{(s)}_{0}=\delta,\qquad
    G^{(s)}_{t+s}=G^{(s)}_{t}*G^{(s)}_{s}\quad (t,s\ge 0).
    \end{align}
    However, with no converse inverses in general, this means the convolution algebra defines a semigroup in phase space. 
    
\end{proofprop}

\begin{proofprop}[Dilation in phase-space]
We resolve the identity with the s-parametrised kernel basis and switch the Stratonovich Weyl trace with the extended system resolution of the identity (marginalization). That is:
\begin{equation}
    f_{\hat\rho_A}^{(s)}(\Omega) = \operatorname{Tr}_{A}\left[\operatorname{Tr}_B(\ket{\psi_{AB}}\bra{\psi_{AB}})\right] \hat{\Delta}_A^{(s)}) = \operatorname{Tr}_{AB}\left[\ket{\psi_{AB}}\bra{\psi_{AB}} \hat{\Delta}^{(s)}_A \otimes \frac{\hat{\mathbb{I}}}{d_B}\right],
\end{equation}   
where $d_B$ is the environment's system dimension so that $\dim(AB) \geq 
\dim(A)^2$. Now, since $ \forall s$  $\hat{\mathbb{I}} = \int_{S^2} \hat{\Delta}^{(s)}(\theta, \varphi)\, d\mu(\theta, \varphi)$.
\begin{equation}
\begin{split}
    \operatorname{Tr}_{AB} \bigg( \ket{\psi_{AB}}\bra{\psi_{AB}} \, \hat{\Delta}^{(s)}_A &\otimes \frac{1}{d_B} \int_{\mathcal{M}_B} \hat{\Delta}_B(\theta, \varphi)\, d\mu(\Omega_B) \bigg)  \\
&=\int_{\mathcal{M}_B} \operatorname{Tr}_{AB} \left( \ket{\psi_{AB}}\bra{\psi_{AB}} \, \hat{\Delta}^{(s)}_A \otimes \frac{1}{d_B} \hat{\Delta}_B(\theta, \varphi) \right) d\mu(\Omega_B),
\end{split}
\end{equation}
which gives the desired result.
\end{proofprop}
\begin{proofprop}[Single qubit MGF]
    Notice that for $\omega \in \mathbb{R}^3$ and $n = (\sin \theta \cos \varphi, \sin \theta \sin \varphi, \cos \theta)$ we have
    \begin{equation}
        \frac{\partial}{\partial \omega_{\mu}} e^{\omega \cdot n} = n_{\mu} e^{\omega \cdot n}.
    \end{equation}
    Thus 
    \begin{equation}
        \left. \frac{\partial^k}{\partial{\omega_{\mu_1}} \ldots \partial \omega_{\mu_k}} e^{\omega \cdot n }\right|_{\omega = 0} = \prod_{j = 1}^k n_{\mu_j} e^{\omega \cdot n},
    \end{equation}
    which implies
    \begin{equation}
        \begin{split}
             \left. \frac{\partial^k}{\partial{\omega_{\mu_1}} \ldots \partial \omega_{\mu_k}} \chi_{\hat{\rho}}^{(s)}(\omega)\right|_{\omega = 0} &= \left. \frac{\partial^k}{\partial{\omega_{\mu_1}} \ldots \partial \omega_{\mu_k}} \int_{S^2} d\Omega f^{(s)}_{\hat\rho} (\Omega) e^{\omega \cdot n} \right|_{\omega = 0}\\
        &= \left. \int_{S^2} d\Omega f^{(s)}_{\hat\rho}(\Omega) \frac{\partial^k}{\partial{\omega_{\mu_1}} \ldots \partial \omega_{\mu_k}} e^{\omega \cdot n} \right|_{\omega = 0} \\
        &=\int_{S^2} d\Omega f^{(s)}_{\hat\rho}(\Omega) \prod_{j = 1}^N n_{\mu_j}(\Omega)
        \end{split}
    \end{equation}
    Now inserting Eq.~(\ref{eq:bloch_phase_function}) as our construction for $f^{(s)}_{\hat\rho}(\Omega)$ we see
    \begin{align}
        \left. \frac{\partial^k}{\partial{\omega_{\mu_1}} \ldots \partial \omega_{\mu_k}} \chi_{\hat{\rho}}^{(s)}(\omega)\right|_{\omega = 0} = \frac{1}{4 \pi} \int_{S^2} d \Omega (1 + \lambda(s) r_{\hat\rho} \cdot n(\Omega)) \prod_{j = 1}^N n_j (\Omega).
    \end{align}
    What remains is to simplify the inner products over harmonics. For this, we note that 
    \begin{equation}
        \int_{S^2} d \Omega n_{\mu} = 0, \quad \int_{S^2} d \Omega n_{\mu} n_{\nu} = \frac{4 \pi}{3} \delta_{\mu \nu}
    \end{equation}
    from the spherical harmonic definitions, and that the measure $d\Omega$ is invariant under $SO(3)$ action, which implies
    \begin{equation}
        I_{\mu \nu \gamma} = \int_{S^2} d\Omega \;n_{\mu} n_{\nu} n_{\gamma} = 0\;\forall \mu,\nu,\gamma \in \{x,y,z\},
    \end{equation}
    as the only totally symmetric rank-3 tensor invariant under all rotations is zero. This means the relation is true at first order, where we find
    \begin{equation}
        \begin{split}
            \frac{\partial \chi_{\hat{\rho}}^{(s)}}{\partial \omega_{\mu}} &= \frac{1}{4 \pi} \int_{S^2} d\Omega \; \left(1 + \lambda(s) \sum_i r^i_{\hat\rho} n_i(\Omega)\right) n_{\mu}(\Omega) \\
            &=\frac{\lambda(s)}{4 \pi} \sum_{i=1}^3 r_{\hat\rho}^i \int_{S^2} d \Omega \; n_{i}n_{\mu} = \frac{\lambda(s)}{4 \pi} r_{\mu} = \frac{\lambda(s)}{4 \pi} \langle \hat{\sigma}_{\mu} \rangle,
        \end{split}
    \end{equation}
    since the $\mu$ component of the Bloch vector $r$ is the expectation value $\langle \hat{\sigma}_{\mu} \rangle$. Hence we may write for first order moments that 
    \begin{equation}
        \langle \hat{\sigma}_{\mu} \rangle = \frac{3}{\lambda(s)} \left. \frac{\partial \chi_{\hat{\rho}}^{(s)}}{\partial \omega_{\mu}} \right|_{\omega = 0}.
    \end{equation}
    For higher-order moments, the Pauli algebra guarantees that $\hat{\sigma}^n = \hat{\mathbb{I}}$ for even $n$ and $\hat{\sigma}^n = \hat{\sigma}$ for odd $n$. Thus only first order moments matters. For products like $\hat{\sigma}_{\mu} \hat{\sigma}_{\nu}$ on one qubit, the $\mathfrak{su}(2)$ Lie algebra means that this may always be written as a first order moment. Therefore the first order MGF is sufficient to completely characterise the correlations on a one-qubit system.
\end{proofprop}
\begin{proofprop}[Many qubit MGF]
    For $N$-qubits recall that 
    \begin{equation}
        f^{(s)}(\boldsymbol{\Omega}) = \frac{1}{(4 \pi)^N} \sum_{\hat{P}_1\ldots\hat{P}_N} \langle \hat{P}_1 \otimes \ldots \otimes \hat{P}_N \rangle_{\hat\rho} \prod_{i = 1}^N Y^{(s)}_{\hat{P}_i} (\Omega_i),
    \end{equation}
    where $Y^{(s)}_{\hat{P_i}}(\Omega_i) = \lambda(s) n^{(i)}_{\hat{P}_i}(\Omega_i)$ for $\hat{P}_i \in \{\hat{\sigma}_x^{(i)},\hat{\sigma}_y^{(i)},\hat{\sigma}_z^{(i)}\}$ and  \( n^{(i)}(\Omega_i) = (\sin \theta_i \cos \varphi_i, \sin \theta_i \sin \varphi_i, \cos \theta_i)\). Notice again that 
    \begin{equation}
        \frac{\partial}{\partial \omega_{\mu}^{(i)}} e^{\boldsymbol{\omega} \cdot \boldsymbol{n}} = n^{(i)}_{\mu} e^{\boldsymbol{\omega} \cdot \boldsymbol{n}},
    \end{equation}
    so we see
    \begin{equation}
        \left. \frac{\partial ^N \chi_{\hat{\rho}}^{(s)}}{\partial \omega_{\mu_1}^{(1)}\ldots \partial \omega_{\mu_N}^{(N)}} (\boldsymbol{\omega}) \right|_{\boldsymbol{\omega} = 0} = \int_{(S^2)^N} d \boldsymbol{\Omega} f_{\hat\rho}(\boldsymbol{\Omega}) \prod_{i = 1}^N n^{(i)}_{\mu_i},
    \end{equation}
    so using our definition for $f_{\hat\rho}(\boldsymbol{\Omega}$ we find
    \begin{equation}
        \begin{split}
            \left. \frac{\partial ^N \chi_{\hat{\rho}}^{(s)}}{\partial \omega_{\mu_1}^{(1)}\ldots \partial \omega_{\mu_N}^{(N)}} (\boldsymbol{\omega}) \right|_{\boldsymbol{\omega} = 0} &= \int_{(S^2)^N} d\boldsymbol{\Omega} \frac{1}{(4 \pi)^N}\sum_{\hat{P}_1\ldots\hat{P}_N} \langle \hat{P}_1 \otimes \ldots \otimes \hat{P}_N \rangle_{\hat\rho} \prod_{i = 1}^N Y^{(s)}_{\hat{P}_i} (\Omega_i) n_{\mu_i}^{(i)} \\
            &=\frac{1}{(4 \pi)^N}\sum_{\hat{P}_1\ldots\hat{P}_N} \langle \hat{P}_1 \otimes \ldots \otimes \hat{P}_N \rangle_{\hat\rho} \int_{(S^2)^N} d\boldsymbol{\Omega}\prod_{i = 1}^N Y^{(s)}_{\hat{P}_i} (\Omega_i) n_{\mu_i}^{(i)} \\
        \end{split}
    \end{equation}
    Inspecting the integral, notice that it is in fact just $N$-copes of one-qubit inner products of harmonics. Indeed,
    \begin{equation}
        \begin{split}
             \int_{(S^2)^N} d\boldsymbol{\Omega}\prod_{i = 1}^N Y^{(s)}_{\hat{P}_i} (\Omega_i) n_{\mu_i}^{(i)} &= \prod_{i =1}^N \int_{S^2}d \Omega_i Y^{(s)}_{\hat{P}_i} (\Omega_i) n_{\mu_i}^{(i)} =  \prod_{i =1}^N \lambda(s) \int_{S^2}d\Omega_i n_{\hat{P}_i} n_{\mu_i} \\
        &= \prod_{i=1}^N \lambda(s) \frac{4\pi}{3} \delta_{\hat{P}_i,\mu_i} = \left( \frac{4 \pi}{3} \lambda(s)\right)^N \prod_{i = 1}^N \delta_{\hat{P}_i,\mu_i}.
        \end{split}
    \end{equation}
    If we insert this back into the full expression we find 
    \begin{equation}
        \begin{split}
             \left. \frac{\partial ^N \chi_{\hat{\rho}}^{(s)}}{\partial \omega_{\mu_1}^{(1)}\ldots \partial \omega_{\mu_N}^{(N)}} (\boldsymbol{\omega}) \right|_{\boldsymbol{\omega} = 0} &= \frac{1}{(4 \pi)^N}\sum_{\hat{P}_1\ldots\hat{P}_N} \langle \hat{P}_1 \otimes \ldots \otimes \hat{P}_N \rangle_{\hat\rho}  \left( \frac{4 \pi}{3} \lambda(s)\right)^N \prod_{i = 1}^N \delta_{\hat{P}_i,\mu_i} \\
             &= \left( \frac{\lambda(s)}{3} \right)^N \langle \hat{P}_1 \otimes \ldots \otimes \hat{P}_N \rangle_{\hat\rho}.
            \end{split}
    \end{equation}
    Hence we have
    \begin{equation}
        \langle \hat{P}_1 \otimes \ldots \otimes \hat{P}_N \rangle_{\hat\rho} = \left( \frac{3}{\lambda(s)} \right)  \left. \frac{\partial ^N \chi_{\hat{\rho}}^{(s)}}{\partial \omega_{\mu_1}^{(1)}\ldots \partial \omega_{\mu_N}^{(N)}} (\boldsymbol{\omega}) \right|_{\boldsymbol{\omega} = 0}
    \end{equation}
    as required. We note like above that this is true for Pauli strings which are first-order in each subspace, and that these are the moments that completely characterise a many-qubit system since powers of local operators collapse to the identity or first order moments, as do local products.
\end{proofprop}

\begin{prooflemma}
Let \( \hat{A} \in \mathcal{B}(\mathcal{H}) \) be a Hermitian operator and \( \hat{P} \in \mathfrak{su}(2) \). Consider the unitary conjugation
\begin{equation}
    \hat{A}_\varepsilon = e^{-i\varepsilon \hat{P}/2} \hat{A} e^{i\varepsilon \hat{P}/2}.
\end{equation}
By the Baker-Campbell-Hausdorff formula, we have:
\begin{equation}
    \hat{A}_\varepsilon = \hat{A} - i\frac{\varepsilon}{2} [\hat{P}, \hat{A}] + \mathcal{O}(\varepsilon^2).
\end{equation}
Applying the Stratonovich-Weyl (SW) correspondence and using linearity, we obtain:
\begin{equation}
    f^{(s)}_{\hat{A}_\varepsilon}(\Omega) = f^{(s)}_{\hat{A}}(\Omega) - i\frac{\varepsilon}{2} f^{(s)}_{[\hat{P},\hat{A}]}(\Omega) + \mathcal{O}(\varepsilon^2).
\end{equation}
On the other hand, using the covariance property of the SW kernel,
\begin{equation}
    \hat{U} \hat{\Delta}^{(s)}(\Omega) \hat{U}^\dagger = \hat{\Delta}^{(s)}(R \cdot \Omega), \quad \forall\, \hat{U} \in \mathrm{SU}(2),
\end{equation}
we have:
\begin{equation}
    f^{(s)}_{\hat{A}_\varepsilon}(\Omega) = \mathrm{Tr}[\hat{A} \, \Delta^{(s)}(R_{-\varepsilon} \cdot \Omega)] = f^{(s)}_{\hat{A}}(R_{-\varepsilon} \cdot \Omega),
\end{equation}
where \( R_{-\varepsilon} \cdot \Omega \) is the infinitesimal rotation of \( \Omega \in S^2 \) generated by \( -\varepsilon \hat{P} \in \mathfrak{su}(2) \).

Thus, Taylor expanding:
\begin{equation}
    f^{(s)}_{\hat{A}_\varepsilon}(\Omega) = f^{(s)}_{\hat{A}}(\Omega) - \varepsilon \, \mathcal{J}_{\hat{P}} f^{(s)}_{\hat{A}}(\Omega) + \mathcal{O}(\varepsilon^2),
\end{equation}
where \( \mathcal{J}_{\hat{P}} \) is the Killing vector field on \( S^2 \) generated by \( \hat{P} \).

Comparing both expansions, we get:
\begin{equation}
    - i\frac{\varepsilon}{2} f^{(s)}_{[\hat{P}, \hat{A}]}(\Omega) = -\varepsilon \, \mathcal{J}_{\hat{P}} f^{(s)}_{\hat{A}}(\Omega),
\end{equation}
which yields:
\begin{equation}
    f^{(s)}_{[\hat{P},\hat{A}]}(\Omega) = 2i \mathcal{J}_{\hat{P}} f^{(s)}_{\hat{A}}(\Omega).
\end{equation}
Therefore, the action of the commutator with \( P \in \mathfrak{su}(2) \) is realised in phase-space as the Lie derivative along the vector field \( \mathcal{J}_P \). Since this construction depends only on the covariance of the kernel and the group action, it holds independently of the \( s \)-parametrization. Therefore, the sine bracket is independent of s.
\end{prooflemma}

\end{appendices}

\end{document}